# Protein bioelectronics: a review of what we do and do not know

Christopher D. Bostick,[#1,2] Sabyasachi Mukhopadhyay,[#3‡] Israel Pecht,[3*] Mordechai Sheves,[3*] David Cahen,[3*] David Lederman*[4]

[1] *Dept. of Pharmaceutical Sciences, West Virginia University, Morgantown, WV 26506, USA*

[2] *Inst. for Genomic Medicine, Columbia University Medical Center, New York, NY 10032, USA*

[3] *Departments of Materials & Interfaces, of Organic Chemistry and of Immunology, Weizmann Institute of Science, Rehovot, Israel 76100*

[4] *Department of Physics, University of California, Santa Cruz, CA 95060, USA*

[#] These authors contributed equally to the work

* Corresponding authors, email addresses: david.cahen@weizmann.ac.il, dlederma@ucsc.edu, mudi.sheves@weizmann.ac.il, israel.pecht@weizmann.ac.il

## Abstract

We review the status of protein-based molecular electronics. First, we define and discuss fundamental concepts of electron transfer and transport in and across proteins and proposed mechanisms for these processes. We then describe the immobilization of proteins to solid-state surfaces in both nanoscale and macroscopic approaches, and highlight how different methodologies can alter protein electronic properties. Because immobilizing proteins while retaining biological activity is crucial to the successful development of bioelectronic devices, we discuss this process at length. We briefly discuss computational predictions and their connection to experimental results. We then summarize how the biological activity of immobilized proteins is beneficial for bioelectronic devices, and how conductance measurements can shed light on protein properties. Finally, we consider how the research to date could influence the development of future bioelectronic devices.

---

‡ Current affiliation - Department of Physics, School of Engineering & Applied Sciences, SRM University-AP, Amaravati, Andhra Pradesh, India - 522502

## Glossary:

AFM: Atomic force microscopy
apo: protein without removed cofactor
CP-AFM: Conductive Probe Atomic Force Microscopy
CD: Circular Dichroism
Cofactor: non-peptide organic or inorganic group or ion in protein, necessary for protein's
biological activity; can have/ be aromatic rings, conjugated chains, metal ions
CPR: Cytochrome P450 Reductase
CV: Cyclic Voltammetry
CYP2C9: cytochrome P450 2C9
EC-STM: Electrochemical Scanning Tunneling Microscopy
EDC: 1-Ethyl-3-[3-dimethylaminopropyl]carbodiimide Hydrochloride
ET: Electron Transfer
ETp: Electron Transport
FAD: Flavine Adenine Dinucleotide
FMN: Flavin Mononucleotide
FR: Flickering Resonance
GCE: Glassy Carbon Electrodes
$H_{ab}$: interaction Hamiltonian
Holo: Fully intact protein, including cofactor
HSA Human Serum Albumin
$K_a$: Binding Association on rate
$K_D$: Binding Dissociation Constant
$K_d$: Binding Dissociation off rate
$k_{ET}$: Electron transfer rate
LbL: Layer-by-Layer
LOFO: Lift-Off Float-On
Mb Myoglobin protein
MUA: 11-mercaptoundecanoic acid
NADPH: Nicotinamide Adenine Dinucleotide Phosphate
NHE: normal hydrogen electrode
NHS: N-hydroxysuccinimide
P450: Cytochrome P450
PDMS Polydimethylsiloxane
PFV: Protein Film Voltammetry
PG: Pyrolytic Graphite
Prosthetic group: tightly bound cofactor
PSI: Photosynthetic Protein I
QCM: Quartz Crystal Microbalance
RCI: Photosynthetic Reaction Center I
SAM: Self Assembled Monolayer
SCE: Saturated calomel electrode
SE: Superexchange

SERRS: Surface-Enhanced Resonance Raman Spectroscopy
SPM Scanning Probe Microscopy
SPR: Surface Plasmon Resonance
STM: Scanning tunneling Microscopy
STS: Scanning Tunneling Spectroscopy
turnover rate: rate of protein biological activity for small molecule transformation (exp. Oxidation or reduction)
UV-Vis: Ultraviolet-Visible Spectroscopy
YCC: yeast cytochrome c
WT Wild Type, i.e., natural form of the protein
β: Distance decay factor in tunneling transport [$\ln\left(\frac{I}{I_0}\right) \propto -\beta d$ ; $d$ is distance]
γCPR: Yeast Cytochrome P450 Reductase
ϵ: Optical extinction coefficient

# 1. Introduction

Recent advances in the development of biocompatible electronics have been motivated by applications where interfacing with sensory function in humans (e.g., artificial retinas) is desired, some of which can have direct applications for improving human health (e.g., *in situ*, real time monitoring of glucose or drug concentration in the blood). A fundamental understanding of the properties of biomaterials at the nanoscale is essential for progress towards interfacing with sub-cellular organelles or other nanoscale structures present in living systems. Moreover, because many biologically relevant reactions occur at substrate surfaces and interfaces, exploring interactions between individual biomolecules and various substrate surfaces is of fundamental importance.

Proteins have evolved over billions of years into structures capable of precise reactions, including highly specific substrate recognition, analyte binding, and facile and directional electron tunneling.[1] In addition, proteins are essential to many chemical processes essential to life, such as photosynthesis, respiration, water oxidation, molecular oxygen reduction, and nitrogen fixation.[2–7] Due to their ability to catalyze a vast number of reactions, there has been much interest in harnessing their abilities. One application is in bio-catalysis in synthetic organic chemistry. The high chemo-, regio-, and stereo-selectivity of enzymes negates the need for protecting groups, and thus reduces synthetic steps and simplifies work flow.[8] As such, the pharmaceutical industry is interested in proteins in bioreactors (vessels in which chemical reactions are carried out by biological components) for production of fine chemicals [9–13] and to filter out toxic or hazardous substances by naturally occurring organisms.[14] The high specificity of proteins also makes them ideal candidates for use as sensors. In many sensor applications, proteins are immobilized onto a solid support and act as transducers of optical[15,16] or electronic[17] signals that correlate with specific protein-ligand interactions. Interfacing with solid state electronics has demonstrated that this approach allows biosensing at low voltages with high sensitivities and low detection limits, and offers long-term stability for measurements.[18–20] Protein-based biosensors are currently being developed for use in drug monitoring,[21] environmental toxin detection,[22,23] and early disease detection.[24,25]

Using single organic molecules as electronic components[26] has been suggested as a possible solution to the physical limitations of micro- and nano-scale electronic development. Proteins could provide a way to realize this, because they are small (nm-scale) and capable of undergoing chemo-mechanical, electromechanical, opto-mechanical, and optoelectronic processes.[27] Proteins have been studied for use in the design of bio-electronic devices, including field effect transistors for sensing,[28–32] bio-molecular transistors for data storage,[33] bio-molecular circuits,[34] bio-hybrid solar cells,[35] bio-computers,[36] and enzymatic biofuel cells.[37–39] Biomedical applications exist in devices such as implantable sensors for monitoring chemicals[40,41] and diagnostics,[42] as well as the creation of artificial retinas[43] and noses.[44] There has been extensive research done on electron transfer in proteins[45–47] especially in well-characterized model systems such as cytochrome *c,*[48] plastocyanin[49], and azurin.[50–53] Other studies have explored the solid state electronic transport properties of proteins whose mechanisms of action rely on electron transfer, ET.[54,55]

Several books,[28,54–59] recent special issues of journals (from which we cite results published therein[51,60]), and reviews[51,61,1,47,62,63] have appeared over the years, going back to the early[64] and late 1990s (1st edn. of reference 55). All deal with various aspects of the topics considered here and, taken together, they reflect to some extent the evolution of the idea of protein bioelectronics. Recent interest in possible applications of new protein electronics have laid bare practical problems and stimulated interest in their fundamental scientific issues.

In parallel, the area of ET has remained very active, and we rely heavily in this review on our present understanding of ET to discuss what can be learned from protein bioelectronics devices. We therefore start with a presentation of fundamental concepts related to the electronic properties of proteins as they relate to ET and electron transport, ETp, (cf. next section for an explanation of the difference between ET and ETp) and compare the two processes. Because until now all ETp measurements involve immobilizing proteins on an electronically conducting solid surface used as an electrode, we review the current state of the field regarding protein immobilization techniques, including characterization of the resulting biomolecular films. Subsequently, we discuss the results of measurements on immobilized proteins and their impact on our fundamental understanding of protein electronics. Finally, the implications of the research to date and their possible implications for future protein bioelectronic devices are noted and reviewed.

# 2 Electron Transfer (ET) and Electron Transport (ETp) in Proteins

Electron flow through a protein molecule involves intramolecular charge transport and electron exchange with the surroundings.[65] These two distinct, and widely studied processes cannot be unambiguously separated since the interface of a molecule with an adjacent electrode or ionic solution can have a profound influence on the molecular properties, and hence on the molecule's electrical conductance.[66] The process of directed electron motion, or electron flow, through molecular structures is termed electron transport or electron transfer, depending on the environment in which it occurs. In the discussion below, we denote as *electron transfer (ET)* the electron flow process with all or part of the protein in direct contact with an electrolyte, which is ionically conducting and which can function as an electron sink or source via a redox process. The electrolyte also provides charges, which screen the change in electrical potential due to the electron flow. Electron flow in which an electrolyte is absent, or does not participate in the electron flow process, with electrodes that are not ion conductors, is termed *electron transport (ETp)*. ETp is usually measured in a solid state configuration.[67] In ETp normally there is no possibility for charge balance via ions from an electrolyte, while ET involves a surrounding medium with mobile ions which can provide such charge balance.

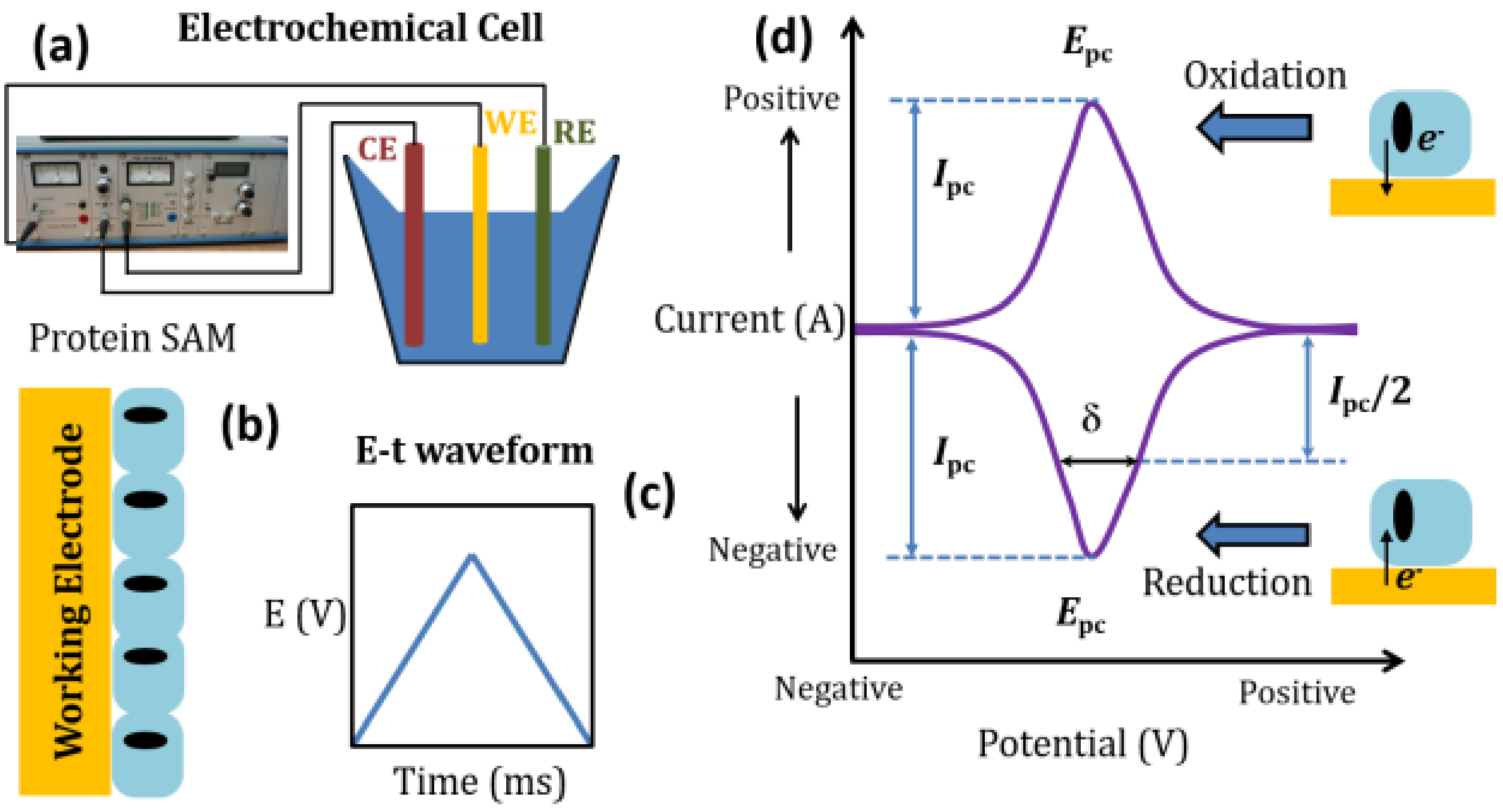

Figure 1: (a) Schematic experimental setup for electrochemical charge transfer studies across self-assembled monolayers of redox-active proteins on conducting electrodes as shown in (b). (c) Time varying electric field applied between counter (CE) and working (WE) electrode with respect to reference electrode (RE). (d) Electrochemical current profile during oxidation and reduction process at the surface of working electrode (details in reference[45,50,60,68] ).

ET consists of electrons moving from specific electronically localized donors to acceptors, which can be domains in the protein, redox agents, or one electrode (in an electrochemical cell) in contact with the protein. ET is driven by a difference in redox potential between donor and acceptor.[48,69–71] ET *within* proteins occurs between donor and acceptor sites. In model systems, one of these is chemically conjugated to the protein, located at a known position, which therefore allows studying ET between specific positions of the protein.

If the proteins are immobilized on an electrode, ET occurs between the electrode and a redox-active protein cofactor, and between the cofactor and the protein surface exposed to the redox electrolyte used for the experiment (Figure 1). Thus, in this case, there is a transition from electronic to ionic charge transport, as is also the case in nature. Within the protein, charge is transferred by electrons, even if a redox center is involved, as that redox-active species does not itself move within the protein. However, charge transfer out of the protein requires reduction /oxidation of a redox-active species in the electrolyte that contacts the protein. This means that the redox-active species is now ionized differently than before, and thus the charge will flow with this ionic species until the next step in the overall process, as is the case in electron transport chains (discussed further below). The change of charge state of the redox-active species in the electrolyte is accompanied by a change in the protein's electrical charge state. This change induces electrical screening to reduce the electrostatic energy barrier for the subsequent steps. In nature and in an electrochemical experiment, this can occur because of the surrounding electrolyte. Naturally, the polarizability of the protein itself also affects the ET process.

Multiple ET processes occur in proteins, such as in respiration or photosynthetic complexes, that involve *electron transport chains*.[72] Proton-coupled electron transport (PCET) can also occur, which is especially prevalent at metal cofactors that activate carbon, oxygen, nitrogen, and sulphur atoms of enzyme substrate molecules.[73] Protein cofactors are organic or inorganic components, with conjugated chains, or metal ions that assist the protein's biological activity. As noted above, redox reactions with biomolecules such as proteins can be studied with electrochemical experiments as shown in Figure 1, i.e., with one electronically conducting, ionically blocking contact (the electrode) and one ionically conducting, electronically blocking contact (the redox electrolyte). ETp, however, requires electronic conduction across proteins set between two electronically conducting, ionically blocking electrodes; it is defined as the flow of electrons across/through the protein that is contacted by these electrodes (Figure 2).[74]

Electric charges in biological systems are generally transported by ions, and thus in nature ET is always coupled to ionic transport where the redox processes serve as donor (Red→Ox) and acceptor (Ox→ Red) electrical "contacts". Thus, from a device perspective, there is no *a priori* reason for electrons involved in (solid-state) ETp to use the same conduction mechanism as those involved in ET processes. Nevertheless, there is experimental evidence that a protein's ETp characteristics are correlated with its ET and redox potential,[75,76] and therefore it is likely that there is a fundamental connection between the ET and ETp processes and mechanisms.

From an experimental point of view (Figure 2), measuring ETp across bio-molecules in dry, solid state condition is fundamentally different from measuring ET. In ETp, the absence of a liquid electrolyte (with its above-noted ability for electrical screening) will normally drive electrons from source to drain electrode, without intra-protein redox process, via some electron transport mechanism (to be discussed later in this review).

ETp is measured on a single protein or on ensembles of proteins (such as monolayers) between the two electronically conducting electrodes (cf. ref.[77] for a discussion of the differences between molecular junctions with single and many molecules). While this type of junction can be thought of as a donor-bridge-acceptor junction, the driving force of the transport is the electrical potential difference between the electrodes (if the electrodes are made of the same material[78]) rather than a difference in chemical potential of the electrons between two regions. In solution experiments, the chemical potential differences of the redox-active ions need to be included.[47]

For practical purposes, the redox potential can be identified with the electrochemical potential of the electron, $\tilde{\mu}_e$ ($\simeq E_F$, Fermi level),[79–83] where $\tilde{\mu}_e = \mu_e + q\phi$, the sum of the contribution of the electrical potential, $\phi$, ($q$ is the electron charge) and that of the electron's chemical potential, $\mu_e = \mu_{0,e} + kT\ ln\ n_e$. Here $\mu_{0,e}$ is the standard chemical potential of the electron, $kT$ is the thermal energy, and $n_e$ is the electron concentration, defined by the difference in activities of the reduced and the oxidized form of the redox species. The latter can be identified as the difference in chemical potential of the (mostly ionic) reduced and oxidized species, and it can drive ET, but not ETp (with electrodes of the same materials, i.e., with the same electron chemical potential).

An additional important difference between ET and ETp is the fact that ET occurs in an electrolyte solution, which allows for electrostatic screening around (parts of) the protein. This is what makes the change in charge of the protein, which otherwise would carry a very high electrostatic energy price, possible. Except for rare cases, no such change in charge state occurs in ETp and electroneutrality is preserved. These differences allow the characterization of ETp distances not considered possible in biological ET. Gray, Winkler, Dutton, and co-workers have estimated the upper limit of D-A separation distances in biological ET (pure quantum mechanical tunneling) to be ≤20 Å and ≤14 Å, respectively.[61,84] However, recent ETp measurements across proteins yield measurable currents over electrode separations of up to ~100 Å.[85]

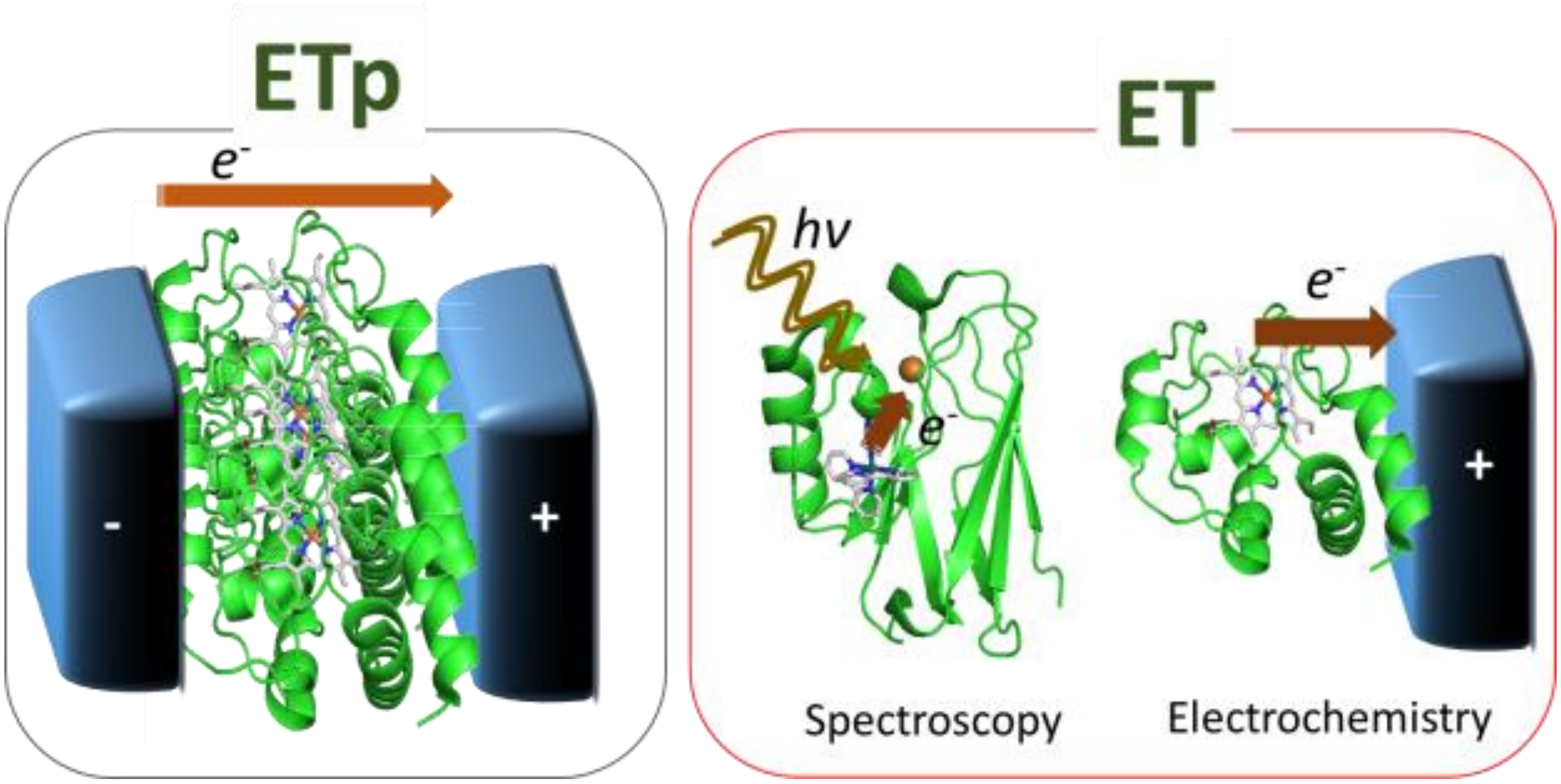

Figure 2: Diagram representing solid state electron transport (ETp) and electron transfer (ET) measurements using spectroscopy and electrochemistry (see introduction to section 2 for more complete distinction between ET and ETp). The very important counter charges around the protein are not shown in the ET configurations.

Cofactors are responsible for the functional properties of proteins because they enable redox activity, including the metal center of metalloproteins, such as $Cu^{2+}$ in the blue copper proteins, $Fe^{3+}$ in the iron-sulfur proteins and cytochromes, or flavins found in flavoproteins.[64] Some redox-inactive proteins, such as the light-driven transmembrane proteins bacteriorhodopsin and halorhodopsin, which contain a retinal cofactor, also support ETp, which, remarkably, is at least as efficient as that of redox-active proteins.

In measurements of the differences in ETp observed between the holo- and apo-forms of proteins, their electrical conduction properties were shown to depend strongly on having the cofactors. Such experiments have been performed with the Cu redox protein azurin,[1,53] the heme-redox protein cytochrome C,[86] the $O_2$-binding protein myoglobin,[32,87] and ferritin.[88] Similarly, changes in transport mechanisms have been deduced from temperature-dependent ETp studies on the proton or $Cl^-$ pumping photoactive membrane proteins, bacteriorhodopsin, bR,[89] or halorhodopsin, phR,[90] and their derivatives lacking retinal, for bR, or retinal and/or carotenoid for phR.

As noted above, multiple ET processes, often involving several proteins and other redox molecules, carry electron transfer over relatively large distances. In the respiratory[91,92] and photosynthetic chains (> 50 Å),[93–96], ET occurs sequentially across several proteins (inter-protein, or protein-protein ET[97,98]); in hydrogenases (FeS clusters, ~ 52 Å),[99] carbon monoxide-dehydrogenases [100,101] and other enzymes,[102] the processes are within one protein complex. This complicated and diverse behavior has made it clear that it is important to develop a theoretical understanding of biological ET and its possible implications to solid-state ETp.

In the following sub-sections, we review the mechanisms that have been developed to explain ET essentially in terms of the Marcus model and various mechanisms that rely on it. Then we briefly discuss ETp in terms of the Landauer one-dimensional conduction model, which has been applied to molecular electronics, while keeping in mind that once electrons enter the protein, ET-like mechanisms may play an important role. ETp is discussed in more detail in terms of specific experiments in Section 4. We end this section by discussing the effects of protein structure and protein immobilization on ET and ETp.

## 2.1 Electron Transfer

In 1956, Marcus presented a model describing ET reactions from a donor to an acceptor that is close to it.[103,104] The Marcus theory was originally developed to explain outer sphere ET, where participating redox centers are not linked by any bridge, and electrons "hop" from the reducing center to the acceptor. It has since been extended to cover inner sphere ET, where two redox centers are linked covalently during transfer, and heterogeneous ET, where an electron moves between a biomolecule (protein, small molecule, cofactor, etc.) and an electrical contact. In the Marcus description, nuclear motions of the reactant and the surrounding environment are approximated by a simple harmonic oscillator potential along a reaction coordinate (Figure 3).[105,106] The left parabola in Figure 3A represents the Gibbs free energy surfaces for the nuclear motion of the reactants prior to ET, and the right parabola represents this free energy for the nuclear motion of the products after ET. The two driving parameters for the ET reaction to overcome the activation free-energy barrier ($-\Delta G^{\ddagger}$) are the driving force, characterized by the Gibbs free energy of activation ($-\Delta G^{o}$), determined from the difference in oxidation potentials of the donor and acceptor, and the reorganization energy (λ) needed for the nuclear rearrangements that accompany ET.[107] The rate of electron transfer, $k_{ET}$, depends on -ΔG° relative to λ. The basic expression for $k_{ET}$ in the Marcus theory is

$$k_{ET} = \frac{2\pi}{\hbar} H_{AD}^2 \frac{1}{\sqrt{4\pi\lambda k_B T}} \exp\left(\frac{-(\lambda+\Delta G^o)^2}{4\lambda k_B T}\right) \quad (1)$$

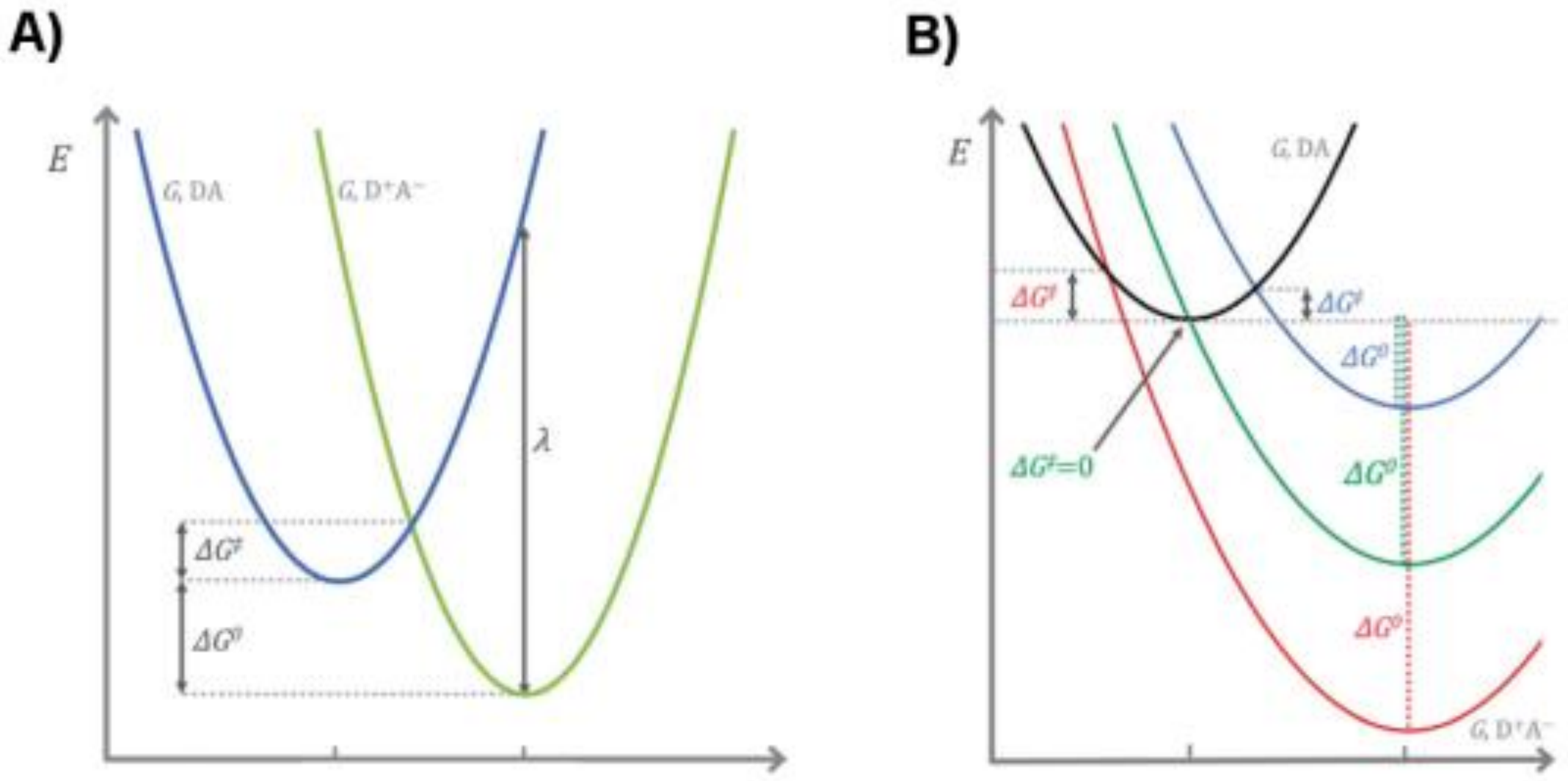

Figure 3. (A) Schematic diagram of the Marcus model for electron transfer, showing the Gibbs free energy (y-axis) surface for the nuclear motion along a reaction coordinate (X-axis) of the donor and acceptor in the initial state, prior to ET (G, DA in blue), and post-ET from the donor to the acceptor (G, $D^+A^-$, green). ΔG‡, $\Delta G^0$, and λ are the activation energy, the free energy driving force, and the outer shell reorganization energy of the ET event, respectively. B) A diagram of the Gibbs free energy surface as in A) showing the effect of increasing $\Delta G^0$ on the ΔG‡. As $\Delta G^0$ increases, a decrease in ΔG‡ is seen until ΔG‡=0 (green curve). After this point increases in $\Delta G^0$ (from the green to red curve) lead to decreases in ΔG‡ as energy must be dissipated for an ET event, which is known as the inverted region. Note that λ does not change for the different scenarios in B), because the nuclear coordinate (X-axis) of the post-ET state does not change. Adapted with permission from reference [106]

where $H_{AD}$ is the quantum mechanical electronic coupling between the initial and final (donor and acceptor) states, $k_B$ is the Boltzmann constant, and $T$ is the temperature. A maximum, or optimal ET rate occurs for activation-less ET (-ΔG° = λ) that decreases with decreasing driving force. The Marcus theory also predicts an inverted region (-ΔG° > λ) where energy must be dissipated to allow ET and where the rate decreases with increasing driving force. According to the semi-classical Marcus' theory, electron transfer takes place when a thermal fluctuation of the solvent or of the nuclear coordinate of the donor and acceptor shifts the donor and acceptor electronic states into resonance. The quantity $H_{AD}$ in eq. (1) originates from Fermi's Golden Rule. $H_{AD}$ contains the interaction overlap integral (matrix element) between the donor (reactant) and acceptor (product) electronic states and the density of states, and thus depends on the distance between the states (cf. also eqn. (2) below).[72,108] The interaction between the donor and acceptor in a polarizable solvent is known as the outer-sphere interaction. When electron transfer occurs within molecules where it causes changes in bond lengths and local symmetries, a similar result is obtained, although the origin of the interaction energy is more complicated. This result related to the so-called inner-sphere interactions was developed by N. Hush,[109], and the joint theory of the inner- and outer-sphere interactions is sometimes called the Marcus-Hush theory (section 4).

Over the years, several theoretical frameworks and computational tools have been developed to explain the biological electron transfer process.[110] In particular, there has been much interest in understanding how the protein structure determines the protein redox function, the implication being that enhanced comprehension could allow for control and tailored biological electron transfer by targeted mutants. The field has evolved significantly since the early ideas of bridge-mediated electron tunneling, first proposed by Halpern et al. in the early 1960s, and the square barrier-tunneling model that Hopfield proposed in the 1970s. [108,111]

**Table – 1: Summary of experimentally obtained ET rate constant for different metalloproteins (see text for discussion)[a]**

| Protein | Electrochemical rate constant (ET) $k_{ET}$ ($s^{-1}$) | References | Spectroscopy rate constant (ET) $k_{ET}$($s^{-1}$) | References |
|---|---|---|---|---|
| **Cyt-C** | 0.2[(b)] | 112 | $9.4\times10^{5}$[(b)] | 61 |
| | 0.63[(c)] | 112 | $2.7\times10^{6}$[(c)] | 61 |
| **Myoglobin** | 60 | 113,114 | $2.3\times10^{6}$ | 61 |

[a] *The difference between the two entries for Cyt-C result from experiments with different donor-acceptor separations (different mutations), ~14 Å [(b)] and ~17 Å [(c)].*

Many experimental studies have been carried out to understand how long-range biological ET (over, at least, several direct bond distances) is accomplished.[45,46,61,84,103,115] ET rate has been determined in cytochromes from line broadening, magnetization transfer, or relaxation measurements in NMR.[116,117] In addition, $k_{\rm ET}$ has been extracted from electrochemistry of a monolayer of a protein (azurin, cytochrome C, myoglobin and other metalloproteins) attached to a working electrode (Figure 1d).[118] Until now, most conclusive and systematic ET rate ($k_{\rm ET}$) data have been obtained via electrochemical[112–114,119] or spectroscopic flash quench studies.[45,61,65,97,119–121]. Table 1 gives some experimental data for two well-studied proteins, obtained by using mutated proteins to vary the donor-acceptor separations.

Gray, Winkler and co-workers have used modified metalloproteins to understand distance-dependent ET between a dye molecule and a metal ion that acts as quencher in Azurin and cytochrome b562 proteins (Figure 4) using flash-quenching, where a foreign dye is introduced on a protein surface and ET processes between redox centers of the protein and dye molecules are monitored.[122].[65,102] By introducing ruthenium and rhenium complexes as donors which are known to attach to certain histidine residues at known distances from Cu (azurin), Fe (in the heme group in cytochrome) or Zn (modified cytochrome) cofactors, they were able to show experimentally (Figure 4) that

$$k_{ET} \propto e^{-\beta L}\,, \qquad (2)$$

where $L$ represents the donor-acceptor distance and $1/\beta$ is a characteristic decay distance. Studies by Gray and Winkler[123] demonstrated that out of nine $k_{\rm ET}$ values for histidine-modified cyt $b_{562}$ derivatives, seven were accurately fit by the tunneling model, but two showed $k_{\rm ET}$ slower than predicted by Equation 2, as shown in Figure 4b. This implies that a different transfer mechanism is at work for those residues. Surprisingly, protein electron transfer rates deduced from electrochemical and spectroscopic methods were significantly different (Table 1). This difference may indicate that protein immobilization on solid surfaces used in the electrochemical methods constrains some of the electron transfer pathways, while in spectroscopic methods performed in solution, more protein conformations and transfer pathways are available. Alternatively, or additionally, the switch from electronic to ionic conduction (assuming the redox center to solution step is the rate-determining one) may affect the measured rate.

Another useful approach to investigate ET in proteins has been pulse radiolysis, primarily those containing transition metal ions in the active sites. There were two main, complementary objectives guiding these studies. One was attaining an understanding of the electron-transfer process within the polypeptide matrix separating the redox centers and defining the parameters controlling their rates. The second was resolving the detailed mechanism of the function(s) performed by the protein, notably enzymes.[122] Azurin has been a model system for pursuing the first objective and was extensively investigated using pulse radiolysis. Triggering an intramolecular ET from the residual ion to the Cu(II) site over a 2.7-nm separation turned out to be useful for examining the impact of specific structural differences introduced by mutations on ET rates. Parameters examined ranged from changes in the medium separating the redox sites to those in the actual coordination of copper site, which drastically affected its reorganization energy.[124–127]

Over the years, more sophisticated theoretical models have been developed to understand ET in proteins, including bridge-mediated electron tunneling[111] and square barrier tunneling.[108] These models establish the theory of ET between fixed sites within the proteins through electron tunneling. In the case of redox proteins, multiple electron transfer pathways often exist, and pathways can be highly dependent on protein structure and the surrounding environment. Refinements by Beratan et al.[128,129] of these early models show that electrons can tunnel across proteins through favorable pathway(s), which include mostly bonded groups, with less favorable non-bonded interactions being important when the through-bond pathway is prohibitively long and a shorter through-space path exists.[130]

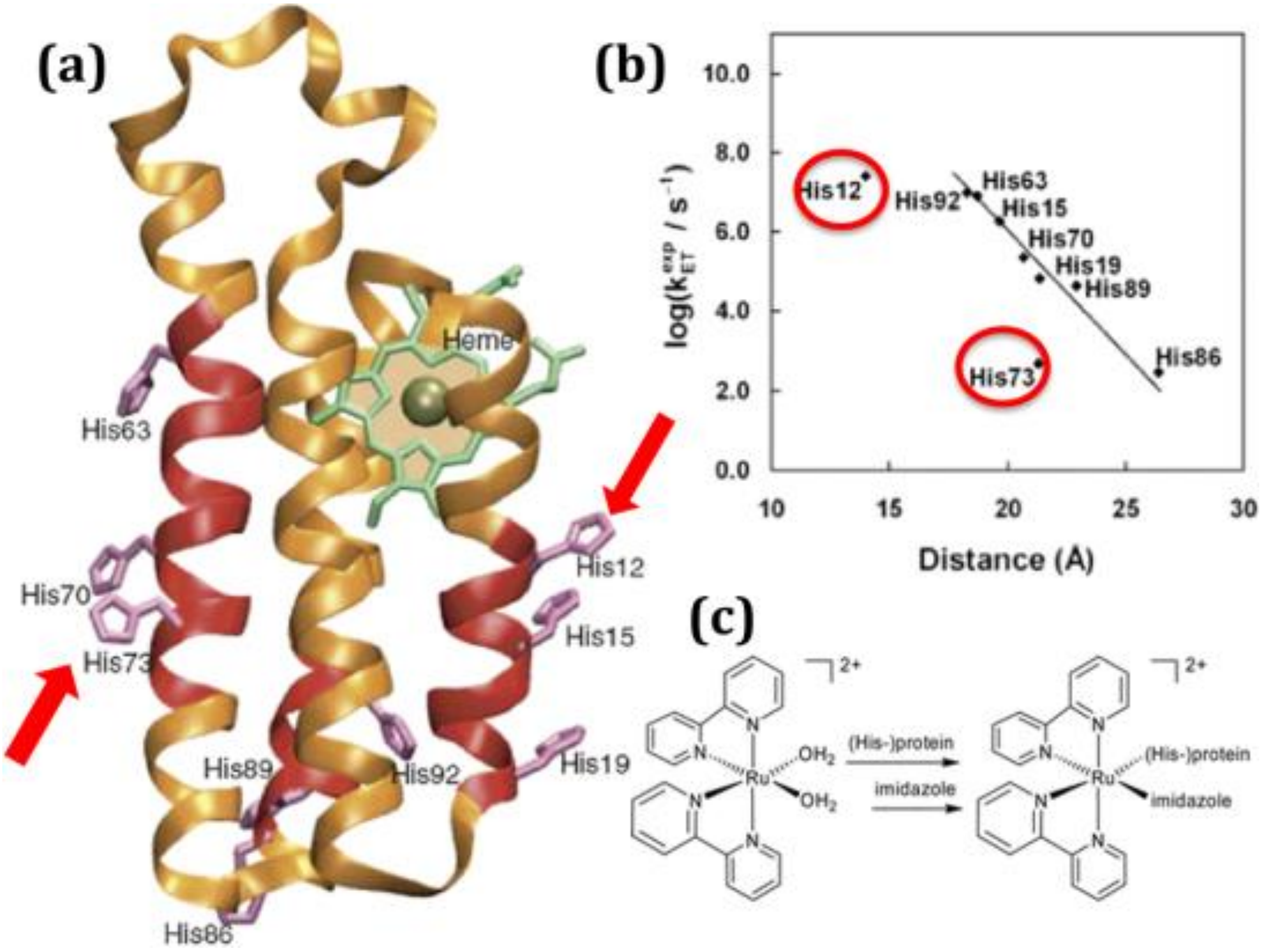

Figure 4. (a) Ribbon diagram indicating the positions of the nine histidine sites on cyt b562, where a Ru complex was connected to different histidine groups in different mutants, to measure variations in electrochemical electron transfer rates between each of these sites and the heme group. (b) Seven ET rates follow the exponential distance dependency of Equation 1, whereas in two cases slower rates than predicted by Equation 1 were measured. (c) Chemical reaction to connect the Ru complex with different histidines in different mutants. Published with permission from ref. [131].

A square-barrier tunneling model, combined with a suitable decay constant, has given a surprisingly good description for experimentally obtained biological ET (Figure 2).[132] This model was subsequently refined by Beratan and co-workers, who proposed that electrons tunnel along specific pathways, connecting electron donating and accepting cofactors.[132] The theory of ET in biochemical systems with several intermediary tunneling (bridge) states was recently reviewed by Blumberger ,[110] who summarized several models that are currently considered viable for explaining electron transfer in biomolecules, all based on the Marcus theory. These are the super-exchange (multi-step tunneling), flickering resonance (proposed by Beratan, Skourtis, and coworkers [133]; *vide infra*), and hopping models, based on either hopping and/or tunneling as mechanisms of electron transfer (Figure 5).

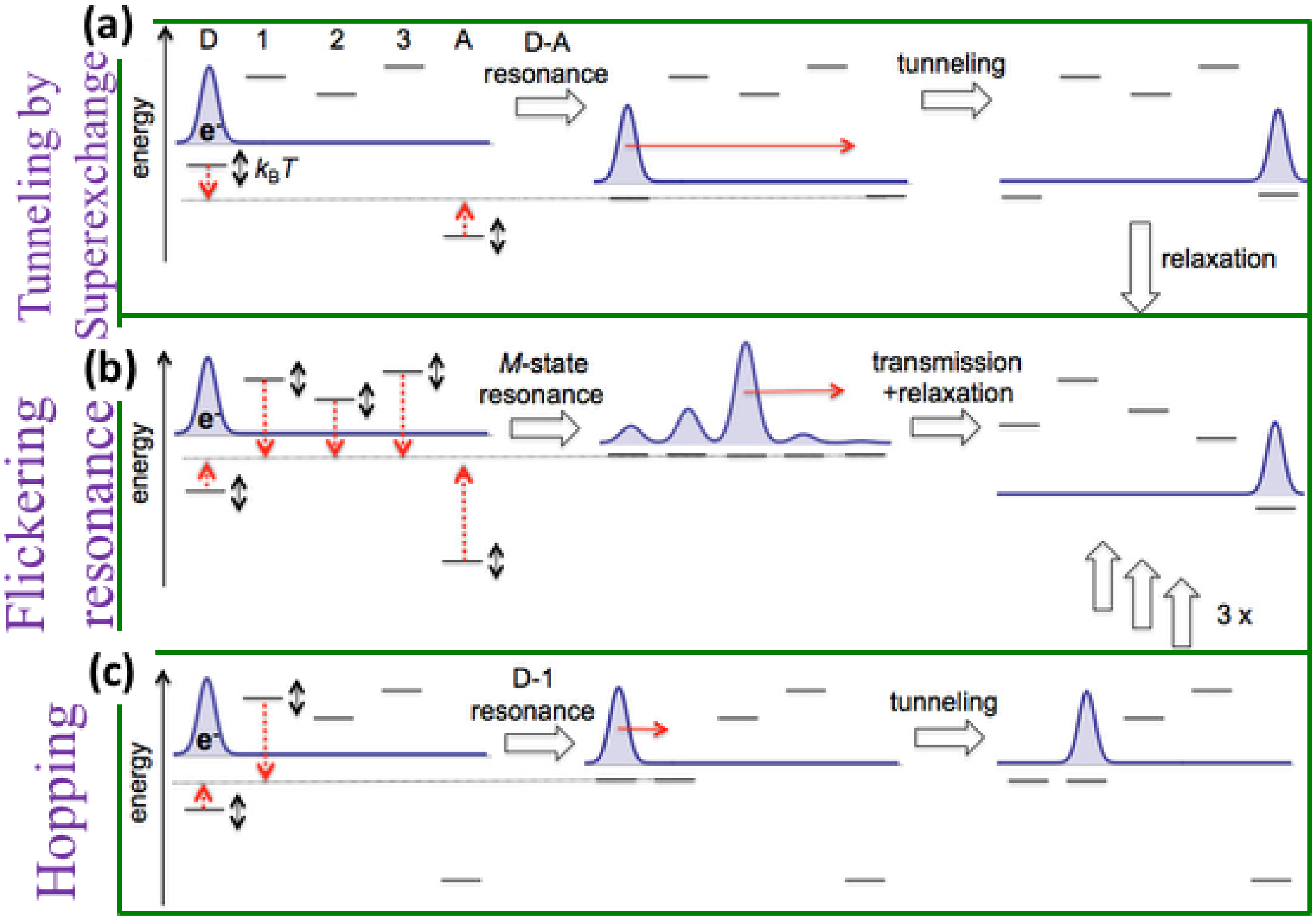

Figure 5: (a) Illustration of protein-mediated electron transfer models, tunneling (via super-exchange), flickering resonance and hopping, along a chain of 5 redox active sites (as could be the case in a multi-heme or multi-copper protein), where D is the electron donor and A is the acceptor (A), leaving three sites (1, 2, 3) between D and A. One-electron energy levels are drawn as black lines for each site (cf. Fig. 5.2 for inclusion of vibrational broadening) The electron that is transferring from D to A is shown as a Gaussian. In (a) thermal fluctuations bring D and A levels into resonance (middle scheme), which allows tunneling from D to A. Also during these process sites 1,2, 3 are non-resonant; they enhance tunneling but are not significantly occupied by the tunneling electron at any time, i.e., there is no nuclear relaxation as a result of electron occupation. In (b) ET occurs only when all levels are in resonance and the tunneling electron transfers ballistically (with tunneling probability of 1). In (c) [133] D and nearest neighbor site 1 become resonant, allowing efficient electron tunneling from D to 1, and so forth till A. Adapted with permission from reference [110], Blumberger *et al*. Copyright © 2015, American Chemical Society.

In the *superexchange model*, the donor and acceptor energy levels are brought into resonance via bridge energy states, and then direct tunneling occurs, bypassing the intermediate energy levels completely. Superexchange therefore provides a model for direct tunneling, with a dependence of the electron transfer rate on distance between donor and acceptor, $R$, given by

$$k_{ET} = Ae^{-\beta(R-\Delta R)} \qquad \textbf{(3)}$$

where $\Delta R$ is the distance between intermediate states, and $1/\beta$ is a characteristic decay length. The proportionality constant $A$ depends on temperature as $A \propto T^{-1/2}\exp(-b/k_B T)$, where $b$ is a constant that has units of energy, but is independent of $R$ and $\Delta R$, while $\beta \propto 1/\Delta R$ and is temperature-independent.

The *flickering resonance (FR) model*, proposed by Beratan, Skourtis, and coworkers, considers the dynamical nature of the system (cf. also discussion in section 4.1).[133] For flickering resonance to occur, the energies of the donor, acceptor, *and* intermediate energy levels need all to be brought into resonance by thermal fluctuations to allow coherent tunneling (≠hopping) between the donor and acceptor sites. In this case, the transfer rate has the same distance dependence as in Eq. 3, with the same $\beta \propto 1/\Delta R$ dependence. The proportionality constant $A$ has essentially the same temperature dependence as for the superexchange mechanism, but with a $A \propto (R/\Delta R)^{-1}$ dependence for its upper bound value. Although it has not been possible so far to unequivocally identify flickering resonance in experimental data, this model might in principle be applicable to ETp in biomolecules if it is sufficiently efficient over long distances (10s of nm to μm). Moreover, biological ET often involves multiple groups and redox cofactors in van der Waals contact with each other, and coupling rates are highly sensitive to conformational fluctuations. Thus, the flickering resonance model may be relevant to understand the electrical conductance in bacterial nanowires and multiheme proteins, with strongly coupled porphyrin arrays with closely packed (≲15 Å) redox groups.[134,135]

For larger distances, ET is assumed to occur primarily via hopping. In this mechanism, the electron *incoherently* hops from site to site in a sequential fashion, with a certain probability of hopping backwards and forwards. The transfer rate for this model is expected to depend on $R$ as

$$k_{Hop} = A/(c + R/\Delta R), \qquad \textbf{(4)}$$

where $c$ is independent of temperature and $A \propto \exp(-b/k_B T)$. The hopping mechanism is thought to be especially relevant for longer biomolecules, such as membrane proteins (Figure 6),[51,61] where the electron transfer distance between donor and acceptor is more than ~2 nm. The rate of hopping can also be described by the relation

$$k_{ET} \propto N^x \,, \tag{5}$$

where $N \sim R/\Delta R$ is the total number of steps and $x$ depends on whether the steps are reversible or irreversible, and has a value between 1 and 2.[107,136] It has been proposed that in ribonucleotide reductases, responsible for catalyzing the conversion of nucleotides to deoxynucleotides in all organisms, the long ET distance of 35 Å is covered via short electron hopping steps between conserved aromatic amino acids, rather than a single super exchange step (as illustrated in figure 7).[137]

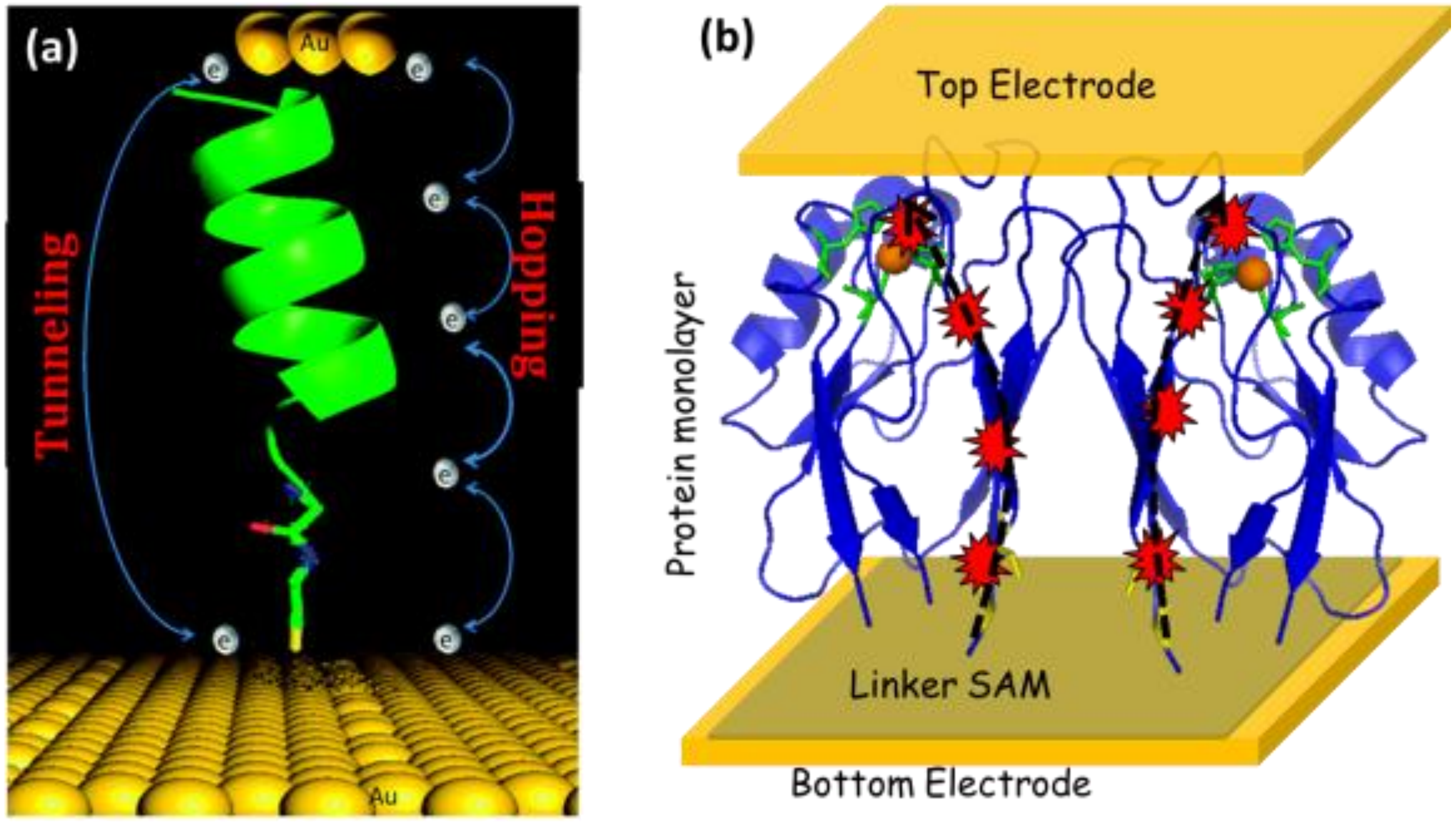

Figure 6: Diagram comparing electron transport by superexchange and hopping across single α-helical peptide at nanoscopic configuration. Reprinted with permission from reference [138] with permission of the Royal Chemistry Society. (b) Diagram of electron transport process across dehydrated protein monolayer sandwiched between two macroscopic metal electrodes. SAM: self-assembled monolayer; linker: small molecule with terminal groups to bind to the protein and to the substrate. The protein shown is Azurin.

## 2.2 Electron Transport

The fundamental mechanisms of ETp via proteins are less understood than those of ET processes. One important reason is that, as discussed above, ETp measurements are not done in solution, and therefore there are no ionic charges in the medium surrounding the protein to screen charging as the electron moves across the protein. Another reason is that the interaction between the metal electrodes, which effectively act as initial and final donor and acceptor states, are defined by the Fermi level (electrochemical potential) of the electrons of the metals used, rather than by discrete molecular orbitals. Nevertheless, some of the ET ideas and models discussed above (Figure 5) have been used as starting points to describe ETp from a microscopic point of view. Some of the most widely accepted ETp models based on ET ideas are illustrated in Figure 7. The top panel describes a scheme with donor and acceptor states (as in ET) that can be the two electrodes of a molecular junction in ETp. Different amino acid residues, which may also include a cofactor, affect the potential that the electron encounters as it is transported through the protein. The lower four panels in Figure 7 are highly schematic illustrations of one-dimensional modes of ETp under an applied electrical potential between the right and left electrodes, according to different proposed model. In the *hopping* model, transport between electrodes occurs via intermediate sites that are potential wells for the electron, requiring activation energy to escape, i.e., this process is always incoherent and dissipative. We note that this model is not the same as the Hubbard model for the hopping mechanism in solid state physics, which relies on an effective repulsive potential between sites.

Using the Marcus model as a starting point (Eq. 1), the conductance in the hopping model is temperature dependent: $\propto T^{-1/2}\exp(-b/k_BT)$. In the *tunneling via superexchange* model, transport is coherent in a similar fashion to the superexchange model used in ET, but because the Fermi levels of the electrodes are tuned by the externally applied electric potential, it does not rely on a spontaneous thermal alignment of D-A levels. As a result, the temperature dependence related to the D-A energies is different in ETp, and the tunneling barrier is influenced by the energy levels of the intervening medium.

The term *sequential tunneling* has been and is used in different publications for different processes and, as a result, its use is ambiguous. It has been stated that sequential tunneling refers to hopping (cf. e.g., ref. 72) or that it is equivalent to resonant tunneling.[139] Büttiker considered the issue, and concluded that tunneling probability through a barrier can in general consist of coherent and an incoherent parts.[140] The incoherent part, where inelastic processes occur, destroys phase coherence. In a completely incoherent tunneling process, the electron can scatter backwards or forwards. According to Büttiker, purely sequential (incoherent) tunneling occurs if tunneling occurs solely through an inelastic channel; on the other hand, if tunneling occurs through the coherent channel, *coherent resonant tunneling* occurs. In general, a tunneling process with intervening energy states is composed of both incoherent (sequential) and coherent tunneling channels. In this situation, Büttiker has demonstrated that if the two channels are connected in series, they are not independent of each other. In other words, increasing the sequential tunneling probability affects the coherent channel probability.

Whether the transport process is coherent or incoherent depends mainly on the residence time of the electron at the intermediate electronic states available for transport in the medium between the electrodes. For example, if the electron spends enough time in the resonant state so that a nuclear relaxation accompanies this process (a time $t \sim 1/f$, where $f$ is the characteristic frequency of the vibrational energy associated with state), inelastic tunneling will occur. It is also important to note that while inelastic transport is always incoherent, elastic transport can in general be coherent or incoherent. Randomization of the phase of the electron during elastic, incoherent tunneling can occur, for example, because of elastic scattering that randomizes the electron's momentum direction without changing its magnitude, as in a diffusive process. Another example is when the resonant energy level is sufficiently narrow that the electron is effectively localized for a significant amount of time during which memory of the phase is lost. Thus, depending on what is meant by sequential tunneling, the process can be temperature dependent (hopping) or not. Therefore, whenever the term "sequential tunneling" is encountered, *caveat lector*!

The *flickering resonance* model was described in the ET discussion above and is further discussed in section 4.1. The main difference between the *FR* and sequential tunneling models is that flickering resonance is a coherent tunneling process involving several intermediate resonant energy states, where the electron does not spend much time within each intermediate energy state, and the entire process is elastic and is described by a single wave function within the barrier.

In all four cases, the energy levels of the intervening states inside the protein are generally considered to be thermally broadened, instead of the sharp resonance states used for the ET models shown in Figure 5.

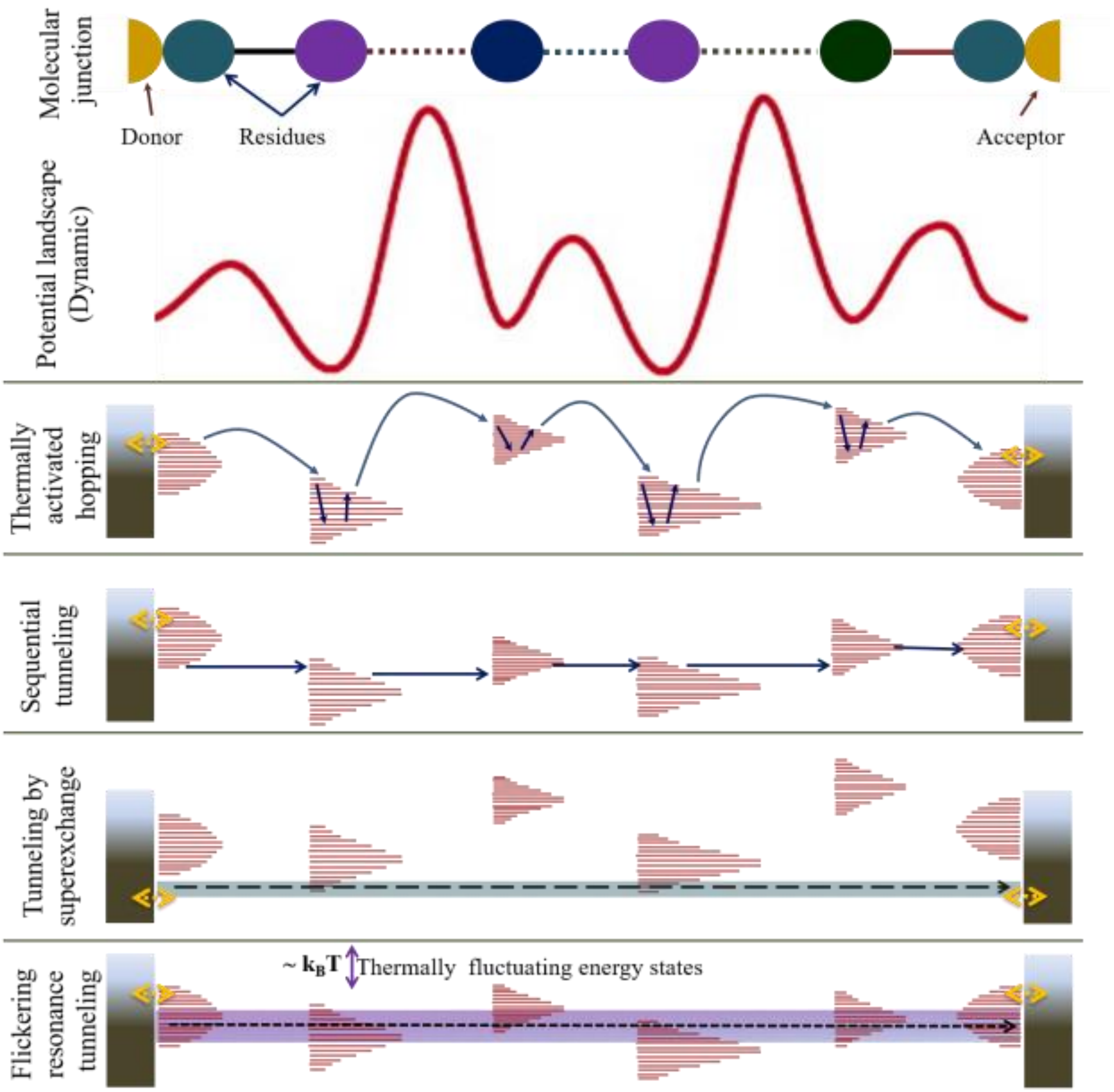

**Figure 7**: Illustration of how ideas and models from ET can be used as bases for ETp models, using Fig. 5a as starting point. The top panel describes a scheme is given with donor and acceptor states (as in ET) that can be the two electrodes of a molecular junction (as in ETp). Different amino acid residues, which may include also possible cofactors, are indicated. Below the top panel a possible 1-D snapshot of the dynamic, varying electric potential profile in the protein, is shown between the electrodes of a molecular junction. The lower four panels are highly schematic illustrations of one-dimensional transport modes under an applied electrical potential between the right and left electrodes, with the y-axis being an energy scale. The Fermi energy distribution of the metallic contacts (dark-gray rectangles) and the broadening of the intervening electronic states within the protein (in red) are indicated with Gaussian(-like) distributions.

Proteins in molecular electronic devices have also been modeled as one-dimensional solid-state conductors (such as carbon nanotubes), in which case the Landauer formalism has been used to describe their ETp characteristics (Figure 6b). This model ignores inelastic interactions and only takes into account elastic scattering at interfaces. If the molecule is considered a tunneling barrier, and the temperature is low, the conductance is described by the Landauer formula

$$G = G_0 T(E_F), \qquad \textbf{(6)}$$

where $G_0 = \frac{2e^2}{h} \approx (12.9\ \text{k}\Omega)^{-1}$ is the quantum unit of conductance (taking into account spin degeneracy) and $T(E_F)$ is the transmission coefficient of the barrier evaluated at the Fermi energy of the metallic contacts.[56,141] The Landauer model can be extended to include molecular energy levels by taking into account resonant tunneling, in which case the transmission coefficient can be modeled by an effective Lorentzian width of the molecular energy level given by

$$T(E) = \frac{4\Gamma_L\Gamma_R}{(E-\varepsilon_0)^2+(\Gamma_L+\Gamma_R)^2} \ . \qquad \textbf{(7)}$$

In Eq. 7, known as the Breit-Wigner formula, $\Gamma_L$ and $\Gamma_R$ are the coupling energies of the molecular orbitals to the left and right electrodes, respectively, $E$ is the energy of the tunneling electron, and $\varepsilon_0$ is the energy of the molecular orbital (which in general can depend on the coupling of the orbital to the electrodes). This approximation is valid when $E \approx \varepsilon_0$ and the separation between molecular energy levels is greater than $\Gamma_L + \Gamma_R$. Note that if $E = \varepsilon_0$ and $\Gamma_L = \Gamma_R$, $T(E) = 1$, that is, perfect resonance occurs which results in ballistic transport.

We reiterate that in this review we focus on describing protein ETp (electron transport occurring in dry, solid state conditions) and on making further connections with ET (electron transfer that occurs in conditions where electrolytes play a role in the process) and protein biological function. It is important to note that in order to make real ETp measurements, it is necessary to immobilize the protein (unlike in measurements made in solution). This adds complexity to the problem because attachment of the protein to an electrode can itself change its electronic structure. This can occur as a result of hybridization with the metallic electrical contacts and/or the formation of Schottky barriers and image charges in the contacts. These interactions can change the electronic structure directly or by causing a configurational change that in many cases can render the protein biologically inactive. Therefore, understanding the effects of immobilization on proteins on their electronic structure is crucial to the interpretation of real experiments where ETp is measured. We discuss the effects of protein structure and immobilization of proteins in general on their ET and ETp properties below.

## 2.3 Protein Structure, Electron Transfer (ET) and Electron Transport (ETp)

Biological functions of proteins are dependent on their structure in a fundamental way. As a result, preserving protein structure and chemical functionalities upon immobilization on a solid surface is key for studying both ET and ETp. In this section, we discuss some fundamental concepts related to protein structure that are relevant to ET and ETp.

Protein structure can be understood at several levels, the first being the primary structure, which is the linear sequence of amino acids, 2HN-(H)C(R)-CO(OH), held together by covalent, amide (or peptide) bonds, -(H)N-C(O)-, forming a polypeptide chain. The next level is the secondary structure, where the polypeptide chains form higher order three-dimensional features (e.g. α-helices and β-sheets or turns) based on hydrogen bonding between amino acid peptide bonds. Lastly, a tertiary structure is formed by additional secondary structure elements that undergo the necessary folding by specific interactions, including formation of salt bridges, hydrogen bonds, and disulfide bonds, to achieve free energy optimization (lowest entropy) in the solvent, giving rise to a compact structure. Proteins are generally amphiphilic, i.e., they contain both hydrophobic and hydrophilic parts, with hydrophilic amino acids forming a protective shell via folding to minimize the exposure of hydrophobic amino acids to the solvent. In the case of multi-subunit proteins, where the sub-units are stabilized by the same interactions as the tertiary structure, a quaternary structure refers to the global structure of the connected subunits. Multi-subunit proteins are multi-protein complexes, with different complexes having different degrees of stability over time to achieve a specific set of functions. A well-known example is hemoglobin, which is a tetramer composed of four sub-units with a high degree of stability.

The oxidation-reduction potential of redox-active proteins is extremely sensitive to the structure of the polypeptides. For example, a single point mutation or minimal alteration of the secondary or tertiary structure of redox proteins can change the redox potential by 100 mV or more.[142] Conditions that break bonds or disrupt electrostatic interaction within proteins may result in a loss of the structural order of proteins, a process known as denaturation. Denatured proteins display some or complete loss of protein function.

*Enzymes* are proteins that act as macromolecular biological catalysts, and thus accelerate chemical reactions without themselves being altered by the reaction, often by binding to small molecules. An example of how enzymes work can be seen in the cytochrome family when discerning differences among members of the cytochrome P450s. Cytochrome P450 is denoted by "CYP" followed by a number placing it in a gene family, a letter that denotes a sub family. An individual gene receives a second number. Thus, the name CYP2C9 represents a cytochrome P450 in the '2' family, the 'C' subfamily, and individual gene '9'. Both cytochrome c and cytochrome P450 are heme-containing metalloproteins (with Fe as the metal ion) that have comparable redox potentials for ET. P450s contain an active site that allows binding of small molecules, while cytochrome c does not. Different small molecules bind specifically to different P450s, coordinated to the heme group, and are known to alter the spin state of the heme Fe.[143,144] Studies have shown that the presence of small molecules in the active site alters the rate of ET in multiple P450s[144] and our research has demonstrated that changes both in ET rate and ETp efficiencies of P450 from CYP2C9 occur when small molecules are bound inside the active site.[145–147]

Electron transport across dry protein monolayers has been studied using solid-state protein-based molecular junctions, with macroscopic and nanoscopic current-voltage (I-V) methods.[1,47,54,148,149] Results have been reported for several protein types, such as those functioning as ET mediators, azurin (Az) and cytochrome c (CytC), or bacteriorhodopsin (bR), a light-driven H+ pump protein, and proteins lacking any bound cofactor (bovine and human serum albumin, BSA and HSA). Despite some general similarities, these proteins differed in their ETp behavior. For instance, proteins with bound cofactors produce higher currents than those that lack it. These observations indicate not only that the primary sequence of the amino acid, which is mainly responsible for secondary structure is important, but also that the presence of a cofactor creates relatively efficient current transport paths and affects ETp in solid-state measurements.

The importance of the cofactor has also been demonstrated in studies of proteins for which the apo form, from which the cofactors was removed, can be generated while retaining structural integrity. Specific examples are azurin, myoglobin and CytC, which have undergone rigorous studies in different forms. In the apo form, electrochemical redox properties are absent for all metalloproteins, and electron conduction efficiencies are reduced by only one to a two orders of magnitude compared to the holo-protein, when both are measured in the dry solid state monolayer form.[87] For CytC, removing the heme group leads to denaturation, but it is possible to remove only the Fe (from the heme group), and this was found NOT to affect ETp at all, in contrast to its effect on ET, which showed loss of redox activity.[150] Cofactors often serve as the elements that facilitate superexchange-mediated tunneling for electronic charge carriers and protein-electrode electronic coupling, and alterations of these groups can affect ETp significantly.[151] Nanogap measurements of single myoglobin molecules in solid state demonstrated resonant tunneling through the heme group with a strong dependence on gate voltage only for the holo- but not the apo-protein.[32] In section 4.3.2 of the impact of cofactors on ETp is discussed further.

For electrochemical studies (Figure 1), intramolecular electron transfer in redox proteins specifically relates to the binding sites of redox prosthetic groups and redox-active amino acids and their locations with respect to the working electrode, which influence the electronic coupling between the redox-active center and the electrode contact point.[48,50,150,152] "Doping" human serum albumin protein with hemin (porphyrin, protoporphyrin IX, with $Fe^{3+}$; in heme, the metal ion is $Fe^{2+}$) or retinoic acid enhances ETp efficiencies by more than an order of magnitude in solid state molecular junctions; with hemin, this protein also becomes redox-active in an electrochemical cell. The natural heme-containing protein, Cyt C, shows a symmetric cyclic voltammogram with an ET rate constant of ~18 $s^{-1}$ (measured between the heme and the Au electrode), while the HSA–hemin complex exhibits an asymmetric curve with a somewhat lower ET rate constant of ~ 5 $s^{-1}$.[150,153] In case of redox proteins we note that the cofactors carry out the redox reactions, which need not be so for solid-state ETp (details in Section 4). [154]

In solid-state ETp across protein monolayers, electronic coupling ($\Gamma$, cf. eq. 7) to electrodes dramatically influences its efficiencies and the electron flow mechanisms.[155,156] Experimentally, the thermal activation energy for ETp is found to be governed by $\Gamma$ for all proteins that we have studied. This is illustrated by results on immobilized cytochrome c junctions, for which the thermal activation energy is reduced by approximately a factor of two, if it is covalently rather than electrostatically bound to the electrode.[156] In the case of metalloproteins, electronic coupling is mainly controlled by the relative position of the redox cofactor w.r.t. the electrodes, and by its oxidation state.[154,156] Covalent protein–electrode binding (hybridization) increases $\Gamma$, in the tunneling transport regime (30 – 150 K). Proteins that are covalently bound to the electrode via a cysteine thiolate yield ten times higher ETp efficiency than those that are electrostatically adsorbed on the surface.[47,86,156] Coordination changes of metal groups in different oxidation states also affects ETp. For example, the Fe in the heme group in myoglobin and cytochrome P450 (P450) undergoes a change in coordination upon ET that leads to a change between $Fe^{3+}$ and $Fe^{2+}$ oxidation states, which have different spin states. This type of coordination change is common in heme-containing proteins, and is known to have profound effects on both ET and ETp.[51] Another important consideration is how the cofactor is coupled in the ET pathway. Prytkova *et al.*[131] proposed that the anomalously slow ET rates seen by Gray and Winkler in their histidine-modified cyt $b_{562}$ derivatives (His 12 and His 73, see Figure 4) could be explained by different coupling pathways to the heme group. In the case of these two slow ET rates, there is a pathway coupled to the ligand, axial (perpendicular to) the heme plane, as opposed to the other seven predicted rates with a pathway coupled to the heme edge.

An increase in transport efficiency was observed across a myoglobin monolayer on silicon if Mb was oriented by chemical binding of its hemin group to the substrate (via protein reconstitution) compared to a monolayer of Mb that was randomly oriented, i.e., its heme group was randomly oriented with respect to the substrate and, thus, with respect to the contacts.[87] This ability to change ET and ETp by binding and orientation could allow use of proteins with different properties in a precise way in bioelectronics.

One method for achieving desired characteristics is the design of metalloproteins *de novo*[157] or by utilizing native protein scaffolds.[3] *De novo* design has allowed the creation of metalloproteins with ET, redox-coupled proton exchange,[158] hydroxylase,[159] peroxidase,[160] and oxygenase[161] activity. Metalloprotein design, using native protein scaffolds, has also been useful because the current incomplete understanding of protein folding mechanisms limits *de novo* design capabilities. Using techniques such as site-directed mutagenesis[162–164], it is possible to introduce new functions, metal specificity, and substrate specificity into existing metalloproteins.[165–169] A better understanding of protein function will allow greater control of protein design for use in bioelectronics.

## 2.4 General Properties of Immobilized Proteins

The ability to immobilize proteins allows for a direct study of their electrical transport properties, including ETp (measured in solid state conditions).[47] Protein immobilization is a powerful tool for controlling protein assembly and allows single molecule analysis. Immobilization entails physically localizing proteins in a region of space, usually on a solid-state substrate, while retaining their function for useful ETp studies. Factors that determine protein function after immobilization include amino acid composition of their surface, physical and chemical properties of the solid substrate, and the type of interface between the protein and the substrate ("coupling"),[170] all of which can affect the structure, orientation, and (average) conformation of the immobilized protein.[133]

A variety of immobilization techniques have been used to link proteins to electronic components. For redox-active proteins immobilization should result in the part that is redox-active and the electronic component onto which the protein is immobilized should be appropriately positioned for efficient electronic coupling. For photoactive proteins, immobilization should minimize fluorescence quenching by non-radiative energy transfer to the substrate.

Many proteins spontaneously adsorb on solid surfaces through hydrophobic or electrostatic interactions (physi-sorption), but uncontrolled and likely undesirable orientations of physisorbed proteins lead to multiple types of contacts and interactions with the surface, thus compromising the proteins' inherent functionality, which, for redox proteins often leads to a substantially slower electron transfer than in solution.[170–172] For the creation of bioelectronic devices, achieving controlled immobilization via direct chemisorption of functional proteins on metal or semiconductor surfaces remains an important task. The challenge is to achieve direct ET between the protein and electrode.

Metalloproteins, such as azurin, plastocyanin and cytochrome c, adsorbed on the surface of solid substrates, have been extensively characterized by a combination of techniques, as assemblies or at the single molecule level. These techniques include measurements of their topography, as well as spectroscopic, and ET properties.[28,173–179]

Atomic force microscopy (see section 3.6.2) and scanning tunneling microscopy (STM) have been employed to investigate morphological properties, ETp, and redox activity of individual metalloproteins, chemisorbed on gold substrates.[2,32,49,85,149,180–186] The arrangement and orientation of the proteins on a gold substrate, and their structural and dynamic properties, have been simulated using molecular dynamics.[62,181,187,188] Conductive probe AFM (CP-AFM; see section 3.6.2) experiments and scanning tunneling spectroscopy (STS) have probed ETp across adsorbed proteins. Redox functionality of azurin, cytochrome c, and myoglobin, immobilized on gold, were found to be preserved by cyclic voltammetry measurements of ET.[50,113,118,189–193] Electrochemical STM (EC-STM, see section 3.6.2) allows studying ET in solution with a variable electrochemical potential difference between sample and a reference electrode. In STM-based molecular junctions, redox-active cytochrome $b_{562}$ was engineered by introducing a thiol group, allowing for controlled binding to gold electrodes.[181,194] Various studies using a combination of spectroscopic and scanning probe techniques have found that immobilized proteins retain their structure partially over a range of electrochemical over-potential differences, as shown, for example, by the measurement of the ET properties of azurin immobilized on a gold substrate.[177,179,195–198]

Proteins can be absorbed on surfaces through electrostatic interactions of charged surface amino acids and the solid surface,[171] hydrophobic interactions if there are such exposed regions, or through tethering with a linker molecule.[199–201201,202] Direct immobilization to the substrate not only gives poor control of the protein orientation, as noted above, but also often results in inactive protein conformations.

Proteins are classified into two groups in terms of rigidity, which is a measure of the protein's ability to resist conformational changes upon adsorption to a surface.[171] The terms "hard" and "soft" are used to characterize flexibility of a protein, as deduced from its molecular adiabatic compressibility.[203] Protein rigidity can affect the viability of the protein if attached to a solid-state surface. Hard proteins, such as horseradish peroxidase, lysozyme, and ribonuclease A, generally go through minimal conformational changes upon adsorption.[204] In contrast, soft proteins, such as bovine serum albumin, myoglobin, and hemoglobin, are usually more susceptible to interaction with the surface onto which they are adsorbed and can show marked changes in secondary and tertiary structure upon adsorption.[205] Strong interactions with a solid substrate can even denature soft proteins that lack a rigid structure.[204] The nominal surface coverage for maximum protein activity correlates with the rigidity of the protein.[206] Although adsorption on a solid surface can lower protein activity due to denaturation, in some instances it increases stability of proteins, and actually enhances activity. Several studies have demonstrated an enhancement of lipase activity if immobilized to hydrophobic supports.[207,208] Enhancement of protein activity and increased stability were also demonstrated on nano-scale platforms, where a possible reason may be nano-structuring, i.e., binding between substrate and protein promotes more active and/or stable conformations than is the case for the unbound protein.[20,209–212]

## 3. Methods of Protein Immobilization

Immobilizing proteins is critical for studying electron flow across proteins, especially those related to ETp, where electronically conducting contacts are required. For ETp studies to be biologically relevant, immobilized proteins must be in their native conformation with structural and biological activities preserved. For electrical conduction studies and other applications, historically the first strategy employed was protein immobilization on bare metal electrodes or metal electrodes modified with small molecules to promote protein adsorption. Direct absorption was found to be ineffective because of loss of biological function, and thus electrodes modified with organic overlayers were developed to address this problem. In this section, we review these attachment techniques and the ways in which the attached proteins can be characterized.

## 3.1 Direct Adsorption

Polycrystalline metals, conducting polymers, semiconductors, especially silicon, or metal oxides, and carbon electrodes have all been used for direct immobilization (cf. ref.[213] for a recent review of semiconductors as substrates/electrodes for molecular electronics).[214–219] As mentioned previously, proteins are immobilized non-covalently by passive adsorption onto the surface through hydrophobic or electrostatic interactions (Figure 8).

Redox-active proteins can be studied by protein film voltammetry (PFV), pioneered by Armstrong, where the protein is adsorbed directly on to the electrode surface and probed electrochemically with techniques such as cyclic voltammetry (Figure 1).[220,221] In PFV it is important to choose an electrode surface that allows adsorption of the protein in an electroactive conformation, usually with the redox center near the electrode (Figure 9). While electrodes like pyrolytic graphite (PG), which can be polished to add oxide functionalities making it hydrophilic, have a distinct advantage over metals in allowing multiple and varied interactions with protein surfaces, metal electrodes are better suited for chrono-amperometric and impedance measurements[69,222,223] due to the absence of slow charging issues with graphite.[224]

Utilizing PFV, Hill, Kuwana, and co-workers demonstrated reversible diffusion-controlled voltammetry of adsorbed cytochrome c.[225,226] This approach was further extended to other proteins, including azurin, to demonstrate electron-exchange studies.[220,224] Direct adsorption of multi-heme proteins, such as cytochrome c nitrite reductase,[227,228] fumurate reductase,[229–231] and flavocytochrome $c_3$,[232] reveal distinct redox peaks for each heme group. These studies enabled interrogation of ET in flavin adenine dinucleotide (FAD) cofactors, covalently attached to the redox enzyme fumurate reductase, and non-covalently attached to flavocytochrome $c_3$. FAD is particularly prominent in voltammetry because of its two-electron transfer center, which can be detected electrochemically easily, compared to other methods like UV-visible spectroscopy, where it is obscured by the intense bands from heme groups.[233]

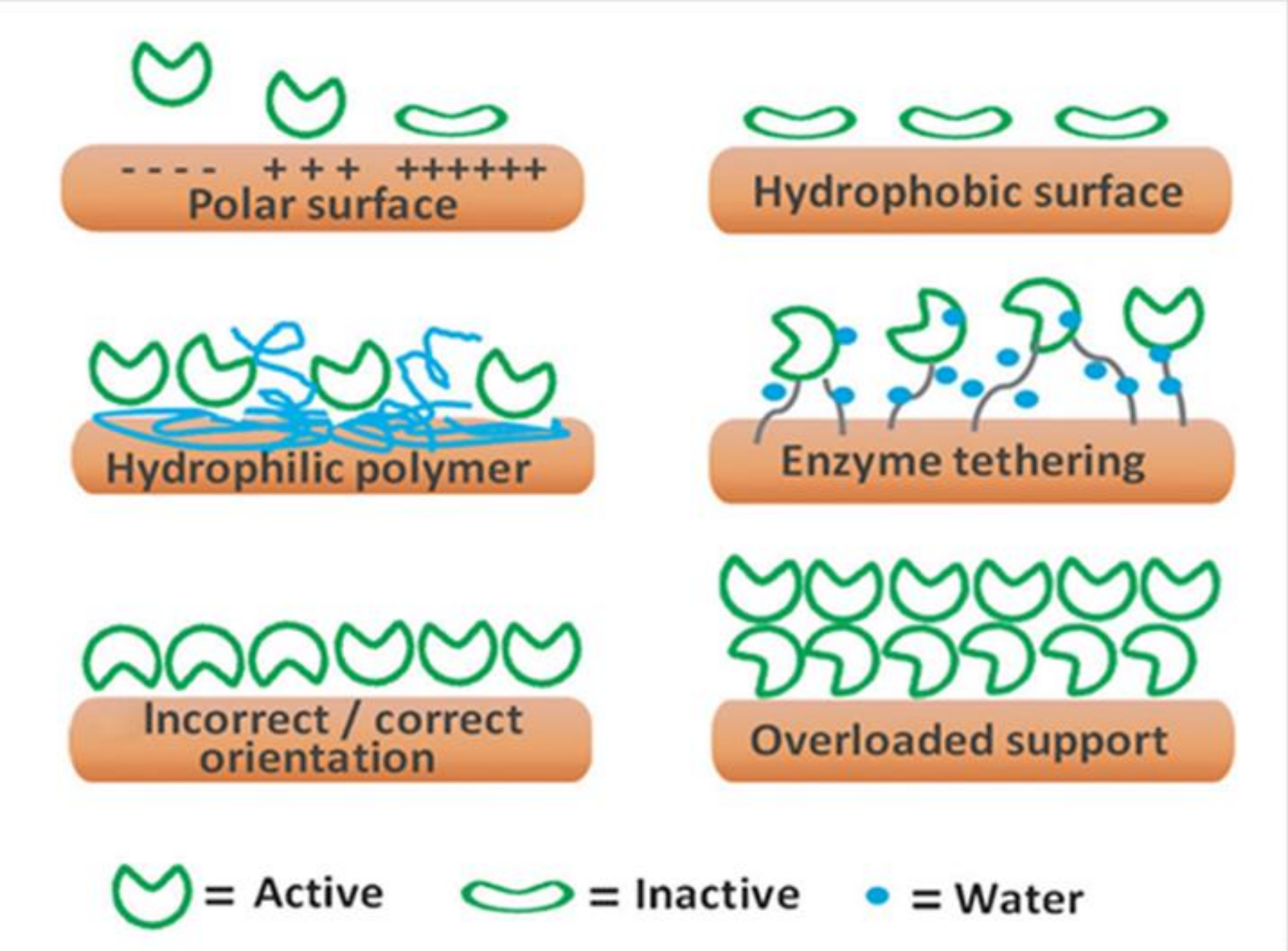

Figure 8. Enzyme immobilization on different interfaces and possible effects on the enzyme orientation/conformation. High charge density (above left) or hydrophobic surface (above right) are possible causes of enzyme conformational changes and inactivation. Enzyme co-immobilized with hydrophilic polymers (middle, left) or tethered (middle, right) can reduce unfolding and inactivating support–enzyme interactions. Incorrect orientation (below, left) and multilayer formation (below, right) may cause reduction of specific activity. Reproduced in part from reference [171] with permission of The Royal Society of Chemistry.

*In situ* vibrational (Raman and IR) spectroscopy measurements have been carried out to examine the retention of proteins structural and conformational after direct immobilization via adsorption or covalent attachment to electrode surfaces.[234–236] ET activity, as measured by electrochemistry, is often hindered by the large distance of redox-active groups (several nm) from the electrode surface upon adsorption (Figure 9). The spectroscopy data, together with large changes in the electrochemical potentials, also have shown that direct immobilization results in denaturation, such as a loosening of the helix packing, including partial dissociation of a histidine ligand in the ferrous state.

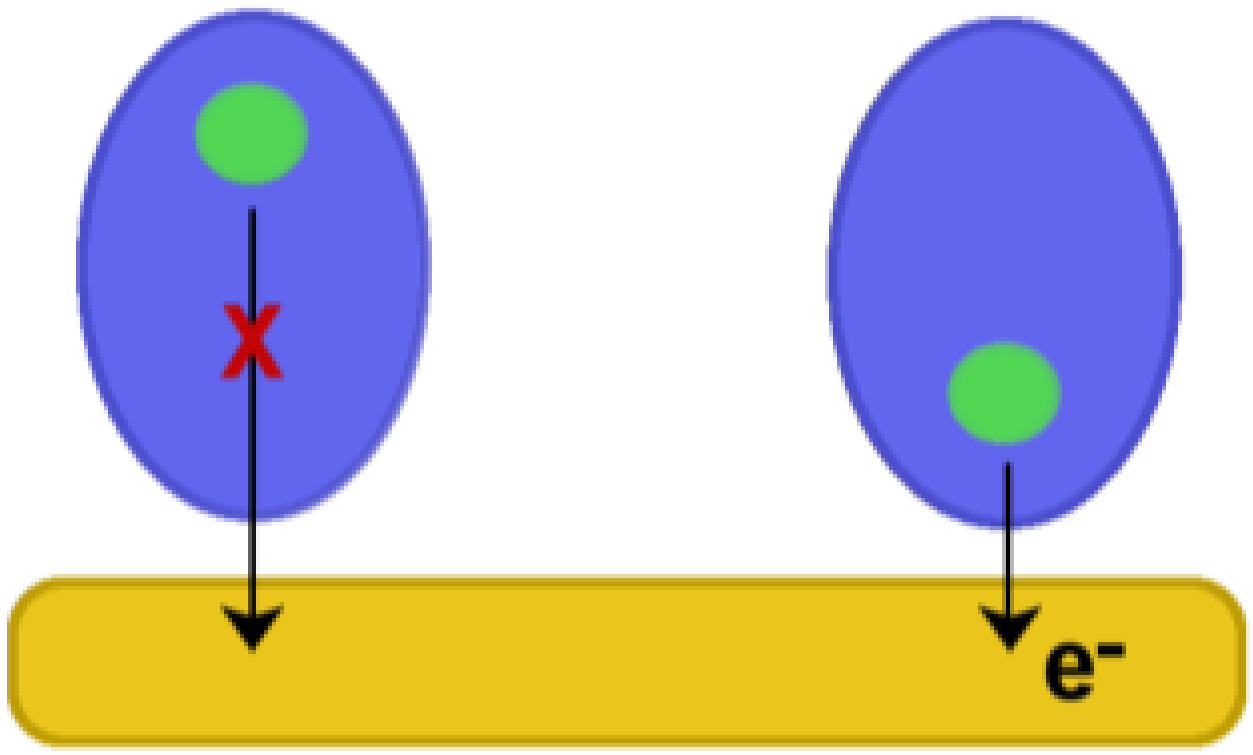

Figure 9. Effect of orientation of immobilized protein (blue) on direct electron transfer with electrode. On the left the redox group (green) is far away from the electrode (yellow), and electron transfer cannot be observed. On the right, the redox group is close to the electrode allowing for efficient electron transfer. Adapted with permission of reference[237]

## 3.2 Modified electrodes

Indirect adsorption of proteins using a chemical linker or cross-linked chemistry[238,239] on modified electrodes allows for better protein structure than direct adsorption. Thiol-containing molecules allow for the formation of self-assembled monolayers (SAMs) on metal surfaces, which provide a convenient and simple system with which to tailor the interfacial properties of electrodes.[214,240–243] Proteins with a natural dipole moment (such as membrane proteins) are electrostatically adsorbed to a polar head (carboxylate, amine) group of a SAM, which in turn is attached to an electrode. Successful surface modifiers are often those that bind the protein in a fashion similar to that of known redox factors, i.e., small molecules or proteins, which mediate ET from a donor to an acceptor, and preferably at the same location.[244] Extensive characterization has revealed that binding of proteins to their redox partners often involves hydrophobic and electrostatic interactions,[245–249] and thus tailoring the choice of surface modifier to mimic natural interaction can position and adjust the (average) protein conformation[133] for optimal electron transfer properties upon adsorption.

A number of surface modifiers have been evaluated for enhancement of heterogeneous ET of immobilized cytochrome *c*, and it has been proposed that successful modifiers allow a cytochrome *c* conformation with the prosthetic heme group close to the electrode.[250,251] Later work demonstrated that electronic coupling greatly impacts ETp, with covalent attachment doubling the current density compared to electrostatic adsorption. No correlation could be found between measured current and electrode-heme distances, although close electrode-heme distances generally displayed more efficient ETp.[156] Electron transfer of immobilized cytochrome c has been accomplished via various small organic linker molecules, especially those containing thiol groups.[252–258]

Chemically functionalized long-chain molecules, such as silanes or thiols, can spontaneously assemble on metal, glass, and carbon surfaces to form stable covalent bonds between their terminal functional group and the activated metal, metal oxide, or silicate. Densely packed, organized molecular layers assemble approximately orthogonal to the surface via extended hydrocarbon chain linkers.[240] Early examples include monolayers of long chain alcohols on carbon glass,[214,259,260] long chain amines on platinum,[261] alkyl trichlorosilanes on silicon and mica,[262] and long chain thiols, thioesters, thiones and alkyl disulfides[263–265] on gold.

Amphiphilic molecules, such as siloxanes, alkanethiols and carboxylic acids, form well-ordered mono-layers on metal and metal oxide surfaces.[240,266–268] Long chain thiols such as the 11 carbon units long 11-mercapto-undecanoic acid (MUA; ~ 1.1.nm), shorter ones such as the 3 carbon unit long mercaptopropionic acid, and others have been used to immobilize proteins.[269] On metal electrodes, the long-chain thiols form a dense self-assembled layer that is effective as a promoter, *i.e.*, an electro-inactive surface modifier that enhances ET of immobilized proteins,[270,271]. Linker length[47,272] and composition[266] have been shown to affect the electron transfer rate.

The control of protein stability and orientation through surface modification has led to the development of many different techniques to enhance ET rates and ETp efficiency across proteins. In addition to the small molecules mentioned above, polymers,[113,201,273–276] surfactants,[188,254,277,278] clay composites,[279–283] and nanoparticles[284–289] have been used to modify electrode surfaces to achieve effective coupling between the electrode surface and immobilized redox proteins. These compounds help to enhance protein immobilization through their entrapment in thin films. Thin films provide a favorable microenvironment for proteins, similar to biomembranes,[290] allowing them to retain their native conformations while enhancing the electron transfer to/from the electrode with respect to systems where the proteins are closer/tighter bound to the substrate. As the adsorbed protein is strongly bound to the film, much lower protein concentrations are required for SAM-modified surface methods than for direct immobilization, where the protein binds reversibly to the surface.[291]

Electrodes modified with surfactants and/or lipids can mimic the membrane environment where proteins are found in nature. This is especially true for enzymes which function in nature while bound to, or embedded in bi-layer lipid membranes (Figure 10).[292] These surfactant/lipid films are generally water insoluble, have polar head groups, and contain one or more relatively long hydrocarbon chains. Lipids and surfactants can form stable multilayers through electrostatic interactions with electrodes maintaining the native conformation.[278,290,293] Thus, the goal in this approach is to create an environment on the electrode that allows proteins to be immobilized in a manner that mimics their natural, membrane-bound environment and maintains their native conformation that facilitate electron transfer across protein-based junctions.

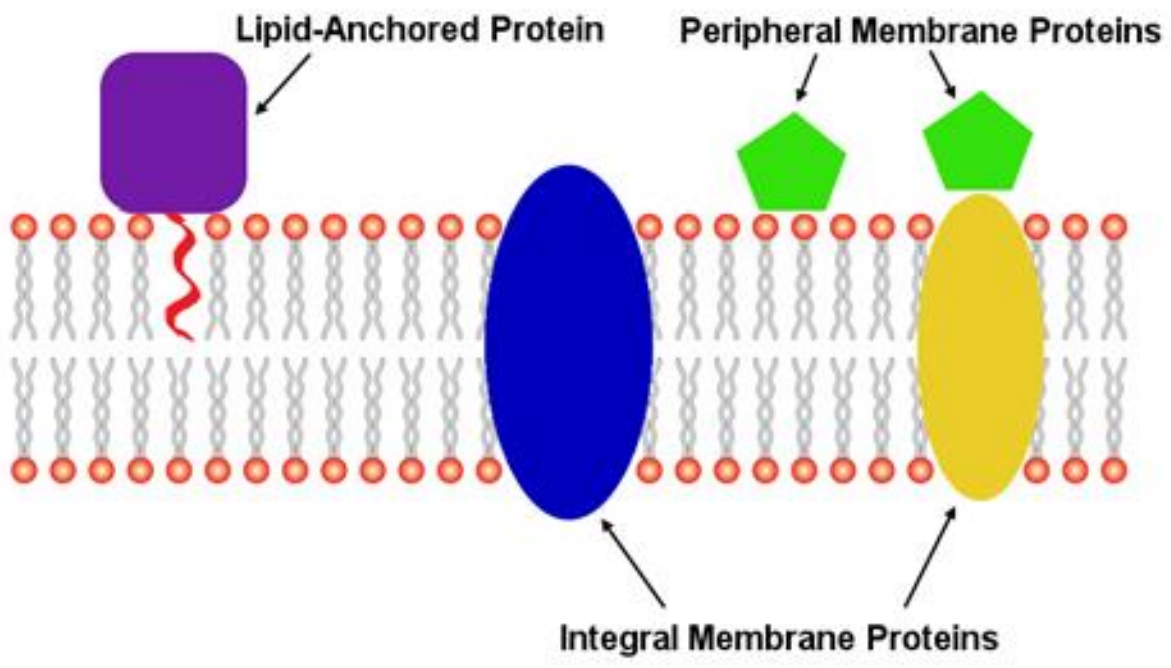

Figure 10. Example of protein interactions with lipid membranes in nature, which are dictated by the proteins' hydrophobic and hydrophilic surface groups. Membranes consist of amphiphilic phospholipids with hydrophilic phosphate heads and hydrophobic tails. Proteins can be embedded within the lipid environment as depicted for integral membrane proteins. They can be bound to lipid head groups or exterior surfaces of embedded proteins as seen for peripheral membrane proteins, or they can be linked to the membrane

via a hydrophobic tail group that buries itself in the lipid membrane.

Work by Murray and co-workers is the first report of reversible ET of a protein incorporated in a polymer film on an electrode.[294] In their work, cyclic voltammograms for cytochrome c, incorporated in gel coatings of poly-acrylamide and polyethylene oxide, yielded reversible peaks, indistinguishable from cytochrome c in aqueous solution. The success of this study led to the investigation of different polymers and methodologies for direct electron transfer studies. Amphiphilic polymers, polymers containing both polar and non-polar groups, have been demonstrated to allow immobilization of myoglobin and hemoglobin, in their native conformations.[295,296] In addition to amphiphilic polymers, there has also been increasing interest in using naturally created biopolymers as immobilization matrices for proteins because of their biocompatibility and non-toxicity, and a variety of natural supports have been used to immobilize proteins successfully.[297] Ultra-thin clay coatings, such as sodium montmorillonite, kaolinite, talc, goethite, and ochre, have also been demonstrated to promote the direct electrochemistry of heme-containing proteins.[281–283,298] An example is sodium montmorillonite, which is used to immobilize redox proteins and enzymes onto glassy carbon electrodes (GCE) for sensitive chemical and gas sensor applications.[299]

Polyelectrolytes, which are polymers with negative charges (polyanions) or positive charges (polycations), have also been used to immobilize proteins on a substrate surface, using layer-by-layer (LbL) deposition (Figure 11). First proposed for the synthesis of charged colloidal particles and proteins,[300] later work established that adsorption at every stage of polyanion or polycation assembly leads to multilayer formation and a reversal in the terminal charge.[301] In LbL growth, multiple layers of oppositely charged films of proteins, which have natural surface charges, and polyelectrolytes of opposite charge, are built up on the electrode surface, providing several active layers of entrapped protein (Figure 11). Layers can be formed by successive submersion/wash/submersion cycles of a substrate in solutions of the relevant polyelectrolytes, each at a pH suitable for the desired ionization; greater layer control and reproducibility can be achieved through electrochemical deposition.[275] The ability to incorporate multiple layers on an electrode increases the effective protein concentration on the surface, and could thus enhance downstream applications such as biosensing sensitivity. It, also enhances control over the film architecture when compared to solution-casting.[293]

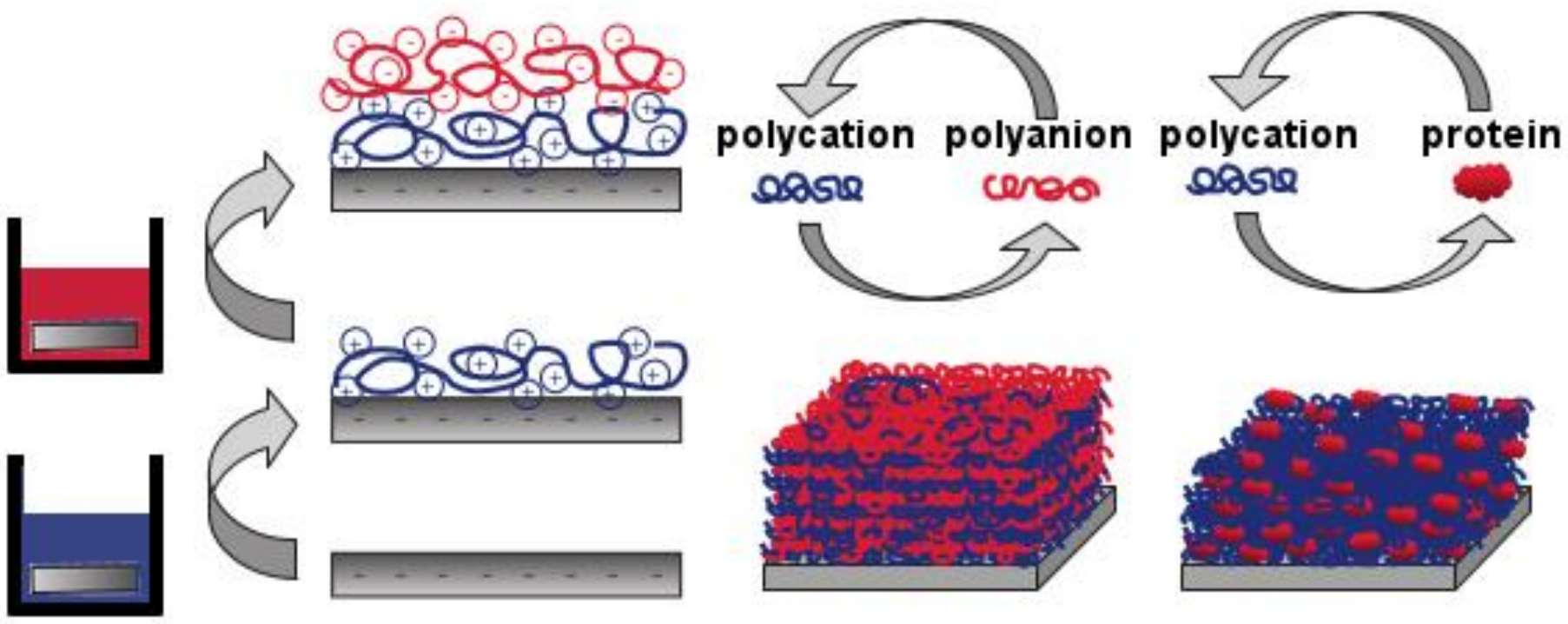

Figure 11. Representation of layer-by-layer (LbL) formation on a substrate using polyanions (negatively charged polymers, shown in red) and polycations (positively charged polymers, shown in blue). Layers are formed by immersion of the substrate in alternating solutions of polycations and polyanions. This allows the creation of a film to entrap a desired protein based on its surface charges, as shown on the right for the positively charged protein. Adapted with permission from Macmillan Publishers Ltd: Nature Mater [273] Copyright © 2009.

In addition to polyelectrolytes, nanoparticles (NPs) have also been investigated to generate ultrathin protein films.[302–304] Inorganic NPs have large surface areas, can have good biocompatibility, and provide a favorable microenvironment for electron transfer from electrodes to proteins. Using NPs in conjunction with polyelectrolytes leads to protein films on electrodes that are more electroactive than what is obtained by layer formation with polyelectrolytes alone. This is evidenced by an increase in how thick a protein layer remains electroactive, when formed with LbL immobilization: from ~1.4 electroactive layers for polyelectrolyte - myoglobin films on gold[113] to 7 electroactive layers of polyelectrolyte - myoglobin films on pyrolytic graphite,[274] and up to 10 such layers with polycations and manganese oxide nanoparticles.[305] He and Hu also demonstrated that LbL films of myoglobin, hemoglobin, and horseradish peroxidase with silicon dioxide nanoparticles retained their electroactivity of samples up to 6 layers thick, and such immobilized proteins can electrochemically catalyze reactions with their respective bound

ligands.[287]

## 3.3. Protein Tethering Through Linker

Proteins attached to substrate surfaces through an organized monomolecular layer with site-specific immobilization should provide better reproducibility and better control over electron transfer and transport measurements than randomly immobilized proteins, by eliminating unpredictable orientations and improving uniformity of protein conformation. Indeed, covalent attachment of proteins to an electrode via a tether has become a robust method to control protein adsorption.[53,184,195,306] This allows for a greater degree of control over protein orientation on the SAM surface, for an appropriate choice of linker and for controlled coupling chemistry.[217]

To overcome the slow ET rates reported in studies of electrostatically adsorbed proteins on modified electrodes, researchers have immobilized unmodified proteins through *covalent* attachment to SAMs on an electrode. There are various techniques that have been used to covalently link proteins to SAMs.[307] One of these uses chemical activation with reagents ("EDC-NHS" method) to bind a protein to a carboxylate ($COO^-$)-terminated SAM via a surface-exposed amine ($NH_2$) group of the protein. In this method, a covalent, amide bond (-(H)**N-C**(O)-) is formed [308] by activating the carboxyl group of the thiol linker molecule to allow bond formation with the amine (-$NH_2$) group of the protein.[87] If the substrate is reflective, then Fourier transform Polarization-Modulation InfraRed Reflection-Absorption Spectroscopy, PM-IRRAS, can serve to confirm bond formation.[309] The EDC-NHS method has been used successfully to covalently attach different proteins to a gold surface.[310] For example, immobilized cytochrome P450(CYP2C9) was shown to retain its enzymatic activity in the presence of its redox partners cytochrome P450 reductase (CPR), and nicotinamide adenine dinucleotide phosphate (NADPH, a biological redox agent)[147] A cyclic voltammetry scan of the Au/Linker/CYP2C9 electrode revealed quasi-reversible reduction and oxidation peaks in the range of known redox and oxidation potentials for P450s (Figure 12a).[311] CYP2C9-modified electrodes were also used to sense small molecules. This is possible as the current ($I$) - concentration ($C$) curves agree with the Michaelis-Menten equation,[312]

$$I = \frac{i_{max}\, C}{K_M^{app} \pm C},\qquad \textbf{(8)}$$

where $i_{\max}$ is the maximum current, C is the concentration of CYP2C9, and $K_M{}^{app}$ is the Michaelis constant which describes how well a small molecule (ligand) binds to the enzyme (Figure 12a). The work showed that concentration-dependent increases in current were observed only when the full P450 reaction cycle was completed, in the presence of both $O_2$ and small molecule binder. More recently, Fantuzzi *et al.* showed rapid determination of small molecule binding to various cytochrome P450s using a SAM linked P450-electrode integrated into a high-throughput plate format (Figure 13).[313,314] Taken together, these results demonstrate successful use of a linker to preserve innate protein functionality, which allow for biologically-relevant ET studies and suggest possible biosensing platforms.

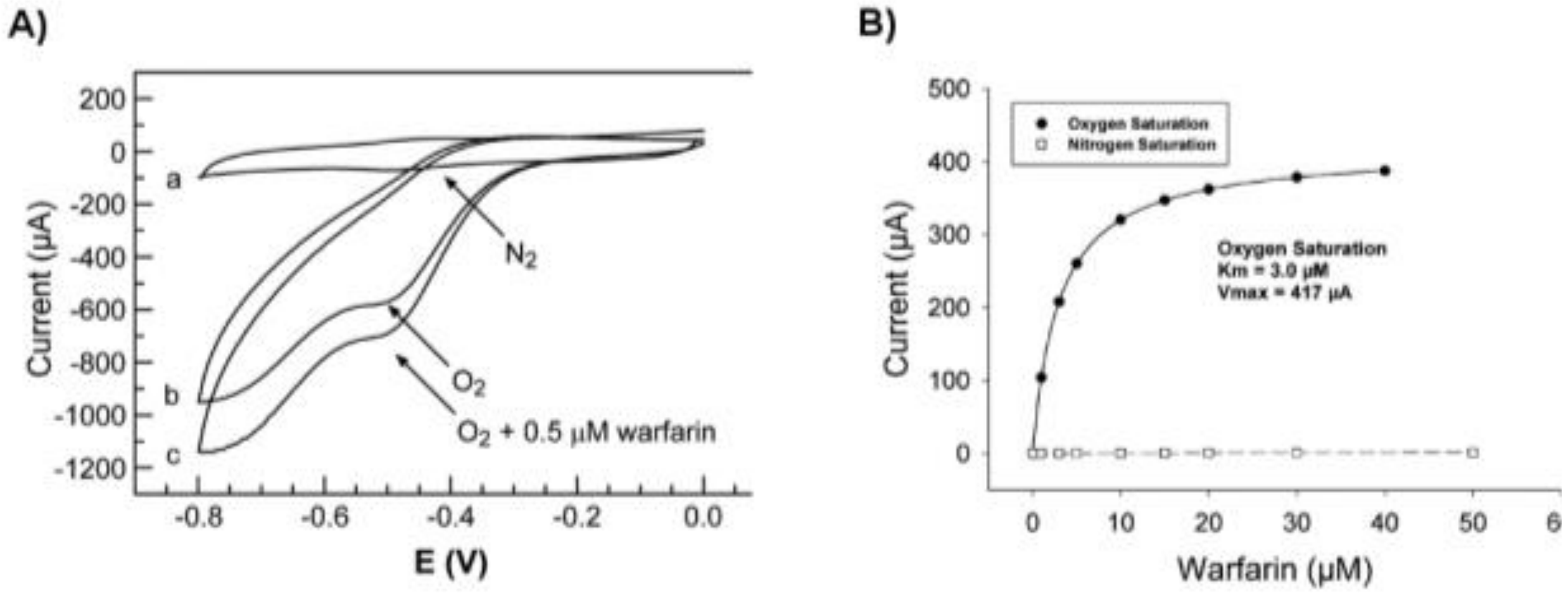

Figure 12. (a) Electrochemical Current-Voltage curves, Cyclic Voltammograms, of CYP2C9, covalently immobilized to Au(111) via a linker SAM at a scan rate of 1.6 V/s. Curves were obtained in the absence (curves a, b) and presence of warfarin (curve c), a small molecule that binds CYP2C9 and coordinates with its heme group. In the absence of $O_2$ (curve a), the reaction stops at the conversion of Fe($^{3+}$) to Fe($^{2+}$) and there is no reduction. (b) Reduction (of P450) current dependence on the -concentration (of small molecule that binds to protein). The solid line is the non-linear regression fit of the oxygen-saturation data, and the dashed line is the linear regression line of the nitrogen-saturation data (saturation was not observed with the nitrogen-saturation data). Reprinted with permission from reference 147.

Non-redox proteins were also successfully attached on Au(111) surface using carboxylic acid-terminated thiol (-SH) molecules, where the thiol group binds to the Au surface following the EDC-NHS reactions discussed above.[315,316,87]

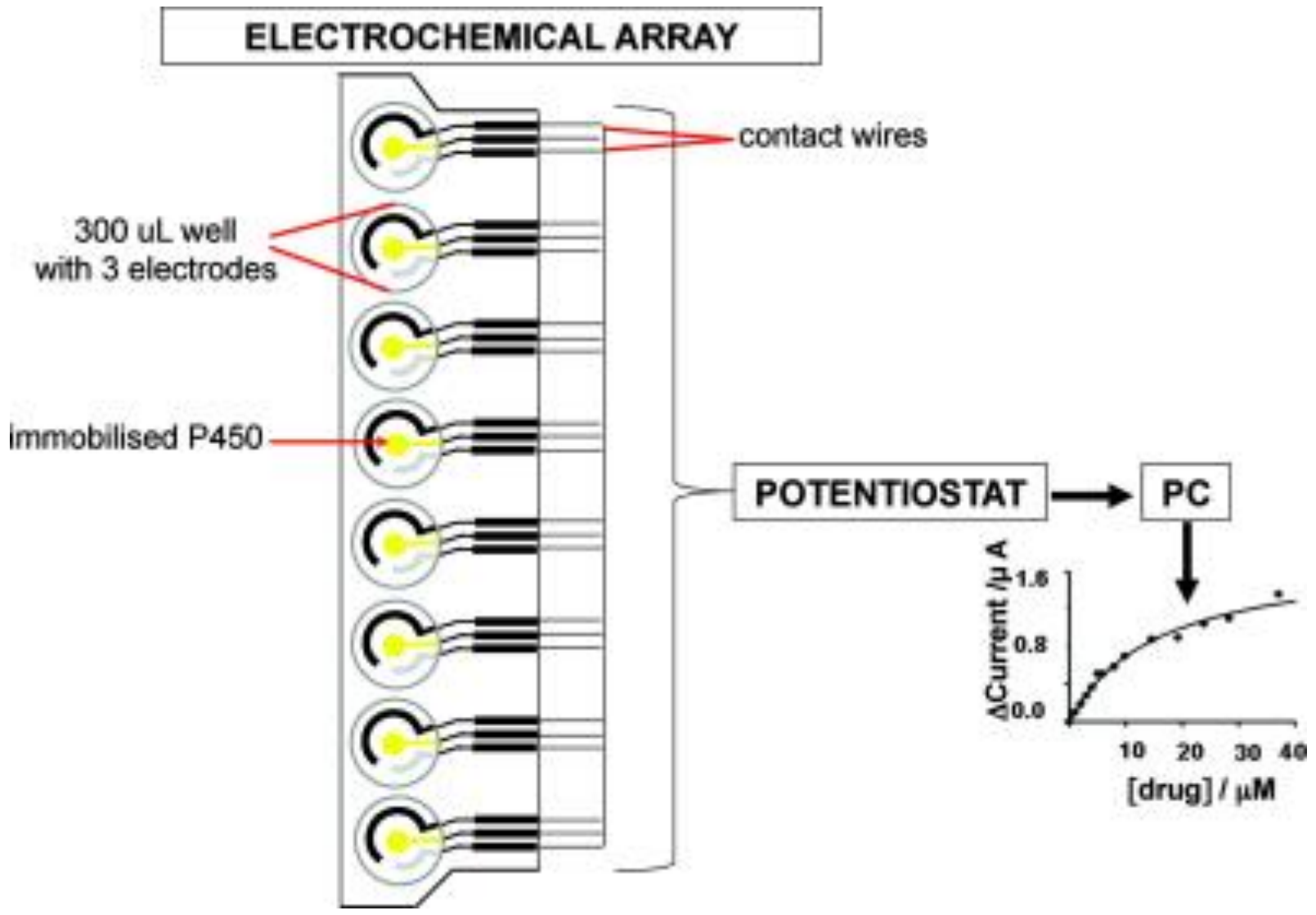

Figure 13. Schematic of a high-throughput drug screening plate with eight parallel cytochrome P450-electrode containing wells. The P450 was covalently immobilized through coupling between a sulfur atom of the linker molecules and the gold electrode (yellow). Each well also contains a Pt counter electrode (black) and carbon reference electrode (gray). Current was measured by a potentiostat as drug is added in increasing concentrations to determine current maxima. Reproduced with permission from reference[314] Copyright © 2011 American Chemical Society.

Protein immobilization by itself does not guarantee that individual proteins can be placed exclusively on specific sites on a device. To achieve placement, multifaceted approaches have been implemented that combine different techniques. For example, a regular array of Au bottom nanowells (100 nm diameter), fabricated using nanosphere lithography, was modified by covalent immobilization of biologically active cytochrome P450 (CYP2C9) via coupling to a SAM of octanethiol.[317] An array of 10 nm diameter Au nanopillars, functionalized with suitable SAMs, allowed reliable determination of β value (cf. Eq. 2) for alkanethiols that agreed with previous studies.[318] With a similar approach, Bostick and co-workers used an indexed array of gold nanopillars as individual pixels for protein immobilization to which CYP2C9 could be linked using a SAM comprised of a mixed thiol layer.[145,146] Indexing of the array allowed multiple studies on the same P450 enzyme exposed to different substrates.

## 3.4 Break-junction, Nanoscopic techniques

To help understand protein structure – function relations, and explore the use of protein properties for bioelectronics, the ability to study single proteins is an important complement to the monolayer studies discussed above. Measurements of monolayers provide average values of a large ensemble of proteins in various conformations, and thus the data represent a statistical average over all proteins that may hide specific transport features. In single molecule break-junction studies, a conductance trace across a single protein captures a time-averaged statistical histogram of conductance distributions over many conformations (> ~ 10,000 traces, depending on the temperature and how fast the data are acquired). Thus, the method can provide information on conductance variations over protein conformation fluctuations.

There have been several investigations of ET and ETp properties of single proteins. One method involves the use of electrochemical scanning tunneling microscopy (EC-STM; see section 3.6.2) to investigate electron transport across a single azurin protein.[179,198,319,320] An advantage of this method is the high spatial resolution of EC-STM that allows visualization of single proteins that are measured (Figure 14a). Although EC-STM allows confirmation of the existence of single proteins, it does not provide a means of controlling aggregation. Another disadvantage of this method is that it does not allow multiple measurements to be conducted on the same protein after removing the sample, which prevents systematic studies to be performed under different conditions. The recently reported possibility to use mechanical controlled break junctions in aqueous environment[321] opens the possibility of forming such junctions with proteins, which may improve the control over the force applied to this floppy system, something that is a problem in other single molecule junction methods (Figure 14b).

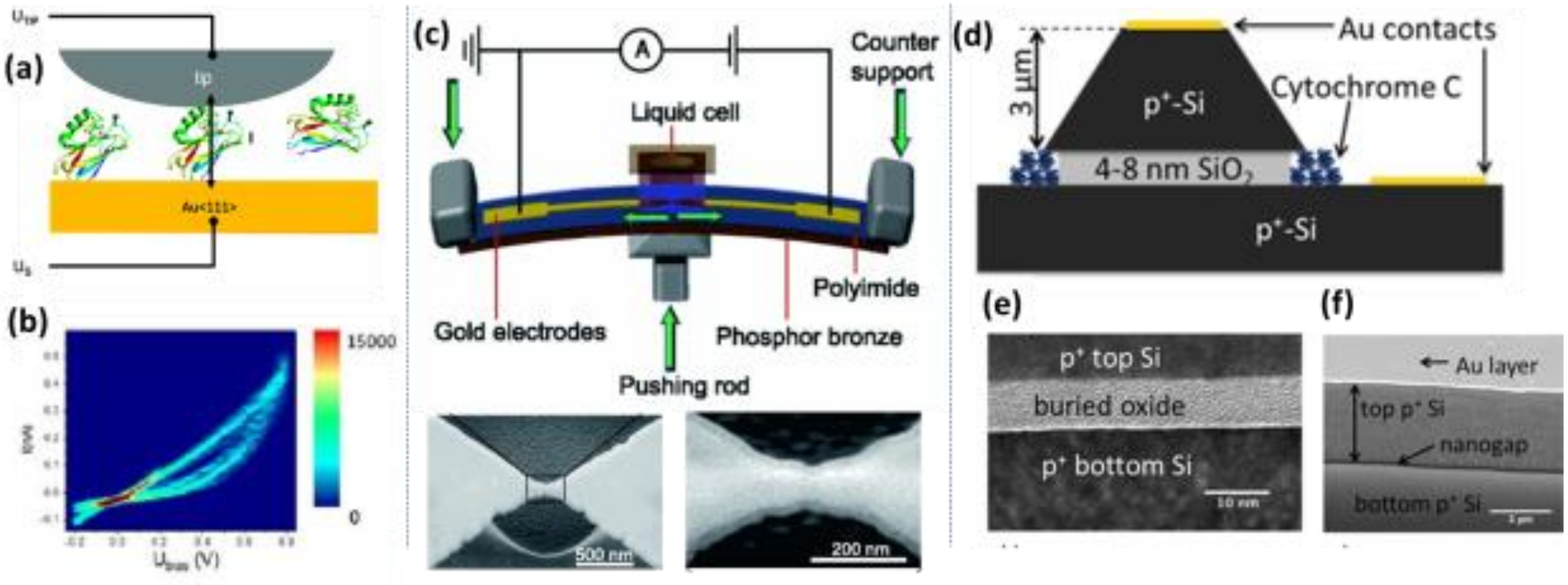

Figure 14. (**a**) Experimental electrochemical tunneling microscope, EC-STM, set-up for single protein junctions consisting of Azurin-bridged between a gold substrate and the probe of an EC-STM and (**b**) 2D I-V histogram showing two populations of curves in Azurin following its orientation in EC--STM junction configuration. (**c**) Schematic of the mechanically controlled break junction (MCBJ) with a liquid cell and scanning electron microscope images of electrodes in MCBJ devices. (**d**) Schematic of vertical nanogap device with protein molecules assembled in the gap (drawing is not to scale); (**e**) and (**f**): corresponding scanning electron microscope images. Images are reproduced with permission from references [31,320–322]

Another nanoscale technique for studying single proteins involves the creation of nano-gap electrodes through mechanical breaking[323,324] or electromigration techniques.[325] Electromigration has been used to create electrodes with 5 nm gaps for protein entrapment between Pt source and drain electrodes deposited on Si oxide/Si substrates.[32] This narrow gap allowed isolation and interrogation of molecular energy levels of apo-myoglobin or myoglobin single proteins.[29,32]

The availability of apo-myoglobin (myoglobin without the heme group) allows performing controlled experiments to determine the relevance of the heme group to ETp. The role of the heme group was studied explicitly by Li et al. using Pt break junctions deposited on a $SiO_2$/doped Si substrate. A drop of the protein solution was placed on the Pt junction, which was then cooled to 6 K and broken by applying a large voltage (electromigration).[32] In a successful sample, a single protein would fall into the ~5 nm gap, and using the Pt contacts as source and drain electrodes, and a bottom contact on the Si substrate as a gate electrode, a transistor structure was formed (Figure 15a). Coulomb blockade triangles were observed at 6 K in the drain-source vs gate voltage conductance maps for myoglobin samples, indicating the presence of single electron transistor (SET) behavior (Figure 15b), but not in any apo-myoglobin samples. During the tunneling process in a SET, the Coulomb repulsion between a tunneling electron and electrons at the contacts precludes simultaneous tunneling of more than one electron at a time, and the transport is characterized by Coulomb blockade triangles in the differential conductance data. The presence of Coulomb triangles in the myoglobin data proved that ETp across the protein can proceed via energy levels that allow resonant tunneling within the protein. We note, however, that the conductivities measured were an order of magnitude larger than those observed in semiconductor quantum dot SETs.[326–328] Inelastic electron tunneling spectroscopy (IETS), where the second derivative of the conductance data is measured to determine if vibrational levels are involved in the tunneling process (IETS is discussed further in Section 4.3.4), indicated that these resonant energy levels are associated with electronic energy levels of the heme group (Figure 15c).[32] The results of this work are consistent with ETp composed of an incoherent sequential tunneling process (cf. section 2.2) through the protein with resonant energy levels associated with the iron heme. The key role played by the heme group energy levels in the enhancement of the (ETp) conductivity in metalloproteins is consistent with the work of Raichlin et al.[87] on myoglobin and apo-myoglobin, and with some of the theoretical models discussed in Section 2 above, which rely on the alignment of internal energy levels for efficient ET and ETp.

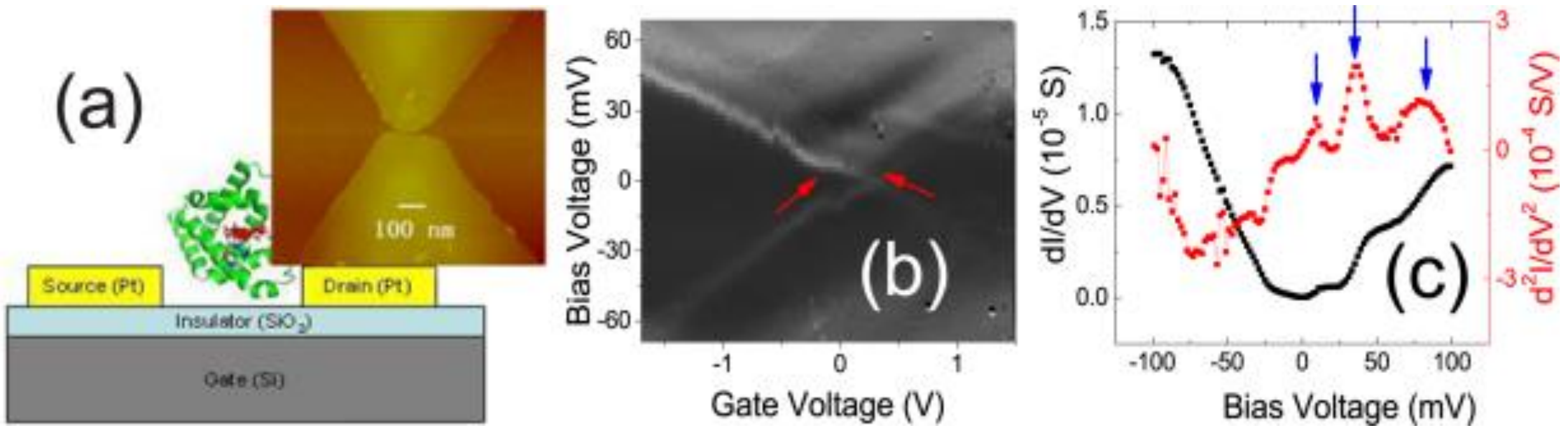

Figure 15. (**a**) Sketch of a three-terminal device and AFM image of a bare Pt junction broken by electromigration. The AFM image indicates that the gap is on the order of 5 nm wide. (**b**) Differential conductance (dI/dV) data from a Mb sample [gray scale from 0 (black) to 5.0 × $10^{-6}$ S (white)] at T = 6 K. The red arrows point to vibration-assisted (IETS) conduction lines. (**c**) dI/dV (black) and $d^2I/dV^2$ (red) spectra at $V_G$ = 11 V for another myoglobin sample at T= 6 K. The vertical arrows correspond to inelastic tunneling peaks (see section 4.3.4). Adapted with permission of reference 32.

In more recent work, ETp in cytochrome c was characterized with a vertical nanogap device.[322] The devices were comprised of all silicon contact and represent the first approach towards nanometer-spaced silicon contacts to proteins. The work demonstrated the ability to conduct nanoscopic solid-state ETp measurements via a protein, revealing thermal activation only above 140 K.

The success of these studies on heme-containing myoglobin and cytochrome c is proof that nanogaps can be applied to study other (metallo) proteins. However, nanogap contact configurations have several drawbacks. Creation of working junctions is inefficient (low yields) and, if not performed correctly, can result in probing of the characteristics of non-molecular components of the junction. For example, if metal nanoscale islands remain in the gap during the electromigration process, they could dominate the measurements instead of the molecule of interest.[32,329] In addition, to date it has been not possible to control the orientation nor to determine the conformation of the protein on the electrodes. More details about the nanoscopic techniques are discussed in Section 3.6.2.

## 3.5. Macroscopic and permanent contact measurement

Studies on the single molecule level are of great scientific importance, but reaching conclusions is limited by the statistics and reproducibility that can be achieved. Ron *et al.* established a broad experimental basis for studying ETp across proteins in a dry state, and thus, undoubtedly, at least a partially dehydrated solid-state configuration.[63,330] Measuring the collective electrical properties of an ensemble of molecules, in the form of a monolayer, led to highly reproducible results, which indicated electrical behavior of these macromolecules, comparable to that of organic molecules. Surprisingly, the ETp efficiencies (measured as currents at a given voltage) were more like conjugated than saturated molecules.[63] One advantage of using macroscopic electrical contacts on protein monolayers over nanoscopic methods is that protein monolayers can be characterized *in situ* by various spectroscopic measurements before a top contact is made (IETS, described in Section 4.3.4, is an exception). Along with obtaining information about protein structure upon assembling proteins into a device configuration, it is possible to verify protein orientation with respect to the electrodes and provide information about *in situ* structural alterations during electron transport. A systematic alteration of protein-electrode combinations allows to obtain an understanding of how protein/contact interactions can affect ETp processes.

For macroscopic contacts, a dense protein monolayer is prepared on conducting substrates, such as a highly doped Si wafer, or template-stripped gold/silver (rms roughness ≤ 1 nm), which can serve as substrates and bottom electrode of protein-based junctions.[89,90,151,219] Low roughness is important for macroscopic junctions, because otherwise the probability for electrical shorts is too high, leading to questionable results and/or low junction yields. Common top contacts are Hg or In-Ga eutectic (E-GaIn) electrodes.[258,331,332] The former has the advantage of being semi-noble, and with a surface that will thus be much better defined and cleaner than those of most other contacts, with the possible exception of evaporated Pb (see below). Its main disadvantage (in addition to its toxicity) is its relatively high vapor pressure, which allows for ready amalgamation with Au, unless an extra molecular layer is adsorbed on it

In-Ga was used recently, to fill PDMS micro-channels, resulting in a contact configuration with a reportedly better defined junction contact area than in earlier experiments, allowing recycling these junctions *with reproducible measurements*.[85,148] Another method for top contact formation is floating ready-made thin metal pads onto monolayers (Lift-Off Float-On, LOFO). The pads (circular Au films with > 0.1 mm diameter) are then transferred onto protein monolayers in a non-destructive way, serving as a top electrical contact. [90,151] This protocol helps avoid possible thermal damage to the proteins and metal penetration through the monolayer, both of which are likely to occur using direct thermal evaporation of most metals. An exception is thermal evaporation of Pb, which can be done under very mild conditions and which was recently shown to allow good contact to bR monolayers (cf. figure. 33), following earlier work with organic molecules.[333,334] These contacting schemes also shield the protein layer from the surrounding atmosphere, which is advantageous because it protects the proteins. The down side, however, is that the proteins are not accessible for chemical interactions with other molecules, which is possible in the nanoscopic contacting schemes discussed above.

To obtain junctions that are intermediate between macroscopic and AFM-type nanoscopic ones, photo-lithographically patterned Au substrates and gold nanowires can serve to form solid-state molecular junctions with protein monolayers.[335] Such junctions are also of interest for monitoring optical plasmonic effects.[336–340,335] Gold nanowires can be trapped on molecularly modified electrodes by AC dielectrophoresis.[341,342] As an alternative geometry, devices with electrode gaps of 10- to 50 nm, developed *via* complex (electron-beam) lithography have been used to trap proteins so as to allow transport measurements while still allowing the proteins to interact with the surrounding.[30,32,343]

Both macroscopic and nanogap junctions demonstrate possible avenues toward the fabrication of permanent electrical junctions, where proteins can be wired without harming them. Such a development will allow studying transport mechanisms via these junctions in detail, combining different experimental techniques, interaction with the environment, and using biological activity as an active electronic function.

## 3.6 Measuring protein adsorption and functionality on a surface

### 3.6.1 Optical techniques

Linking ETp measurements across proteins, especially enzymes, to their biological electron transfer function requires studies of functional immobilized proteins that are in conformations similar to those of their native states. To interpret results obtained from immobilized proteins, it is necessary to know the protein layer thickness, density of adsorbed proteins, their orientation, secondary and tertiary structure, protein stability, and most importantly, the effect of adsorption on their biological function.[172] Many studies of immobilized proteins use electrochemical techniques, in particular cyclic voltammetry, to ensure their electroactivity after immobilization. These studies can show if immobilized proteins are still capable of catalytic turnover, i.e., small molecule transformation. An example can be seen in studies of currents through [Fe-Fe]-hydrogenase, immobilized on Au (111), after adding its specific bound ligand, $H_2$ in this case. The current obtained was directly proportional to the protein's catalytic turnover rate and to the electrode potential.[344] Using this approach, the turnover rate has been determined for various hydrogenases.[345–347] The finding that the immobilized protein still retains its biological activity can also be supported via measurement of products, if electrochemical techniques are coupled to standard detection techniques such as high performance liquid chromatography and liquid chromatography-coupled mass spectrometry.[348]

Although these characterization methods offer no information on the orientation of the immobilized proteins, information can be obtained on the homogeneity, in terms of orientation and conformation of the adsorbed proteins, from the width of voltammogram peaks. For example, peak widths wider than 90 mV, the theoretical value for a typical single electron wave of an adsorbed protein,[349] point to heterogeneous protein orientation.[147]

Circular dichroism (CD) spectroscopy is a well-recognized technique for examining the secondary structure of proteins, especially the degree of helicity (essentially the presence of α-helices).[350] While solid state CD was applied to molecular monolayers recently, intermolecular interactions in the solid state do not give rise to large discrepancies between the L/R absorption spectra.[351,352] Thus, no CD data that provide convincing information on the conformation of the proteins that make up the monolayer have been reported. To overcome this limitation, CD measurements in solution of nanoparticles onto which proteins are adsorbed as monolayers are sometimes used to examine possible structural alterations after immobilization method..[353,354]

Some properties, such as heme spin state and binding of substrates to heme proteins, can be obtained readily with UV-Vis spectroscopy in solution.[355,356] For example myoglobin's optical absorption spectrum is characterized by its very intense Soret band (molar attenuation coefficient $\epsilon \approx 200$ mM$^{-1}$ cm$^{-1}$) at 409 nm[357,358] that can be detected even in a single monolayer with an absolute photoluminescence quantum yield spectrometer,[359] and which can be used to resolve protein aggregates by monitoring peak shifts.[360] Another example is that of cytochrome P450, which exhibits a Soret band at 450 nm ($\epsilon \approx 91$ mM$^{-1}$ cm$^{-1}$) in its reduced form bound to carbon monoxide.[361] This peak shifts to 420 nm ($\epsilon \approx 176.5$ mM$^{-1}$ cm$^{-1}$) if the protein is denatured, and can be used to estimate which fractions of the protein are native or denatured.[362]

A very sensitive technique for measuring attachment of ultra-thin films of atoms or molecules, including proteins, is that of a quartz crystal microbalance (QCM). The QCM technique measures changes in oscillating frequency of the substrate upon mass load to determine layer thickness.[172] Electrodes covering the QCM piezoelectric crystal can be coated with practically any desired thin film, allowing tailoring to the relevant immobilization scheme.

Optical techniques are widely used for monitoring protein structure, conformation and specific functions.[363,364] Proteins with photo-induced conformational changes (bR, Halorhodopsin, photoreceptor-YtvA, Dronpa,) show shifts in the optical absorption band when converted between light- and dark-adapted states, not only in buffer solution but also after immobilization on quartz as a monolayer.[90,365–367] In a few cases, a small shift in the optical absorption has been observed, which has been explained by the absence of hydration shells surrounding the optically active components upon monolayer preparation.[365] Recently, photoswitching was reported for Sn-substituted Cyt C and attributed to photoexcited hole transport.[258]

Different optical techniques are employed to determine monolayer surface coverage on different substrate and protein-electrode interactions. The thickness and quality of protein monolayers, prepared by chemi- or physisorption on different solid substrate surfaces, have been routinely monitored with spectroscopic ellipsometry (SE). SE is typically used for films with thickness in the range of less than a nm up to a μm or more, allowing for single protein layer resolution, and can be tailored by changing the wavelength of incoming light. However, using optical absorption often yields thickness values, that are less than the actual protein dimensions obtained from crystallography.[63,330] The sensitivity of spectroscopy ellipsometry techniques is approximately ± 5 Å, such that changes of protein monolayer thickness ( ~ 18 - 90 Å) on a different substrate surface can be monitored even after multiple layer applications.[367]

Other useful techniques are the earlier-mentioned PM-IRRAS and grazing angle attenuated total reflection (GATR)–FTIR spectroscopy, which can be used to characterize protein structure and orientation in thin films or monolayers on metal substrates. Surface-enhanced infrared absorption spectroscopy (SEIRA) can be used to probe a protein monolayer's functionality, such as label-free *in situ* study of biomolecular interactions.[368,369] All of these techniques can be used to determine if protein conformation and activity are affected by immobilization. Surface-enhanced resonance Raman spectroscopy (SERRS)[370,371] can also be used to study protein conformation and interactions by analyzing various vibrational modes of amino acids and cofactors.[372,373] Raman spectroscopy has been used to study conformation, and cofactor coordination of heme proteins,[374–376] and non-redox proteins.[377–379] SERRS can also yield information about protein stability and retention of native function. An example is a SERRS study conducted on a CY(101) immobilized on a SAM on Ag electrodes which showed that both the ferric ($Fe^{3+}$) and ferrous ($Fe^{2+}$) states of the protein were in an inactive form.[380] A more recent study demonstrated that cytochrome P450 2D6 (CYP2D6) immobilized via a SAM on Ag electrodes was able to reversibly bind its small molecule ligand, dextromethorphan, and retained its native structure.[381] However, Raman shifts were also observed, indicative of the inactive form, and it was not possible to reduce the protein heme group. This result is interesting given recent work, which demonstrated that biologically active CYP2D6 can be adsorbed via a SAM on Au,[382] and may reveal surface-specific differences in immobilized P450 function. Surface plasmon resonance (SPR) spectroscopy has been successfully used to study multi-layer /aggregation formation during immobilization (Figure 16).[383–386] The surface plasmon resonances are highly sensitive to the electrons' local environment. Attachment of molecules to the surface changes the local dielectric constant, and thus alters the resonance angle and finally the shape of the SPR absorption/scattering spectra.

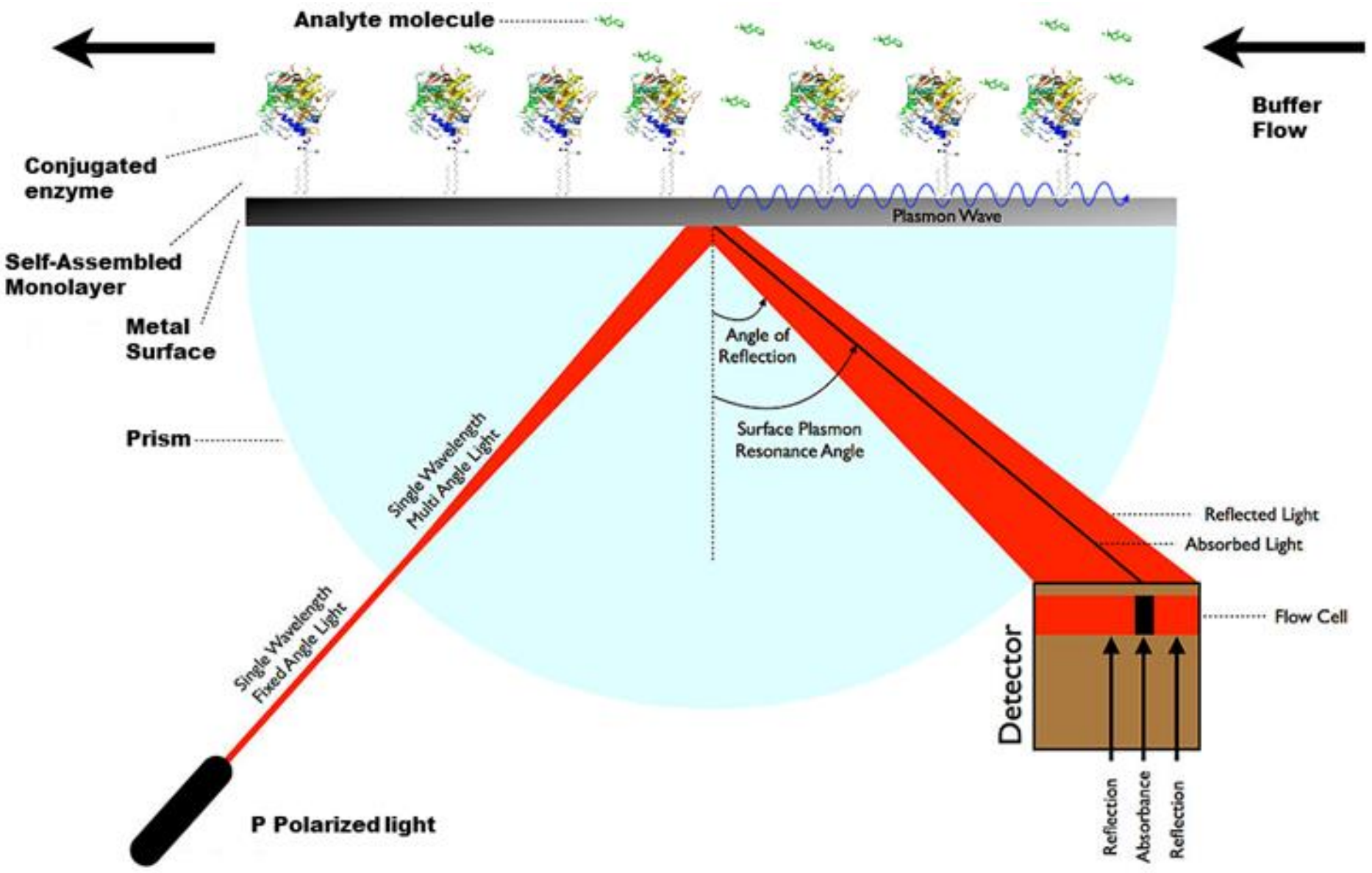

Figure 16. Diagram of surface plasmon resonance, SPR, spectroscopy with a cross-linked enzyme system. The black bar in the detector represents the resonance angle where the light is absorbed by the electrons in the metal creating the SPR dip. Adapted with permission from reference [387].

SPR studies in conjunction with electrochemical studies are used to determine electroactivity characteristics in protein assemblies. Jin *et al.* measured layer formation of alternating ds-DNA and Mb layers, and found that full electroactivity is seen only in the first layer with electron transfer occurring mainly through hopping.[386] In-depth studies by Advincula *et al*. using simultaneous SPR, AFM, and electrochemical measurements allowed for collection of dielectric, surface morphology, and electrical current data (Figure 17).[388,389] Together with electrochemical surface plasmon resonance (EC-SPR), the same group demonstrated protein adsorption of a soluble polypyrrole-terminated with the glucose oxidase ($GO_x$) enzyme (Figure 17B). When exposed to glucose, GOx is reduced and glucose in turn is oxidized and converted to gluconolactone and hydrogen peroxide. Of note in this study is the difference in glucose sensing, determined optically (Figure 17C) and electrochemically (Figure 17D), revealing that the choice of layer composition should be based on the relevant downstream application. In the present case, to create an optical sensor, an undoped polymer would allow for glucose sensing with higher sensitivity, whereas a doped polymer allows for enhanced electrochemical sensing. In other words, even when biological viability upon attachment is demonstrated, the details are important when a bioelectronic device is the desired end product.

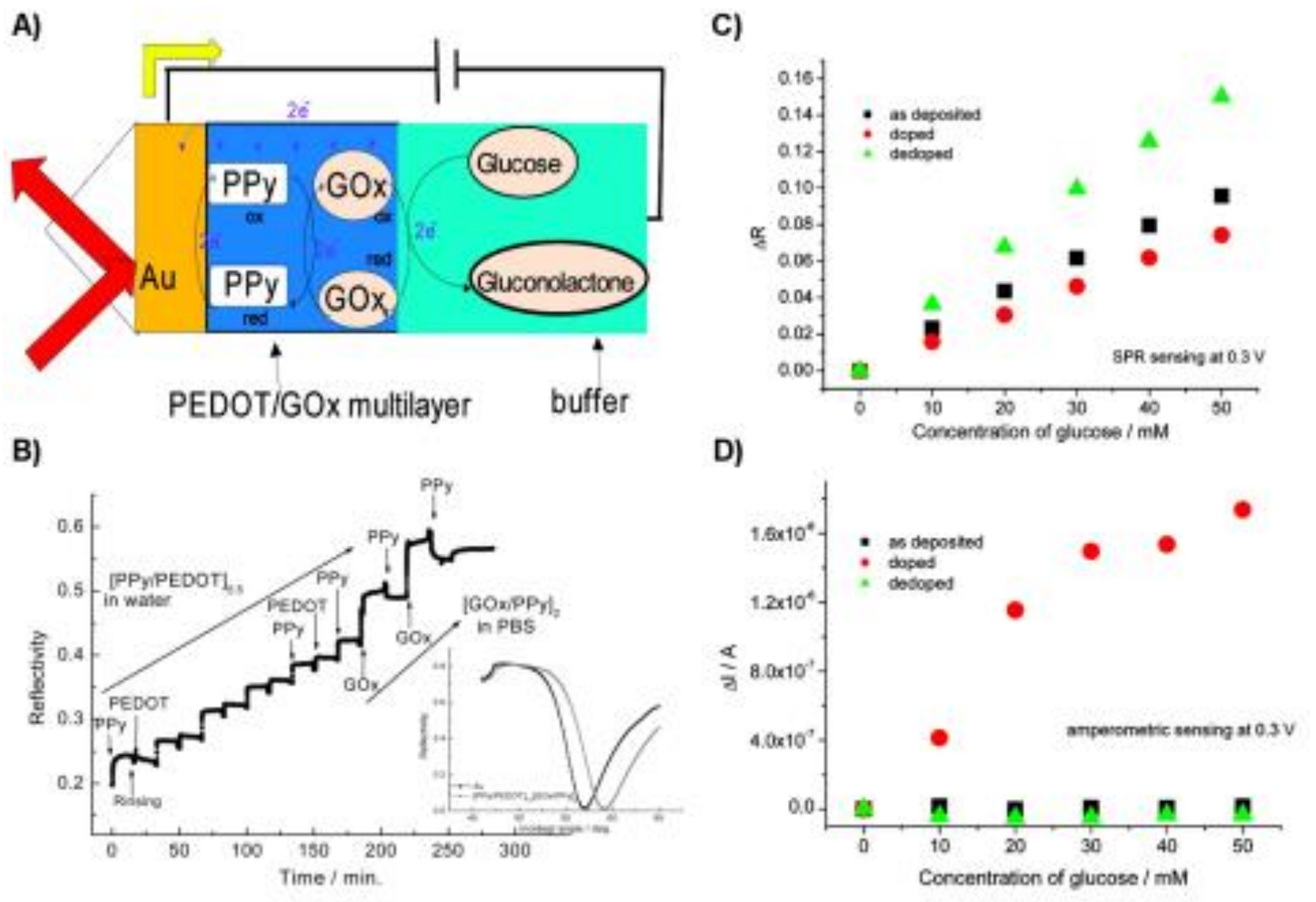

Figure 17: **A**) Setup of EC-SPR, with multilayer of water-soluble polypyrrole poly(N, N-dimethylethyl-3-(1H-pyrrol-1-yl)propan-1-ammoniumchloride) or (PPy), Poly(3,4-ethylene dioxythiophene) (PEDOT), and glucose oxidase ($GO_x$) immobilized onto a dielectric SPR platform (Au on glass. Buffer is phosphate-buffered saline (PBS). **B**) Real-time SPR plot showing changes in reflectivity upon adsorption of layers. *Inset* shows the change in SPR absorption dip, due to layer formation. **C**) SPR glucose sensing as a function of glucose concentration for doped and de-doped states of PPy. **D**) Amperometric glucose sensing as a function of glucose concentration for doped and de-doped states of PPy. Adapted with permission, from reference [389] Copyright © 2010, American Chemical Society.

### 3.6.2 Scanning Probe Technique

Scanning probe techniques have revealed a great deal of information about immobilized proteins due to their high spatial resolution. Scanning tunneling microscopy (STM) has a lateral resolution down to the atomic level, allowing single protein measurements, and in principle can also provide information on adsorbed protein structure. The high spatial resolution has also been exploited for a better understanding of immobilized protein orientation.[31,180,198,320]

Electrochemical STM (EC-STM) combines conventional STM and cyclic voltammetry.[179] The conducting substrate and STM tip act as working electrodes in an electrolyte solution, allowing measurement of ET. Reference and counter electrodes are located in the solution on opposite sides of the tip. The reference electrode is often an Ag wire, whose potential is measured to obtain a measurement that referred to the standard hydrogen electrode potential. A bipotentiostat allows independent control of the potential of the two working electrodes, the substrate and the tip, with respect to the reference electrode, and their potential difference represents the STM's tunneling bias. The counter electrode allows the current to flow.

In EC-STM, electrons tunnel between the working electrodes via the protein due to alignment of energy levels of the protein with the Fermi levels of the contact electrodes. This alignment can be obtained through application of a bias voltage or nuclear fluctuations of the protein.[319,390] The first single molecule studies of a redox-active molecule by EC-STM were conducted by Tao, who investigated the oxidized (hemin) form of the redox-active prosthetic group of heme proteins, Fe(III)-protoporphyrin (FePP), and the redox-inactive protoporphyrin (PP) which lacks the Fe ion.[391] Using PP as a reference, Tao demonstrated the apparent STM-measured height of FePP with respect to PP as a function of substrate potential that gave resonant tunneling-like behavior. In addition, in an STM image FePP had a larger apparent height than the reference PP at a substrate potential near the redox potential of FePP, resulting from the increase in tunneling current of the FePP, demonstrating the usefulness of this technique.[391]

Most extensive work by Ulstrup and co-workers combined cyclic voltammetry and STM to study a diheme protein, cyt c4.[392] STM demonstrated a controlled adsorption of cyt c4 (adi-heme protein, i.e., with one heme on each side) on a SAM in an orientation perpendicular to the substrate, with the heme at the C-terminal (i.e., the carboxyl group peptide's chain end) adjacent to the electrode and the heme at the N-terminal above it, and observed an interheme $k_{ET}$ of 8,000−30,000 $s^{-1}$. These values are much higher than those previously obtained by the same group using standard cyclic voltammetry (1,600 $s^{-1}$), a result that the authors explain was plagued with experimental error due to diffusion of the adsorbed protein, and lower current density due to an inability to generate well-oriented monolayers.[393]

EC-STM measurements have also provided mechanistic information about ET in redox proteins and changes in redox potential and conformation. For example, there have been extensive studies of Azurin using EC-STM, which have yielded mechanistic information on ET.[179] Gorostiza and co-workers used EC-STM to monitor changes in the barrier height for transport in relation to the copper redox state, and to calculate β from current distance measurements (cf. eq. (3) in section 2.1).[320] EC-STM experiments were performed also on other classes of immobilized proteins, including hydrogenases and oxidases.[345,394,395] Madden *et al.*[345] measured the rate of the reduction of a single $H_2$ molecule by a hydrogenase, C*a*HydA, immobilized as SAM onto Au(111), as a function of the applied potential.

Atomic force microscopy (AFM) is another scanning probe technique extensively used in protein surface characterization.[396] AFM measures the topography of a surface using a tip consisting of a small probe attached to a cantilever. Depending on the AFM method, the tip either makes direct contact with the surface (contact mode), or oscillates while driven close to its resonance frequency near the surface so that van der Waals forces cause changes in its oscillation phase and amplitude (tapping mode). AFM has been used to monitor protein interaction with lipid bilayers,[397,398] to determine protein surface coverage,[310,397,399,400] and to demonstrate protein-protein interactions.[401,402]

Conductive probe AFM (CP-AFM) is used to measure the local conductivity of a sample in contact mode. In CP-AFM, an electric potential is applied between the tip and the sample (generally via a conducting substrate for the sample), and the current that flows can then be used to measure ETp locally. Current images can also be obtained if the imaging mode is used. An important feature of CP-AFM is its ability to conduct electrical measurements as a function of applied force, which can be used not only to change the distance between the electrodes, but also to measure the effect of compression on the measured object. Proteins are relatively soft and their stability under compressive (and tensile) stress can affect ETp.[182,186] For example, force-dependent I-V studies have shown that holo-Azurin, i.e., with the Cu redox group present, retains higher stability with increasing force (until ~ 30 nN) than the protein from which the Cu has been removed, apo-Azurin. However, even for holo-Az plastic deformation sets in at an applied force above ~ 10 nN.[186,182,88,53] For a more complex protein such as bR, oriented with its long axis perpendicular to the substrate, and in ether one of the two possible remaining orientations, junction conductance linearly increases in the 3 – 10 nN applied force range reversibly, confirming retention of its native conformation. Under higher applied forces (> 10 nN), the protein exhibits structural deformations and the change in conductance with force is no longer reversible, i.e., plastic deformation occurs (Figure 18).[365] Force-dependent ETp investigations of Az, bR, CYP2C9, and other proteins after immobilizing onto gold substrates, have demonstrated added protein stability in the presence of cofactors or small molecule binders.[146]

The response to an applied force of the conductance of an Az monolayer under cyclic mechanical loading (with CP-AFM) resembled memristive behavior, which could be modeled empirically using the viscoelastic property of the protein. Importantly, the Az conductance was found to depend not only on the force magnitude, but also on loading time. A possible general conclusion could be that conductance, measured by CP-AFM, is process-related and dominated by the time integral of the applied force.[403]

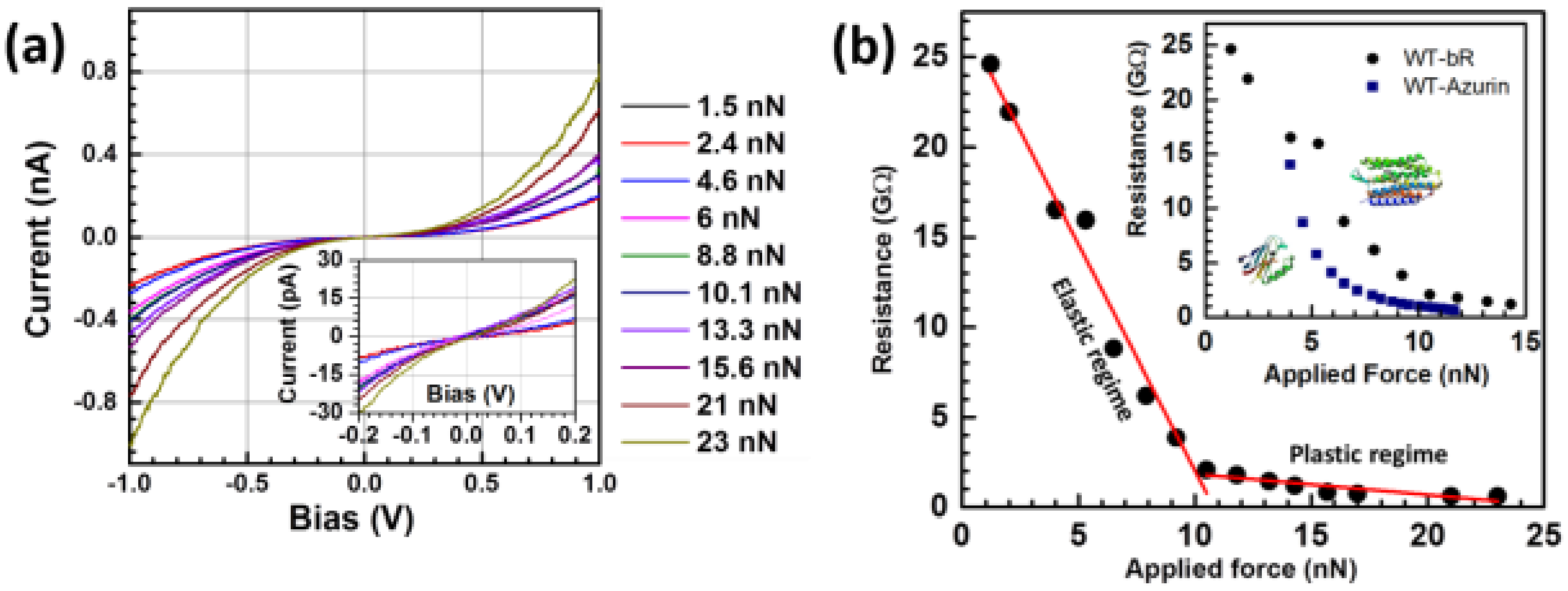

Figure 18. (a) Variation in ETp efficiency expressed by I-V characteristics, with applied force in protein-based conducting AFM junctions. (b) Resistance of bR-protein based CP-AFM junction with applied force (along with additive force of 3 nN). Inset shows same variation with proteins of different secondary structures. Adapted with permission, from reference [365] Copyright © 2014, American Chemical Society

Nanoscopic ETp using CP-AFM measurements showed high reproducibility if applied to the indexed nanopillar structure discussed in section 3 (Figure 19a) even on different days after P450 protein immobilization. The ability to complete repeated studies on the same enzyme demonstrated that ETp correlates with rates of metabolism, *i.e.*, the biological function of the cytochrome P450, and that competitive inhibitors not only block the active site, but also inhibit transfer of electrons through the protein (Figure 19b).[145,146] This exciting technique, which combines the high spatial resolution of CP-AFM with the high reproducibility of the indexed nanopillar architecture, allows for precise measurement of effects of small molecule interactions with proteins on ETp.

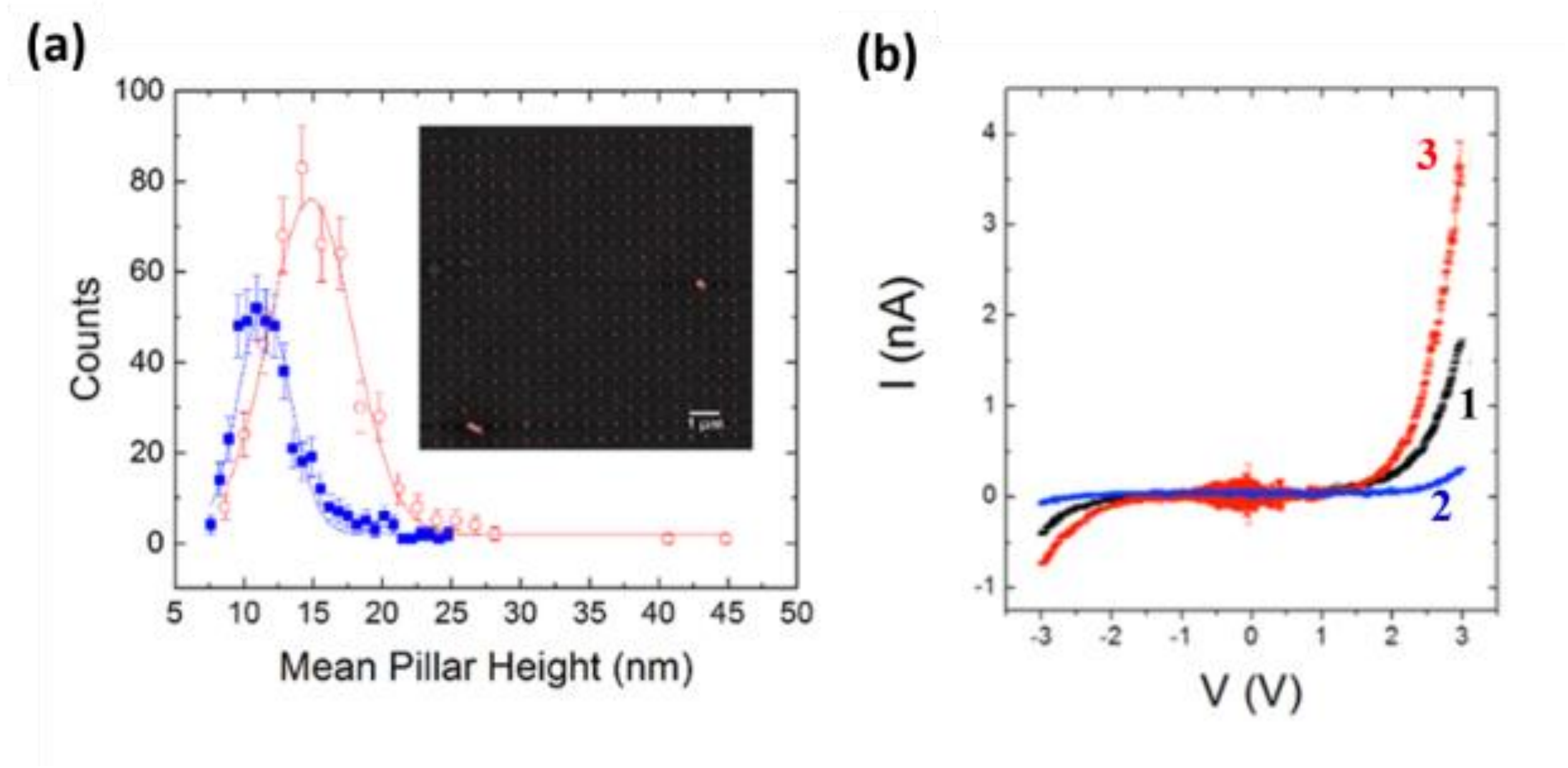

Figure 19. (a) Height distributions (fitted with a Gaussian model) for Au pillars before (blue squares) and after (red circles) cytochrome P450 (CYP2C9) attachment via tethering to a SAM of linker molecules. *Inset*: AFM height image of array with 421 Au pillars (20 - 40 nm diameter), used for protein attachment. The increased peak width indicates an increase in surface heterogeneity.[146] (b) Measured I-V curves (with standard error) for one pillar with (1) depicting nanopillar with attached CYP2C9, (2) representing measurement of the same protein in (1) exposed to sulfaphenazole, a small molecule that specifically binds *and inhibits its function,* and (3) the same protein in (1) exposed to flurbiprofen + dapsone, small molecules that specifically bind to this protein and coordinate with its heme group. A wash step with doubly deionized water was used to remove small molecules in between exposures. Adapted with permission, from reference [146].

### 3.7 Artifacts in ET and ETp measurements

Artifacts are associated with electron transfer and transport measurements, both in solution electrochemistry and solid-state experimental configurations. Artifacts in STM and CP-AFM measurements generally originate from protein denaturation and conformational rearrangements during the force application by sharp probes.[182,186] In some cases it is possible to detect changes in protein conformation *in situ* by IR-absorption measurements, Raman scattering, fluorescence and/or electrochemical characterization.[404,405] Gold substrate topography influences ET rates (investigated with interfacial electrochemical method) across cytochrome c, when adsorbed on carboxylic acid-terminated self-assembled monolayers.[406,407] With smoother topography, SAMs exhibit an increased ability to block a solution probe molecule, which decreases spurious signals from direct contact between the electrode and solution molecules.[406] The extent of physisorption and the magnitude of the electrochemical response of chemisorbed redox-proteins decrease significantly with decreasing substrate roughness. Roughness limits the electrostatically driven adsorption of proteins onto a SAM-covered surface and the effectiveness of the protein's electronic coupling to the SAM-covered substrate.[406] Substrate roughness also leads to artifacts in experimental electrochemical studies from topography-dependent acid/base properties of the SAMs.

In macroscopic transport studies, most of the artifacts originate from defects of the self-assembled monolayer, including pinholes and the random orientation of proteins in monolayers. In general, current density (at high bias) across the chemisorbed linker monolayer is $\sim 10^3$ times larger than that across a monolayer of proteins that is connected to the substrate surface via a monolayer of a < 1 nm long linker molecule, or directly.

Considering simple tunneling transport (Eq. 2), a decay length constant of $\beta \sim 0.12 – 0.3$ Å$^{-1}$ is obtained from solid-state electron transport (ETp) measurements across protein monolayers.[330] We note that this value is several times smaller than that obtained for ET.[65,97,121] Conduction through pinholes in a protein monolayer can only compete with direct transport via the proteins if the gap between the two electrodes is ~3 Å. Given that this will be a vacuum-, air- or maybe (adventitious) hydrocarbon-filled gap, this is extremely unlikely with protein monolayers of ~ 2.5 - 7 nm thickness considering the mechanical stiffness of the top electrode or the surface tension of Hg (inset in Figure 20), as explained in the next paragraph.

In the lower inset of Fig. 20 we show a representation of a monolayer-covered surface, which as an empirical grid, where each pixel represents a unit of a tunneling element, proteins (white) or pinholes (red). This is illustrated more graphically in the upper-right inset of the Figure. The figure itself shows the tunneling currents per pixel, as a function of the tunneling distance (*x*-axis), where the pinholes are filled with hydrocarbon, i.e., the medium with the smallest tunneling decay constant, compared to those of vacuum or air ($\beta \sim 1$ Å$^{-1}$). The plots start at tunneling lengths of 3 Å for pinholes with hydrocarbon impurity (red-white arched region, defined by a range of possible $\beta$ values of 0.6 – 1 Å$^{-1}$) and 20 Å for protein (black-white arced region, defined by β values of 0.12 -0.3).[63] It is clear that currents through hydrocarbon-filled pinholes in a protein monolayer (red-arced) will not be significant for a ~ 30 - 60 Å deep pinhole, compared to the currents through the proteins (black-arced). Thus, most of the current must flow through protein domains rather than through small pinholes with impurities in protein SAMs. If the pinholes are filled with air or vacuum, the probability decreases even further.

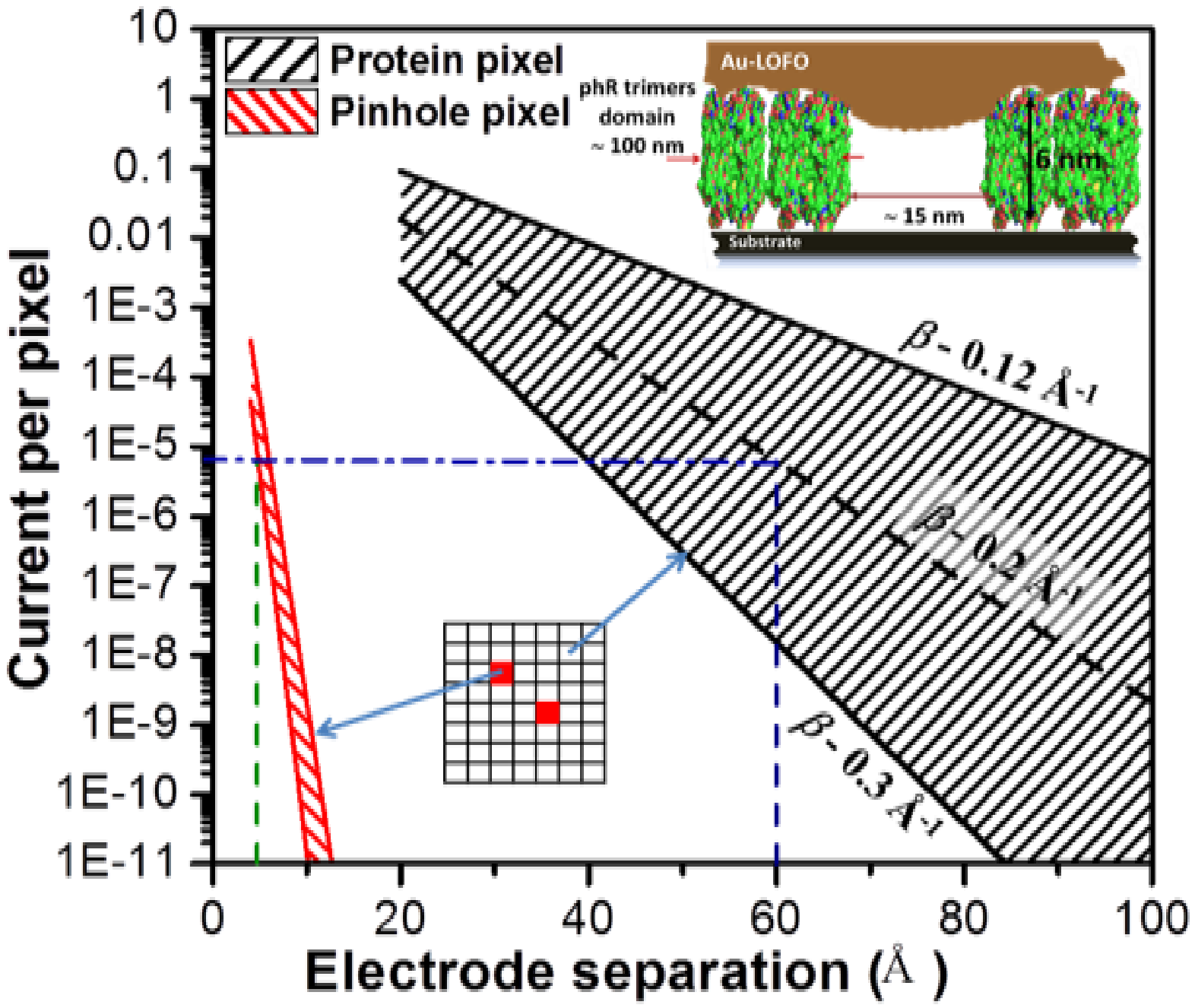

Figure 20: (a) Calculated approximate current (in normalized form) decay profiles over varying tunneling distances (defined by electrode separation) for pinholes and proteins. The *y*-axis is in units of amperes. *Lower Inset* shows an empirical grid where each pixel represents a unit of tunneling elements. A red filled pixel represents a pinhole pixel whereas a white one is a protein pixel. *Upper Inset* shows a schematic illustration of a pinhole as deduced from high resolution AFM scans. Adapted with permission from reference [90] Copyright © 2015, American Chemical Society.

It is relatively straightforward to calculate that the ability of liquid Hg (used for electrical contacts) to penetrate pinholes with diameters of ~ 15 nm between ~ 100 nm protein domains (as found from high-resolution AFM) is negligible compared to the few nm protein monolayer thickness, because of Hg's high surface tension. With solid contacts, such penetration is impossible because of their stiffness.

The next issue to consider for possible conduction artifacts that bypass the proteins is the roughness of the substrate (cf. section 3.5). Here rms roughness is not a good indicator, but rather, maximum roughness information is needed over $\mu m^2$ to tenths of $mm^2$ areas. To assure minimal roughness, highly polished doped Si and template stripped noble metals are desirable conducting substrates.

# 4. Electronic and structural properties of immobilized proteins

## 4.1 What can be expected experimentally from electronic and structural studies of properties by computational methods?

The ETp mechanisms discussed in Section 2.1 and illustrated in Figure 7 suggest that a variety of combined paths may exist in a given protein structure that connects donor and acceptor residues in an ET process. In computational models, after defining donor and acceptor residues in a protein structure, an algorithm is used to calculate the attenuation factors for a large number of paths that connect these two specific residues. The path with the smallest attenuation factor is selected as the preferred path. *In this pathway* method, protein structure and composition are computationally analyzed to determine electron transfer with exponential length attenuation factors that take into account the atomic structure of the protein. The method is derived from the protein physics: electronic interactions decay much more rapidly through vacuum than through chemical bonds, and the protein is viewed as fully coarse-grained, described by way of a dielectric permittivity of the medium and local environment. The simplest structured protein models employ tunneling pathway *and* average packing density analyses to identify specific tunneling paths and transport-mediating protein residues; the outcomes can be experimentally tested by protein mutation studies.[408]

Another interesting question is whether electron tunneling paths, selected by evolution are defined solely by donor-acceptor separation distance, or by other factors. Electrochemical studies on smaller bio-molecules, such as peptides and cysteine-linked peptide nucleic acids (PNA) immobilized on conducting electrodes, demonstrate that peptide backbones, and even via amino acid side chains, affect electron transfer processes. Experiments have verified that backbone rigidity and electron-rich side chains increase ET efficiency across peptide monolayers, supporting the idea that electronic state localization occurs on amino acid side chains.[409] The difference between the Fermi level of the gold electrodes and the nearest peptide orbital energy, determined from density functional theory (DFT) gas phase computations, has been interpreted as a transport barrier height for ETp studies across different homopeptide junctions. A direct comparison was made between a DFT-computed transport barrier and experimental results from a solid state system (monolayer on Au surface) obtained from ultraviolet photoelectron spectroscopy (UPS).[409,410] Detailed ETp studies with a nanowire junction configuration (cf. Fig. 27, below) showed that ETp efficiency across a series of different homopeptides directly correlated with the computed electronic barrier and UPS observations.

Preferred pathways for ET (in biologically-relevant, electrolyte solution conditions) via the peptide backbone might exist, depending on the peptide's secondary structure.[411] The effect of secondary structure on electronic transport has been experimentally demonstrated with molecular junctions of extended helical Ala and Lys homopeptide monolayers. Monolayers of 20-mer alanine and 20-mer lysine (neutral), which have a high propensity to form a helix, are only 4 Å longer than the respective extended hepta-peptides. Nevertheless, ETp comparisons showed that conduction via the helices was substantially higher than via the extended peptides.[155,410] In further work, coupling to the electrode contact was found to be critical for ETp efficiency across Au/7-peptide/Au junctions, as shown by comparing ETp between peptides with tryptophan (an amino acid with an aromatic indole side chain) in different positions along the chain.[403]

In most biological ET processes, the cofactor of the protein is thought to enhance the electron tunneling probability with respect to vacuum tunneling. Protein cofactor-mediated tunneling is generally described by superexchange, with the assumption that electron tunneling between donor and acceptor is mediated by unoccupied energy levels (for electron tunneling) or occupied states (for hole tunneling) of the cofactor, but the electrons that tunnel this way do not have any significant residence time on the bridge at any time. Similar to the pathway model, important amino acids or cofactors between donors and acceptors need to be selected to model superexchange-mediated tunneling in protein ET. This ET mechanism was developed assuming that the energy gaps between highest occupied/lowest unoccupied electronic energy levels of the bridging units and the redox-active donor and acceptor levels are large compared to the electronic coupling energy between donor/acceptor and the bridge.[57] Naturally, charge transfer can also occur via hole transfer (HT) superexchange. As an example, a bridge site close to the oxidized acceptor donates an electron to the acceptor and the resultant hole tunnels through occupied bridge states to the donor. In early studies involving periodic alkane bridges, Beratan and Hopfield showed that both electron- and hole-mediated propagation could be explored to model distance-dependent transfer decay.[412]

Theoretical models of ET processes (in solution) can be used to simulate ETp, in a solid state configuration, across proteins that connect two electronically conducting electrodes. Single proteins or monolayers of azurin, cytochrome-c, and myoglobin were modeled by a square potential barrier with an effective barrier height.[47,87,151,330] Tunneling can describe protein-mediated solid state ETp over length scales up to a few tens of Å, but it is not clear how it can serve to describe electron transport over longer distances (≥ 50 Å).[413] Very long-range ( > 100 Å) ET in biological systems occurs via redox-active cofactors using chains with close cofactor spacing of typically 15 - 20 Å. Transport across photosynthetic and mitochondrial membranes includes protein-protein ET, and, within a protein or protein complex ,is generally described as multi-step hopping between multiple redox cofactors, or via multiple aromatic residues.[97,137] Metal-containing cofactors, protein residues of tryptophan, tyrosine, or cysteines may also act as electron relays, in e.g., ribonucleotide reductase, photosystem II, DNA photolyase, and cytochrome c/cytochrome c peroxidase.[65]

Identifying the transition between coherent single-step transport (tunneling) and incoherent multistep tunneling theoretically and experimentally remains a great challenge. Recent computational studies on charge recombination between hemes in the cytochrome c/cytochrome c peroxidase complexes explored the tunneling vs. hopping transition as function of transport distance.[414] The analysis indicates that even for moderate donor–acceptor separation, hopping involving a tryptophan (residue 191) governs ET.

Experimental ET studies of complex mutants with varying donor–acceptor distance indicate that coupling pathways and reorganization energies dictate ET kinetics and also reveal how hopping through aromatic residues can accelerate ET.[415]

Open questions remain concerning electron transport in biology over much larger distances. As a rather *ad hoc* speculation, band-like coherent conduction was proposed as an alternative to hopping for transport along biological nanowires (see Figure 21).[416] Such an idea is problematic using an electronic band model, because electronic bands originate in solid-state physics for systems that obey the Bloch condition of periodicity, and is described in its simplest (1-D) form by the Kronig-Penney model. Possibly models like those used for liquid metals and semiconductors may need to be revisited, but to the best of our knowledge, that has not yet been done for extra long-range ETp in biology.[417] A very recent report argues, based on experimental evidence, that the mechanisms for all long-range transport reports are electrochemical nature.[418]

Flickering resonance (FR), mentioned in sections 2.2 and 2.2, and in the discussions of Figs. 5 and 7, has been used to describe multistate biological electron transport across a chain of biological redox active units with donor, acceptor, and bridge levels that are similar in energy. The transport efficiencies expected from this model were compared with those of the established superexchange (SE) and hopping models.[110,134,135] To understand FR, we note that site vibrational energies in biological systems can broaden electronic energy levels on a scale of 100s of meV. Thus, short-lived (fs scale) coherences could form among multiple electronic states and create band-like states. Such transient situations would contribute to ET charge transport rates with a soft exponential distance dependence (small $\beta$ values). This is the process that lies at the heart of the flickering resonance idea and could be important in the 1-10 nm range, before incoherent hopping becomes the more efficient transport process. In the FR model, the medium between donor and acceptor is not only a bridge enhancing electronic coupling (as in the superexchange model), but is also a chain of redox sites, each of which can participate in electron or hole transport *when in resonance*. Long-range transport is assumed to take place when thermal fluctuations bring the redox-active energy levels of donor, bridge, and acceptor sites simultaneously in resonance. In contrast to charge hopping, where the carrier moves sequentially from one site to the next with nuclear relaxation at each step, charges in flickering resonance are assumed to move through the 'in resonance intermediate-states' without further nuclear relaxation until they reach the acceptor. (Figures 5 and 7).[110]

The FR model accounts for both exponential distance dependence and partial excess charge location on the bridge. It has been successfully implemented to analyze experimental results obtained for short-range transport in DNA instead of the superexchange model.

The riddle of transport across bacterial nanowires (essentially parts of the *Shewanella oneidensis* membrane)[416] and across pili (hair-like bacterial structures)[419] of *Geobacter sulfurreducens*, inspired a flickering resonance model to explore the physical constraints under which high currents could flow across such wires (Figure 21).[419,420] It was found that a combination of thermal fluctuations and hopping-like transport could reproduce the experimental I–V curves, with plausible values of reaction free energies, reorganization energies, and a packing density for redox cofactors (e.g., hemes), consistent with that of multi-heme proteins found on the bacterial cell surface.[133,135,421,422] However, one could argue that FR should be less relevant for ET along heme wires in proteins because of the relatively small electronic coupling between heme cofactors. While clearly the last word on ET (or ETp?) across bacterial nanowires and pili has not been written,[416] [419,420] we paraphrase Blumberger[110] and note that it is highly desirable to design experiments, motivated and guided by theory, that can elucidate the true mechanism(s) of electron flow in proteins.

There is no *a priori* reason to think that nuclear relaxation could not occur during resonant tunneling processes itself. As an example, consider the resonant tunneling in a metalloprotein discussed in Section 3.4 above. If tunneling occurs via resonance with a heme group, when the electron reduces the Fe atom during the tunneling process, a structural change could occur, depending on how fast is the tunneling process. If the resonant tunneling process is slow, a nuclear configurational change could occur which changes the energy level of the resonance. This is analogous to Franck-Condon transitions in molecules in which electronic transitions (induced optically, for example) result in a change in the nuclear configuration and an excited vibronic state. In ETp, if the occupation of the excited electronic state by the tunneling electron is long enough, this causes the nuclear configuration change. Li et al. have shown that theoretically this could be identified in a single electron transistor (SET) measurement by a gap in the SET Coulomb triangles, that is, there would be a vertical separation between Coulomb triangles at the point where they join to normally give a non-zero differential conductance at zero bias (see Figure 15).[32] However, the experimental data are somewhat unclear about this,[32] and so far this Franck-Condon-like effect in tunneling phenomena in proteins has not been unambiguously identified.

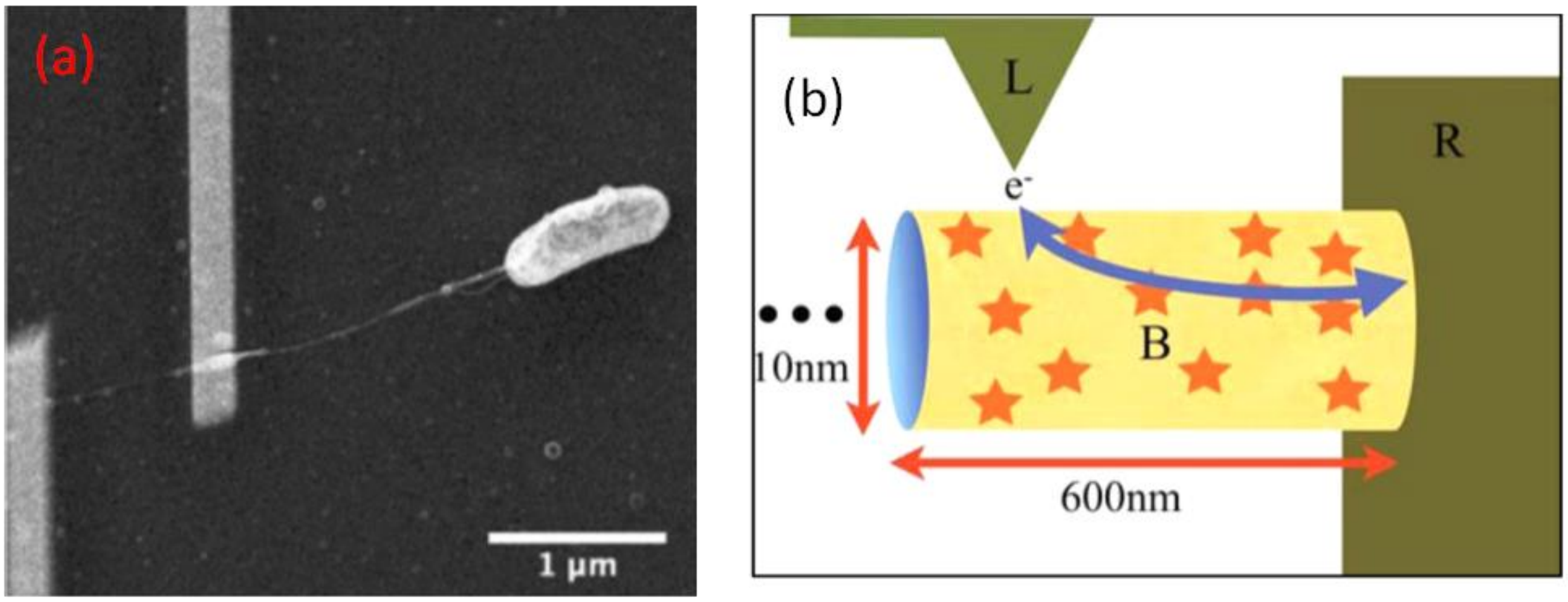

Figure 21. (a) ETp studies across bacterial nanowire (extensions of the membrane of the bacterium) from *Shewanella oneidensis* MR-1 cell with Pt electrodes as macroscopic contact. (b) Cartoon of possible efficient ETp process in bacterial nanowires via hopping sites (star shapes) which mainly originates from closely spaced heme group. Used with permission from ref [135]. Copyright 2012 Royal Society of Chemistry.

Recently, solid state conduction efficiency and existence of multiple transport pathways across bacteriorhodopsin, bR, were investigated theoretically using the extended Marcus–Hush theory.[423] The computed electronic charge mobilities of bR ($1.3 \times 10^{-2}$ and $9.7 \times 10^{-4}$ $cm^2/(Vs)$ for holes and electrons, respectively) based on a hopping mechanism were remarkably similar to the experimentally obtained average electron mobility of $9 \times 10^{-4}$ $cm^2/(Vs)$ for a system consisting of bR, bound to a $TiO_2$ nanowire.[424]

Overall, it seems that only a combination of models described in Section 2 and earlier in this section can quantitatively describe some of the experimental ETp data well, which indicates that our fundamental understanding is incomplete. Further control experiments, together with sophisticated computational work, are needed to improve our understanding of ETp in proteins. Obtaining detailed, microscopic information about the protein's structure in the experiments may be necessary to determine which of the models fits the electrical conductivity data best. The ideal experiment should use samples where the protein structure is well understood, the protein is not denatured, and the protein binding configuration can be measured and varied independently of the ETp measurements. While the various conditions have been met separately, realizing such an experiment remains one of the great challenges of the field. In this section, we describe various efforts aimed at dealing with these problems.

## 4.2 Effects of immobilization on protein structure & electronic properties

Protein immobilization on metal surfaces appears to be advantageous for the study and use of protein's biological activities under solid-state conditions. In most cases, proteins retain their biological functionality when assembled on metal electrodes functionalized with self-assembled monolayers (SAMs) via electrostatic and hydrophobic interactions as well as by covalent cross-linking, and only a few studies report conformational transitions and redox potential shifts of redox active proteins. For example, the redox potential of the electrostatically adsorbed P450 protein decreases upon shortening of the length of the carboxyl-terminated SAMs, i.e., upon increasing the strength of the local electric field.[380] Thus, it is important to summarize how immobilization affects protein structure and electrical properties. Upon immobilization, two competing effects modulate the redox potential of the adsorbed protein. First, the increased hydrophobicity of the redox environment brought about by immobilization on the SAM tends to increase the redox potential by stabilizing the formally neutral form. Second, increasing the electric field tends to stabilize the positively charged oxidized form, thus reducing the redox potential. Unlike solution measurements, immobilization affects solid-state ETp processes. From the device perspective, here we summarize ongoing research of immobilization effects on protein's redox and electrical properties.

### 4.2.1 Effects of choice of modified substrate surfaces on electron transfer

Few proteins directly adsorb on a solid-state substrate without substrate or protein modification. As discussed above, direct binding offers poor control of protein conformation and often results in reduced biological activity due to denaturation, and in the case of enzymes, by blocking active sites. These issues not only hinder protein functionality, but also hamper analysis of results due to the ambiguous nature of the protein state on the electrode surface.

A commonly used method allowing some protein conformational control is the direct immobilization on bare gold using surface-exposed cysteine groups of the protein via the formation of Au-S bonds. This has been used successfully implemented in yeast cytochrome c (YCC) to achieve electroactivity of immobilized species, as measured by cyclic voltammetry on a polycrystalline bare gold wire.[192,425] Studies of cytochrome c immobilized on Au (111) via Cys102 (cysteine at position 102 in the protein amino acid sequence) [426] yielded a $k_{ET}$ =1.8×10$^{3}$ s$^{-1}$; work on the immobilization of cytochrome $C_{555m}$ via Cys18 yielded a $k_{ET}$ =1.4×10$^{4}$ s$^{-1}$. [427] Thus, extremely fast ET, with a large $k_{ET}$, is possible through direct electron transfer between the electrode and the covalently immobilized protein.

Azurin, which has a surface disulfide (Cys3-Cys26) that is suitable for covalent linking to Au (111), has been immobilized on Au via these disulfide groups while retaining its structure.[50,118] This has allowed azurin to be probed electrochemically by both CP-AFM[182,186] and EC-STM.[179] However, several studies have shown that immobilization of azurin on Au (111) by this method produces an electrochemically-inactive protein with $k_{ET}$ values (30 s$^{-1}$ and 300-400 s$^{-1}$)[118,428] that are significantly slower than that of azurin adsorbed onto pyrolytic edge-plane graphite, for which $k_{ET}$ values are as high as 5000 s$^{-1}$.[224] Davis and co-workers have demonstrated the importance of the orientation by measuring electrochemically active azurin on Au (111) with the cysteine residues that bind to Au on different parts of the protein's surface, with limited perturbation of the native structure.[428,429] Of note is that the fastest $k_{ET}$ =300-570 s$^{-1}$, obtained for an Az mutant with a cysteine near the copper center (S118C), is still an order of magnitude slower than for azurin, adsorbed to pyrolytic edge-plane graphite electrodes. These observations indicate that, along with protein orientation, adsorption modes on specific substrates do effect ET.[118]

Direct absorption of cytochrome P450 proteins, first reported by Hill and co-workers who immobilized CYP101 to edge-plane pyrolytic graphite (PG) electrodes,[430] has been less successful. The study concluded that electrostatic interactions by positively charged amino acids (Arg-72, Arg-112, Lys-344, and Arg-364) located near the redox center with the negative charges on edge-plane PG allowed for electrochemically driven catalysis yielding substrate potentials. Work by Fantuzzi *et al.*[431] demonstrated that P450 CYP2E1 adsorbed directly to bare glassy carbon electrodes is more electrochemically active, but with very slow $k_{ET}$ (5 $s^{-1}$). In a recent study, the electrochemical (redox) activity of WT CYP101 was compared to that of mutants with all surface cysteines replaced by inert alanines (surface cysteine-free, SCF) and a mutant containing only one surface cysteine near to the redox center (SCF-K334C).[432] The single cysteine mutant yielded anodic (-50 mV vs. SCE) and cathodic (-170 mV vs. SCE) peaks in cyclic voltammograms (Figure 1), while no difference from background was observed in the WT and SCF mutant. The result indicates that multiple inactive conformations of the WT P450 protein may exist, and that the enhanced electroactivity of SCF P450 may originate from the differences in P450 orientation. In general, protein immobilization in a controlled orientation helps to bypass difficulties such as electroactive prosthetic groups buried within the protein, denaturation of proteins upon adsorption, and unfavorable protein conformations/orientation at electrodes.[221,290,433]

Studies by Waldeck and co-workers have demonstrated that the $k_{ET}$ of cytochrome c varies from ≤$10^{-4}$ to ~$10^{-1}$ cm/s, depending on the type of SAM used.[244,434] Using alkanethiol linker molecules to tether the protein to the Au electrode increased $k_{ET}$ because the orientation of the protein was changed so that the redox center faced the gold electrode upon adsorption. In contrast, when the protein is adsorbed on bare gold, the redox center is close to the protein side opposite to the gold electrode. In addition, by changing the alkanethiol length, the interfacial $k_{ET}$ could be controlled.[50,118,435]Hydrophobic surfaces, obtained from adsorption of methyl-terminated hexanethiol on gold, have been exploited to mimic the interaction of a biological redox partner, binding to the protein, thus preserving native conformations and enhancing conduction.[436]

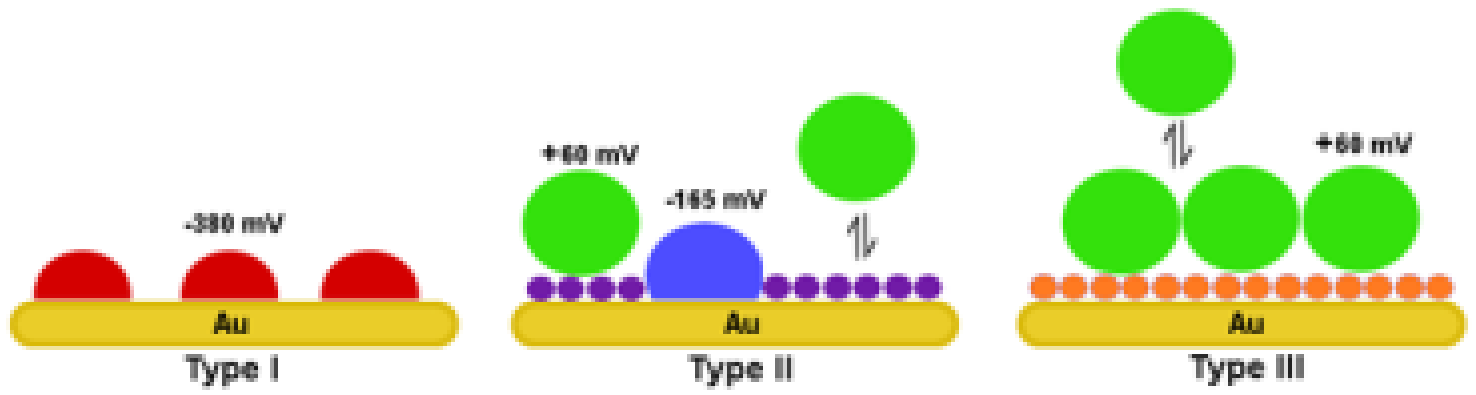

Figure 22. Schematic presentation of Cyt C (circles) adsorbed to a Au electrode without modifiers (Type I), co-adsorbed with 4,4'- bipyridyl (purple, Type II), and co-adsorbed with bis(4-pyridyl) disulfide (orange, Type III). In Type I and Type II adsorption denaturing of the protein (depicted as flattening of the circle that represents the protein) is seen, due to interaction with the electrode. Red depicts highly denatured Cyt C, while the blue depicts partially denatured Cyt C. In Type II we see some, and in Type III complete preservation of the native conformation (green) through interaction with the modifier on top of the electrode. Denaturing of Cyt C strongly affects the protein's redox potential. Adapted with permission, from reference [437] Copyright © 1990, American Chemical Society.

Using various 4-pyridyl derivatives as linker molecules, it was confirmed that two scenarios exist, one in which the protein adsorbs on top of the surface modifier and retains its native confirmation, and another where the protein adsorbs both on the modifier and on the electrode surface[437]. Therefore, denaturation of the protein can be mitigated by modifying the electrode, and a proper choice of modifier is necessary to retain full protein functionality. Also, changes in the redox potential are linked to the amount of unfolding of cytochrome c caused by adsorption (Figure 22).[25] While the $k_{ET}$ of myoglobin on bare PG has been shown to be so small and irreversible that it is almost impossible to measure,[277,438] the reversible ET rate in myoglobin on PG, modified by a positively charged polymer (poly(ester sulfonic acid)), was sufficiently fast to be measured ($k_{ET}$ = 52 ± 6 $s^{-1}$).[296] Using Nafion cast from aqueous mixtures on to the basal plane pyrolytic graphite, reversible CV peaks were measured for cytochrome $c_{551}$, cytochrome $b_5$, and azurin films.[439] This demonstrates that surfaces can be specifically tailored through modification to promote effective conduction in different proteins.

With glassy carbon electrodes (GCE), immobilized CYP101 has a positive shift of the redox potential (-361 mV vs Ag/AgCl) in comparison to CYP101 in solution (-525 mV vs Ag/AgCl). Immobilization on hydrophobic substrate causes partial dehydration of CYP101, excluding water from the heme pocket changing the coordination of the heme iron following shift from low to high spin.[68] Sodium montmorillonite modified GCE with immobilized myoglobin also yielded redox-active protein with quasi-reversible CV peaks, and the ability to detect nitric oxide concentration through changes in heme coordination.[280] The examples discussed above demonstrate heterogeneous values of $k_{ET}$, which were not found to be influenced by the co-adsorption of the second heme protein, leading to the possibility of developing biosensors capable of simultaneously measuring different analytes.

A quartz crystal microbalance method is another important technique to demonstrate regular and reproducible layer formation via alternating layers of proteins utilizing polyions such as cationic poly(ethyleneimine) (PEI) or anionic poly(styrenesulfonate) (PSS) on gold.[113,440] It has been experimentally demonstrated that styrene epoxidation enhances turnover on a carbon cloth electrode when functionalized with protein multilayer, which could be used as sensitive protein multilayer devices.

### 4.2.2 Effects of modified substrate surfaces on protein conformation and electron transport

Two fundamental issues for the field of bioelectronics are (i) how to order proteins on the electrodes with respect to their conformation, position of active sites and (ii) how to affirm electrical contact area between the protein and the electrode. Immobilized protein orientation and electrical contacts significantly alter electron transport mechanisms and junction properties in ETp measurements. Macroscopic and nanoscopic measurements of photosynthesis proteins immobilized on metal substrates show different electrical conductivities based on protein orientation.[85,441,442] Control over the orientation of the photosystem I (PSI) protein complex and especially of its reaction center (RC I; this is the protein part without the majority of chlorophyll molecules, which is used for light-harvesting) on surfaces can be achieved by its direct covalent attachment. This control can be achieved by mutations which provide an exposed cysteine amino acid (has an SH group available for binding) and/or by modification of the surface with functional groups that can bind to exposed amino acids or interact electrostatically with different parts of the protein complex.[441,443]

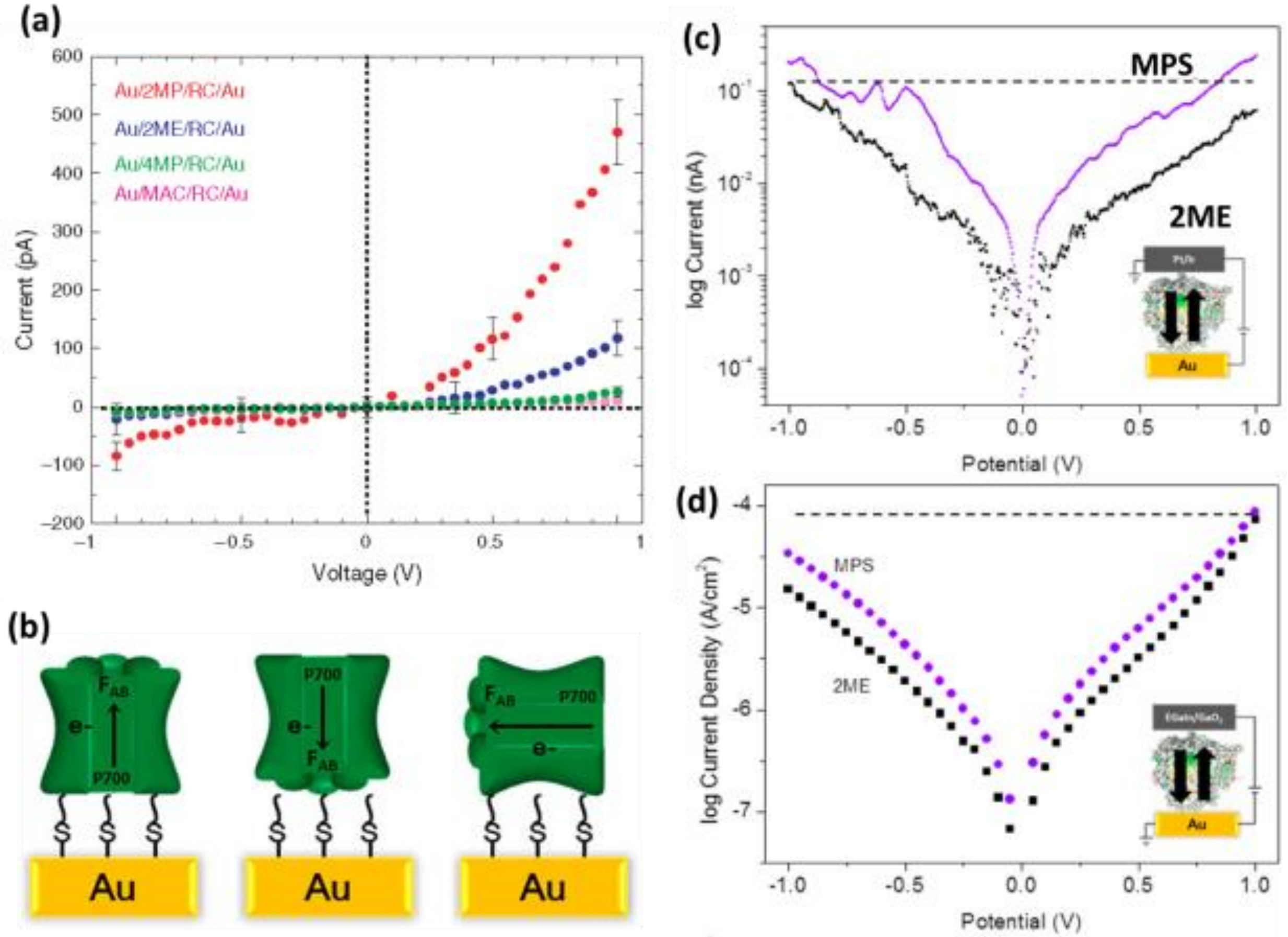

Figure 23: (a) I–V curves of bacterial reaction centers of photosystem I, PSI, deposited onto Au substrates modified with different linker molecules including 2-mercaptopyridine (2MP), 2-mercaptoethanol (2ME), 4-mercaptopyridine (4MP), 2-mercaptoacetic acid (MAC); error bars indicate the variability between different sets of I–V curves. (b) Schematic of possible orientations of PSI on chemically modified Au surfaces. Semi log plots of current (c) and current density (d) versus voltage for junctions measured using CP-AFM (c) and EGaIn (d) for SAMs of PSI on sodium 3-mercapto-1-propanesulfonate, MPS (purple) and 2ME (black), respectively. Black arrows in inset schematic indicate the two different orientations of the PSI complexes on the surface. Reprinted with permission, from references [85,442,444]. (a) Copyright © 2008 American Scientific Publishers.

ETp has been measured across the RC from *Rhodobacter sphaeroides* on SAM-modified Au (111) substrates using CP-AFM. Orientation-dependent ETp efficiency was determined using different linkers to the Au substrate, as shown in Figure 23a and 23b. The large tunneling distance (~7.5 nm) rules out direct tunneling as an ETp mechanism, and the variations in the tunneling current thus support the existence of alternative electron transport pathways and electronic states (localized or delocalized over the entire structure) across the proteins complex. Slight rectification (at negative bias range) in the cases of the mercaptoethanol SAM-modified Au (111) electrode (Figure 23a and 23c) reflects preferential adsorption of the RC such that electrons from the Au-coated AFM probe flow to the chemically modified Au(111).[442] The variations in transport efficiency with PSI and RC orientation mainly originate from electronic interactions between protein complexes and substrate energy levels, which affect the electron injection barrier.[85] For efficient electron injection into proteins, the work function of the electrode has to match the energy level of the lowest unoccupied molecular orbital (LUMO) or highest occupied molecular orbital (HOMO) of the conducting species,[445] taking into account that the work function of the electrode will be affected by the SAM. In these studies, electrode modification was carried out only on the conducting substrates, which seems to contradict the effect of orientation on ETp. Orientation-dependent conductance effects need to be determined by measurements that combine modified-Au (111) substrates and modified-Au-coated probes, which eliminate the effect of lowering carrier injection barrier only at the substrate-protein interfaces.

An investigation of CytC protein monolayers in a sandwich configuration indicates that covalent protein–electrode binding significantly enhances electronic interactions, as currents across CytC mutants bound covalently to the electrode via a cysteine thiolate are orders of magnitude higher than those through electrostatically adsorbed CytC.[150,156] Seven different mutants of CytC, where a cysteine (Cys) residue were incorporated at different locations on the protein's surface and bound covalently to the electrode, forming seven different orientations of CytC with distinct electrode–heme distances (Figure 24). The differences in electrical current magnitude efficiency with different CytC mutants indicate that ETp mostly depends on electrode–heme distance rather than on the separation distance between electrodes.[446,447] A configuration with small electrode–heme distance provides efficient electronic interaction/coupling, allowing for more efficient sequential tunneling from heme to one of the contacts at low $T$ (80 – 130 K) and thermally activated transport dominated via electrode (silicon wafer)-protein coupling at higher $T$ (> 150K).[86,446]

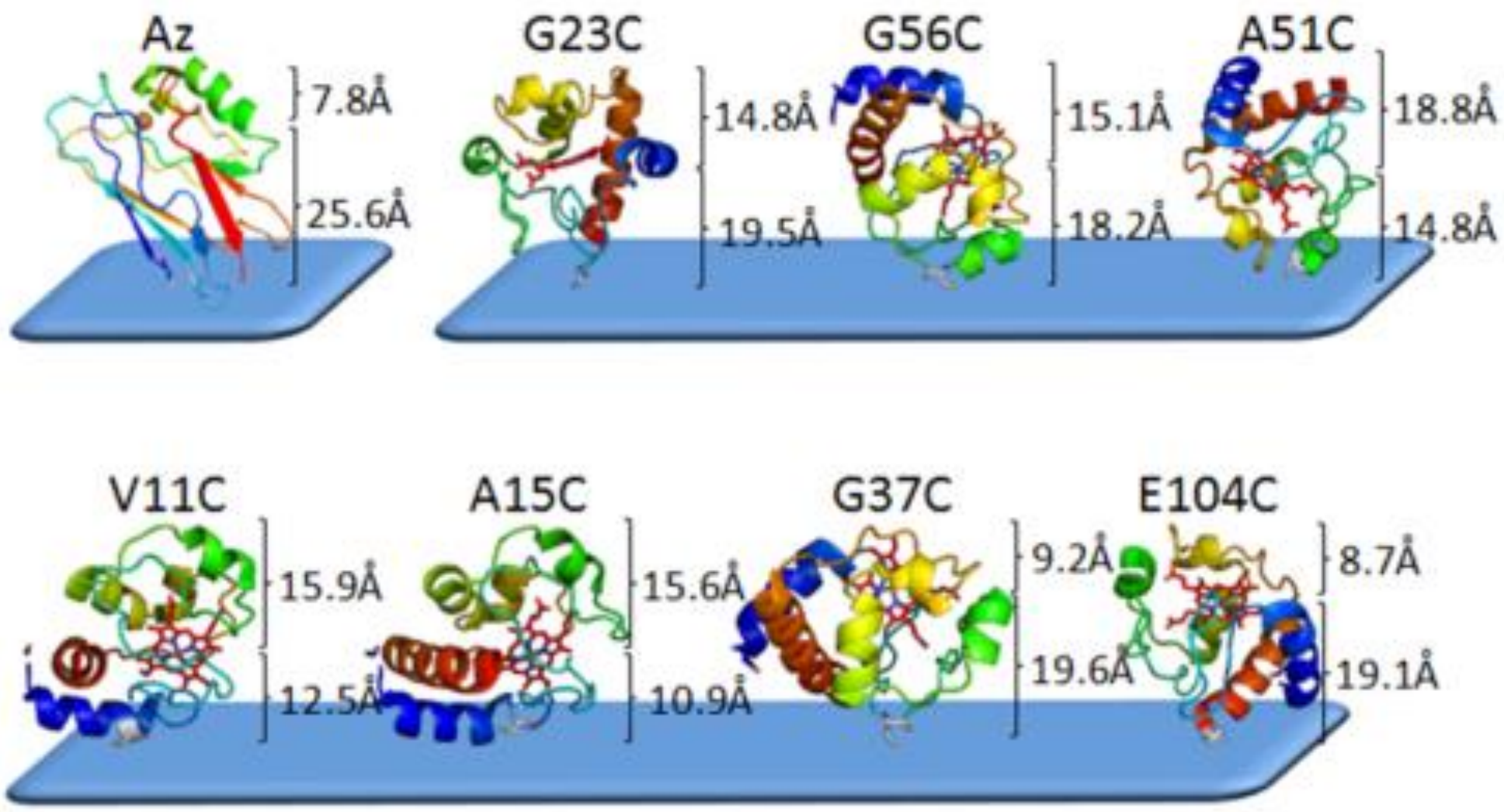

Figure 24: Molecular schemes of the different CytC mutants along with WT azurin, bound to the bottom electrode where the position and orientation of the heme cofactor is systemically modified with respect to the bottom electrode. The calculated distances between where the Cys thiolate is bound to the bottom electrode to the Fe in the heme cofactor, and from there to the other end of the protein (in contact with the top electrode), are indicated. Reprinted with permission, from ref. [156].

## 4.3 Electronic Measurements: Interpreting results in terms of ET & ETp

### 4.3.1 Electron transfer and transport across immobilized proteins

The ability to conduct controlled measurements of ETp on immobilized proteins has allowed mechanistic interrogation of ETp in a variety of proteins with different functionalities. Studies ranging from nanometer to macroscopic scales have demonstrated that ETp occurs mainly by tunneling at low temperatures (< 150-200 K), and can switch to thermally activated transport originating primarily in electrode-protein coupling at higher temperatures (>200 K).[47] By combining these ETp measurements with the structural characterization of immobilized proteins, it is possible to determine at least some of the crucial factors that modulate the electronic transport behavior of these proteins. Aside from the cases discussed in Section 4.2.2 above, another example is work by Rinaldi *et al*.[197] which demonstrates that the orientation of azurin immobilized on a $SiO_2$ substrate strongly affects the measured current intensity. Oriented azurin was approximately ten times more conducting than randomly oriented azurin, as depicted in Figure 25. Remarkably, electrical conductance was measured over a huge 60-100 nm gap. We suggest that this might be explained by the at least 50% relative ambient humidity. Parallel orientations of the azurin Cys3-Cys26 bridge result in a distribution of parallel dipoles that favors conduction, while random electrostatic interactions on the non-oriented layer result in random dipole orientations, and thus a reduced total dipole field and a lower ETp efficiency.[197]

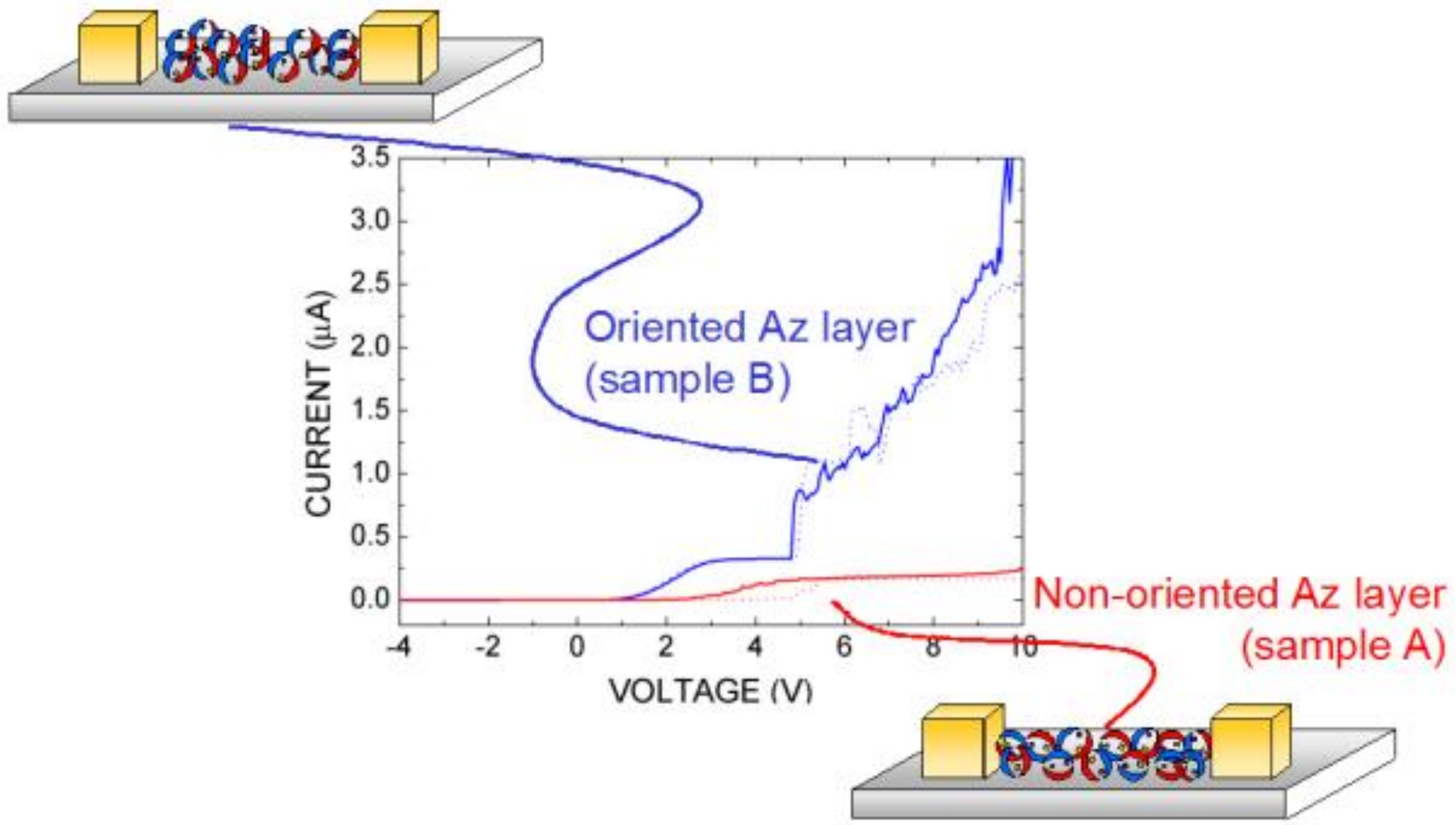

Figure 25. I-V curves of non-oriented azurin bound by electrostatics (sample A) and oriented azurin covalently bonded by surface-exposed cysteins (sample B). Reprinted with permission from reference [197].

In STM-based molecular junctions, the orientation of cytochrome $b_{562}$ was engineered by introducing two thiol groups spanning different axes of the protein for controlled binding to gold electrodes, which yielded a larger conductance than the wilt type (WT) protein immobilized through electrostatic interactions.[181,194] In macroscopic configurations, covalent attachment to both electrodes offers higher conduction and reduced thermal activation energy compared to one-side chemical attachments or electrostatically bound proteins.[156] As in the Cyt C experiments described above, these studies indicate that currents, i.e. ETp efficiencies, are mostly affected by heme-electrode distance rather than by electrode's separation distance, illustrating the importance of redox cofactors in ETp.[156,194] These and the previous results also demonstrate experimentally a connection between ETp and ET (pathways).

Given the importance of redox-active prosthetic groups in ETp, studies of modified proteins which replace redox cofactors with non-redox place fillers, or which remove prosthetic groups/cofactors entirely, have been carried out to understand the role of these metal centers in metalloproteins.[86,150,154,156] Findings from these studies have demonstrated that redox centers are not only important for maintaining protein conformation, but also for mediating the ET and ETp mechanism. Mukhopadhyay and co-workers conducted solid-state electron transport through holo- and apo-ferritin as well as three metal core-reconstituted ferritins including Mn(III)-ferritin, Co(III)-ferritin, and Cu(II)-ferritin using CP-AFM.[149] The study concludes that the metal core has a direct effect on electronic conductivities of the ferritins.[149] In further work, they concluded that the ferritins behave as n-type semiconductors with variable bandgaps (0.8 -2.6 eV) determined from differential conductance data, with the largest gap found in apo-ferritin.[219] ETp studies across holo- and apo-ferritin by the Davis group and Nijhuis group also show increased conductivity of the holo protein over the apo-protein.[88,148] Macroscopic and break junction experiments on apo- and holo- heme-proteins such as CytC and myoglobin conclude that ETp depends on the relative position of the heme group with respect to the electrodes.[29,32,87] In the solid-state, the porphyrin ring of the heme-proteins (with and without iron ion) contributes significantly to determining the protein's ETp characteristics, perhaps acting as a tunneling resonant center. In this scenario, solid-state transport processes differ significantly from electrochemical charge transfer mechanism, which center only on the metal ion for redox reactions. We discuss this issue in more detail in the next section.

### 4.3.2 Biological relevance of electron transfer and transport measurements

Theoretical studies of biological electron transfer aim to understand the effect of molecular and solvent properties on observed rates and yields, mostly following Marcus theory (section 2.1). The connections between molecular charge transfer (ET) kinetics and molecular conductance (ETp) in protein-based molecular junctions may be unravelled by a fundamental understanding of electronic conduction channels (ETp pathways in the pathway model) across macromolecular structures.

ETp in proteins shares common mechanisms with ET, such as superexchange mediated tunneling or hopping via protein residues and/or cofactors (sections 2.1 and 4.1). A hypothetical relation between protein ET and ETp could be obtained by comparing WT and modified forms of proteins with/without their biological activities. ET and ETp of holo- and apo-proteins have been studied in cytochrome c,[86,150] azurin,[53,154] and myoglobin,[87] with increased conductance (G)/turnover rates in the holo proteins. Other relevant results include those from the study of macroscopic ETp of holo-CytC compared to CytC without its Fe ion (porphyrin-CytC), and apo-CytC (without heme).[86]

Chemical “doping” of human serum albumin (HSA) with hemin reveals similar electrochemical and solid-state ETp as CytC.[150] While removal of Fe from heme eliminates ET activity in both CytC and hemin-doped HSA, it has almost no effect on solid-state ETp measurements, demonstrating a clear difference between ETp and ET.[150] A likely reason is that ET requires redox activity of the protein which depends on electrochemical potentials, reflecting the thermodynamic and kinetic dispersion of protein response, whereas ETp does not require it. We compared ET rates with ETp currents (also rates!) after correcting for $\sim 10^3 - 10^4$ ratios between actual and geometric contact area in macroscopic measurements and for protein lengths, solid-state ETp measurements with different proteins overlap, covering a range of rates.[47] The obtained ETp rate is at least an order of magnitude larger than those determined from spectroscopic ET data. Experimentally obtained electrochemical ET rate data are even lower, likely because of the additional electronic to ionic transport step at the protein/solution interface.

Temperature-dependent (~10 -360 K) transport studies can be a powerful tool to identify possible transport mechanisms. Such studies were carried out for ET (mostly on frozen solutions, in a glassy state) some time ago and were surprisingly independent of temperature in some cases.[448,449] Temperature-dependent ETp studies are relatively straightforward if suitable device structures that are stable under temperature cycling are used. Two-electrode, sandwich-configuration devices have been used for such measurements on azurin,[151] cytochrome C,[156] human and bovine serum albumin,[89,153] ferritin,[148] and on membrane proteins such as bacteriorhodopsin,[89] halorhodopsin (phR)[90] and PS-1.[85] Temperature–independent transport over a wide temperature range (160-320K and 200-300K) was found for PS-I, azurin, phR and ferritin proteins,[53,85,90,148,151] although ref.[219] suggests ferritin is a semiconductor, which would be incompatible with temperature- independent transport. Indeed, temperature-independent transport behavior is surprising because while azurin is a relatively small protein (~ 3.5 nm along its largest dimension, some 2-2.5 nm between electrodes because of tilt), all others are large (≥ 6 nm) and form monolayers that are well over 5 nm thick. These distances exceed the maximum length over which ETp across conjugated organic molecules was found to be temperature-independent (via tunneling).[450,451]

Recent data have shown that it is the coupling of the protein to the electrodes that determines if there is temperature dependence, likely because of the absence/presence of an electrostatic barrier.[452] Thus, strong electrode-protein coupling in azurin and halorhodopsin junctions could allow superexchange-mediated tunneling, which is temperature-independent, except for the small temperature dependence due to thermal broadening of the Fermi-Dirac distribution of the electrons in the contacts (although ref. [453] argues to the contrary for strong temperature dependence). This hypothesis is supported by the study of ETp in different CytC mutants (discussed earlier in this review) with different electrode-protein couplings, due to different orientations between the protein and the electrodes imposed by the location of the binding cysteine.[156]

In addition to the redox protein examples, given above, modifications of the non-redox photoactive proteins bacteriorhodopsin[89] and halorhodopsin[90,365] have demonstrated the importance of cofactors for the temperature-dependent ETp, and by inference, for the dominant electron transport mechanism. ETp efficiency decreases for bacteriorhodopsin (bR)-based junctions upon retinal removal, similar to what happens in azurin upon Cu removal. Replacement with retinal oxime (RO) results in a different temperature dependence, suggesting a different dominant ETp mechanism than in WT bR.[89] Halorhodopsin (phR) contains two cofactors, which are chromophores, and which undergo a conformational change following visible light absorption. These chromophore cofactors are directly responsible for halorhodopsin's biological function as a chloride ion pump. In halorhodopsin the removal of conjugated cofactors led to a change of ETp from a temperature independent to a temperature dependent behavior.[89,90] The alterations of the ETp pathway upon removal of cofactors were not the result of conformational changes, as CD measurements showed minimal changes in secondary structure, but, in view of the experimentally found role of electrode-protein coupling for ETp temperature (in)dependence in the above-mentioned recent data, *it is likely that the surface charge density distribution changed*. In the case of phR, it was necessary to have both cofactors for maximal ETp efficiency, and to retain temperature-independent transport.[90]

In conclusion, macroscopic ETp measurements down to low temperatures with different protein monolayers show that temperature-independent transport always dominates < 150 K, though junction currents differ markedly in their magnitude for different proteins.[47,87,89,156] As the temperature increases, currents via different types of protein converge to a similar current magnitude.[154] Temperature-dependent ETp is in keeping with hopping as the dominant mechanism for transport and can occur if a barrier exists at the electrode-contact interface. Thus, given temperature-independent coupling to the electrodes, ETp across proteins is temperature-independent, a conclusion that will require careful experimental scrutiny, as it challenges our present understanding of electron transport.

One can view the α-helices and β-sheets that make up about ¾ of proteins as possible bridges with multiple hopping sites, each site with vibronic broadening. As the temperature increases, the phonon-electron coupling will increase, as well as the level broadening. Thus, coupling between adjacent sites can be strong enough to create efficient conduction paths that do not involve the cofactor moiety. The convergence of the currents at room temperature implies that at elevated temperatures the proteins interact strongly with the electrodes, which dominates the measurements, and there is no dependence on the protein's exact composition. At low temperatures, the molecular vibrations are weak, and therefore other pathways can dominate ETp. Judging from the differences in currents at lower temperature, we presume that those pathways depend strongly on the presence of the cofactors. Cofactors with delocalized electronic states can be viewed as a "dopant" of the peptide skeleton. Such a picture is consistent with super-exchange-mediated off-resonance tunneling as a mechanism for temperature-independent ETp, where the average energy states (less populated) available for tunneling in proteins and the electrodes' Fermi levels is negligible.

At low temperatures, ETp efficiency across a dry protein monolayer that is covalently bound to an electrode is significantly higher if the protein is oriented with respect to the electrode and higher than if it is randomly oriented.[87] This is apparently due to the stronger protein-electrode electronic coupling via covalent bonding of the protein to the substrate, and via cofactors for the case of oriented proteins that may allow highly efficient protein-electrode electronic coupling.

### 4.3.3 Protein structure and electron transport measurements

Conformational changes, especially if they affect a protein's active site, can have large effects on ETp. Stabilization of protein active site prior to immobilization on a substrate is necessary to retain their substrate binding specificity, metabolic functionality, and activity levels comparable to their solution metabolism.[147,310,454] This is important because immobilizing the protein has been shown to alter functionality even while stabilizing the protein's native structure.[310] The method of an indexed array (of gold nanopillars) allows sequential measurements of the same protein under varied conditions (addition and removal of small molecules) when a protein is covalently tethered to the pillars.[145,146] It was demonstrated that the addition of small molecules greatly impact ETp in immobilized P450 CYP2C9 through analysis of I-V plots fit using the Poole-Frenkel (PF) emission model (Figure 26).[145,146] The PF model consists of electrons conducting from one localized state to another within an insulating layer,[455] and in this sense is akin to the hopping mechanism discussed in Section 2.1 (Figure 5). In the PF model, the current I is described by

$$I = CV \exp\left[-q\left(\frac{\Phi_B - \sqrt{qV/\pi d\varepsilon_0\varepsilon_S}}{kT}\right)\right], \qquad \textbf{(9)}$$

where $V$ is the applied (bias) voltage, $q$ is the charge of an electron, $\Phi_B$ is the effective voltage barrier that the electron must overcome to move from one localized state to another, $d$ is the distance across which the voltage is applied, $\varepsilon_o$ is the permittivity of free space, and $\varepsilon_s$ is the relative permeability of the material (immobilized protein) at high frequencies, assuming that there is no local polarization induced, $k$ is Boltzmann's constant, $T$ is the absolute temperature, and $C$ is a constant that depends on the intrinsic mobility of the charge carriers, the effective area of the electrical contact, and the effective distance $d$ across which $V$ is applied. Taking the natural log of both sides results in

$$\ln\left(\frac{I}{V}\right) = \ln C - \frac{q\Phi_B}{kT} + \frac{q}{kT}\left(\frac{q}{\pi d \varepsilon_0 \varepsilon_s}\right)^{1/2} V^{1/2}, \qquad \textbf{(10)}$$

so that a plot of $\ln(I/V)$ as a function of $V^{1/2}$ yields a straight line with an intercept component proportional to $\Phi_B$ (Figure 26). Using this model, changes in $\Phi_B$ were measured in the same CYP2C9 protein after exposure to different small molecules with known impact on CYP2C9 functionality. These studies demonstrated ETp is enhanced by known CYP2C9 small molecule ligands, and reduced by a known CYP2C9 inhibitor. Furthermore, ETp results correlated well with the known biological activity of the protein for activators compared to non-activators.

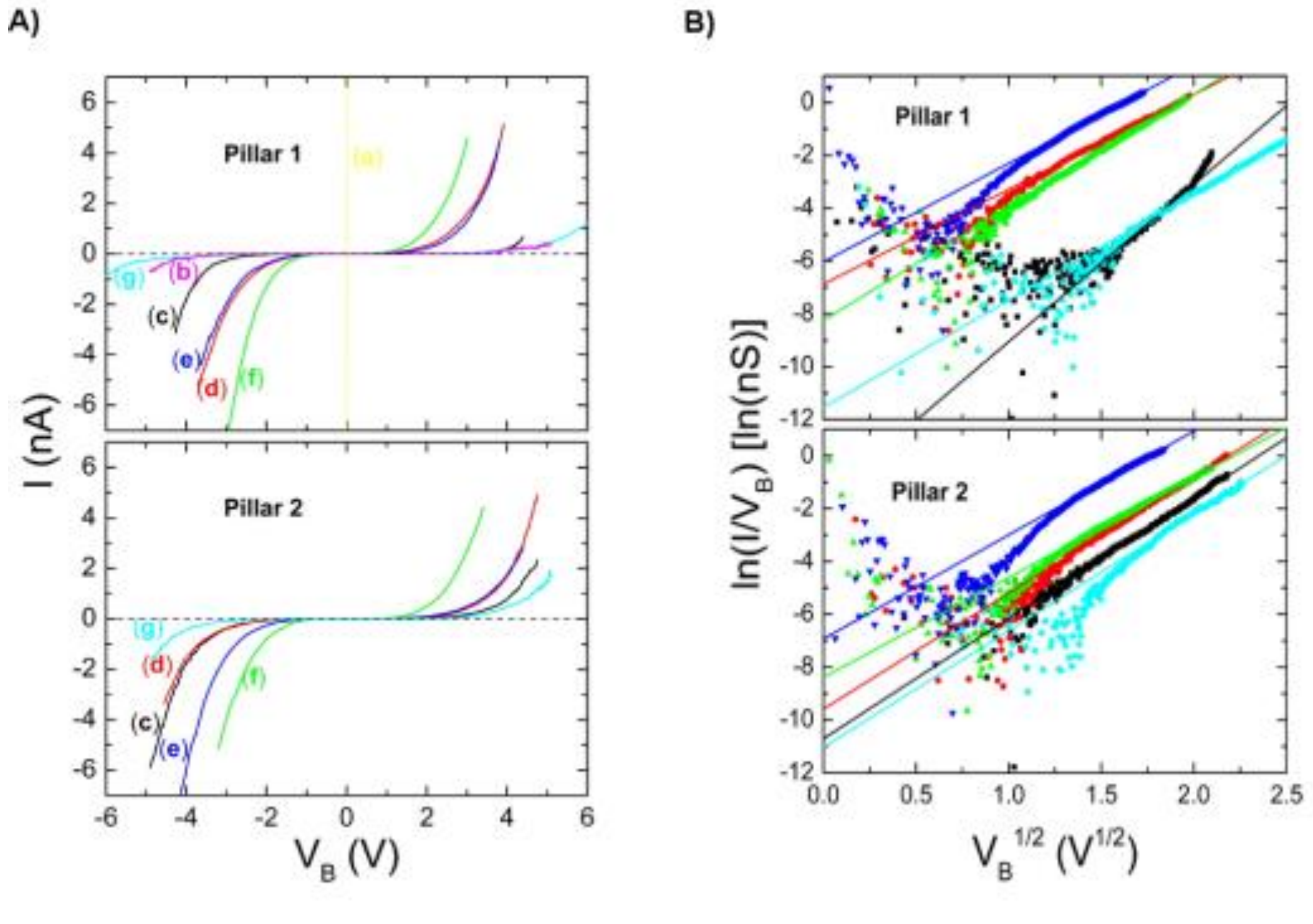

Figure 26. (**A**) I–V curves measured for two nanopillars through the course of an experiment. Curve (a) shows the bare gold nanopillars prior to any attachment, curve (b) is after attachment of thiol SAM (MUA: OT), curve (c) is after covalent bonding of CYP2C9 to SAM. The remaining curves represent covalently tethered CYP2C9 exposed with small molecule binders flurbiprofen (d), dapsone (e), with both flurbiprofen and dapsone simultaneously (f), and inhibitor aniline (g). The gold and SAM data are not shown for nanopillar 2 for clarity. (**B**) shows Poole-Frenkel plot for the data obtained in A) for positive bias voltages using the same color scheme. The symbols are the data, and the lines

are fits to a straight line. Reprinted with permission from reference [145] Copyright © 2013, American Chemical Society.

We previously noted that ETp need not be limited to the same pathways as ET, and thus it is possible that ET and ETp phenomena are not correlated. However, in this study there was a clear correlation of ET efficiency and ETp as related to biological function of the measured protein. This is especially relevant in the case of flurbiprofen and dapsone, two small molecules that are readily transformed by CYP2C9, and which have different effects on the spin state of the heme iron. Flurbiprofen is known to promote a high spin state of the heme iron, while dapsone has no effect on the spin state, but no correlation is observed between ETp and spin state (cf. section 4.3.6 for spintronic effects in proteins). These small molecules have, however, been shown to bind simultaneously and greatly promote CYP2C9 activity,[456] which correlates well with the observed changes in $\Phi_B$. The conductance related to the addition of the small molecules themselves cannot explain differences in the observed electrical conduction, leading to the assumption that induced conformational changes of the protein or heme may lead to altered ETp pathways. Evidence for this comes from I-V studies as a function of tip force which reveal an apparent phase transition at high force (32 nN) in proteins not containing a small molecule within the active site, but absent in the presence of a small molecule possibly due to protein stabilization (Figure 27A).[146] Using the Poole-Frenkel model, changes in the ETp pathway as a function of force can be estimated using a very low force as a baseline. From this analysis, an increase in the pathway length with increasing force was deduced, possibly explained by a zig-zag pathway upon compression of the protein (Figure 27B).

Lastly, these studies demonstrated for the first time that two known inhibitors, thought to function only by blocking the entrance into the protein's active site, both lower ETp conductance.[145,146] This may provide mechanistic insight on how CYP2C9 is inhibited, and could also delineate key structural features of small molecules that keep them from being transformed, which is of great importance in drug design.

Taken together, also these results suggest that functional groups within proteins play an important part in determining mechanisms and pathways for ETp, and those ETp characteristics correlate with protein biological function. Protein conformation is a crucial factor for determining electron transport pathways. When proteins are in known active conformations, studies have shown higher conductivities. This implies that there is some correlation between conformations with higher ETp and conformations for biological activity. The above work is another example of that cofactors impact electron transport with minimal changes in secondary structure, showing that cofactors play a larger role than just stabilizing conformations. The ability to study holo- and apo-proteins, or proteins with known small molecule ligands, will allow better correlation of ETp measurements with biological functionality in the future. These studies provide a glimpse of how results from ETp measurements can inform on mechanisms of protein activity for proteins whose functionality is dependent upon ET.

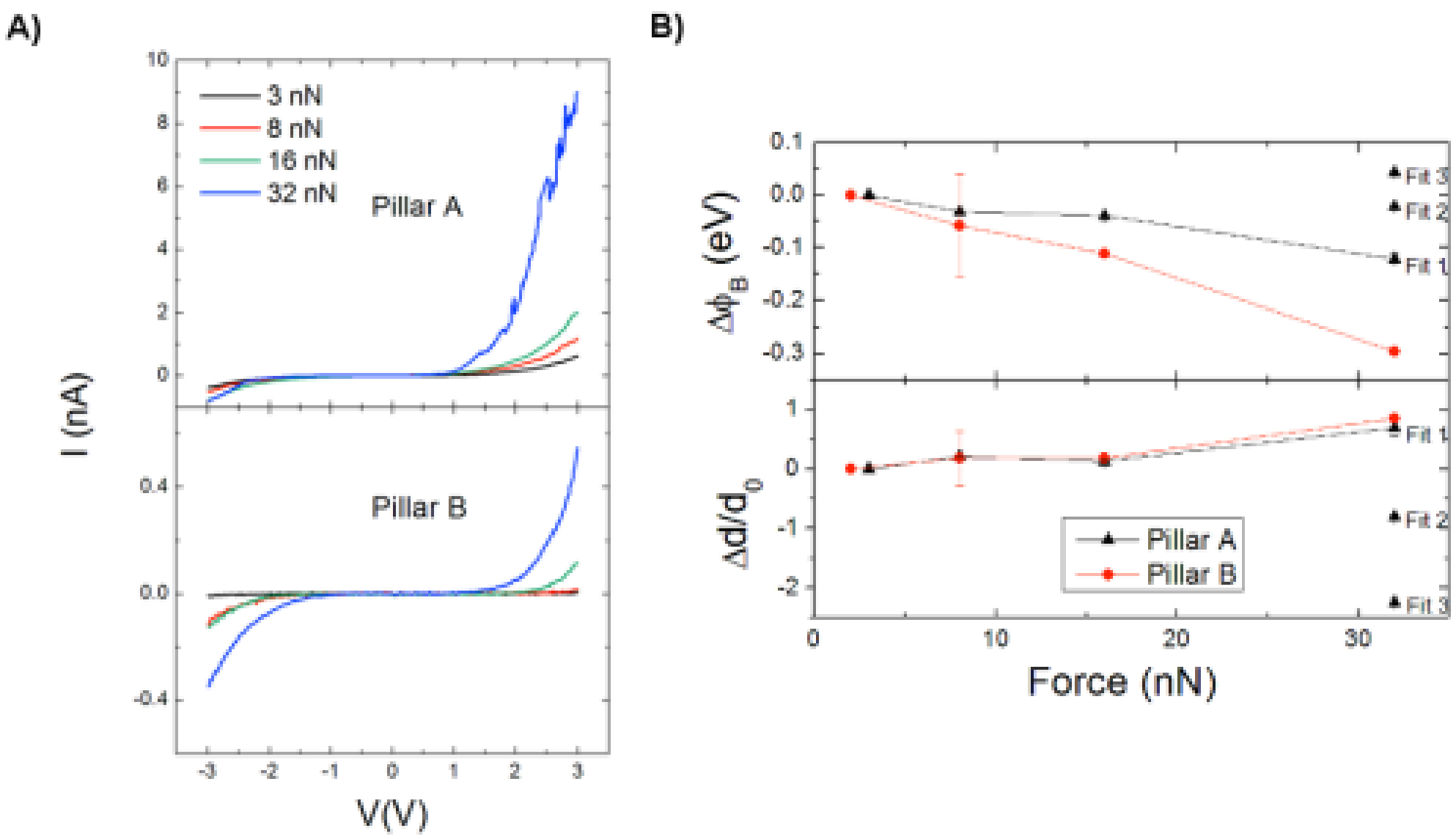

Figure 27. I-V curves for two nanopillars with CYP2C9 immobilized via thiol SAM. A) Curves for a nanopillar with CYP2C9 alone (Pillar A) and CYP2C9 exposed to small molecule binder flurbiprofen (Pillar B). All curves taken on the same pillar and therefore the same CYP2C9 protein. B) Changes in barrier height and $d/d_0$ as a function of force for nanopillars in A) showing decreasing barrier height with force, but increasing $d/d_0$. Reprinted with permission from reference[146].

### 4.3.4 Molecular vibration contribution to electron transport measurements

Molecular vibrations within proteins, which are sensitive to the protein's atomic conformations, sometimes alter the tunneling matrix element during charge transport.[457,458] When a molecular vibrational mode with a characteristic frequency $f_{vib} = \omega/2\pi$ is involved in the transport, depending on the applied bias, the electron can lose or gain a quantum of energy $E_{vib} = \hbar\omega$ to (de)excite the vibrational mode during transport (Figure 28a). This opens an inelastic tunneling channel when the energy difference between the Fermi levels of the electrodes is greater than $\hbar\omega$, which increases the total conductance. Thus, the total junction-current shows a kink as a function of the applied bias for $|eV_{bias}| > \hbar\omega$. This kink results in a step-like signal in the differential conductance ($dI/dV$) plot, and becomes a peak in a $d^2I/dV^2$ plot. Since only a small fraction of electrons are transported inelastically, the conductance step is small. A phase-sensitive ("lock-in") second harmonic detection technique can be used to directly measure the $d^2I/dV^2$ peaks with more sensitivity than taking numerical derivatives of *I(V)* data.

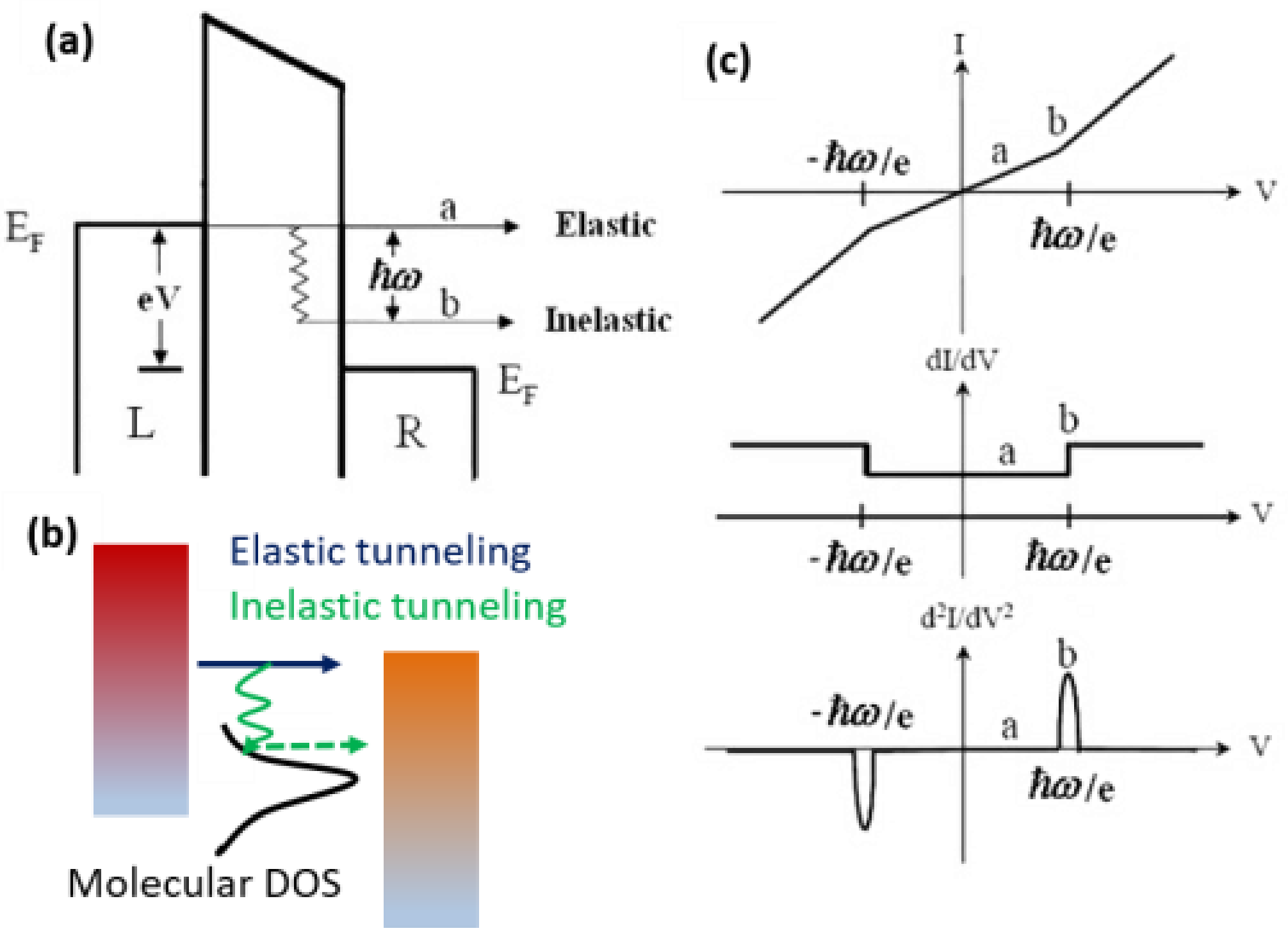

Figure 28: (a, b) Schematic of metal-molecule-metal junctions and energy band diagram with a vibrational mode of frequency ω localized inside the junction. 'a' represents the elastic tunneling process; 'b' is the inelastic tunneling process. (c) Corresponding I(V), dI/dV, and $d^2I/dV^2$ characteristics. Reprinted with permission from reference [459]

Early IETS research included studies of biologically important molecules such as RNA and DNA, and even myoglobin, by Hansma and co-workers.[460] A porous (inorganic) insulator, doped with the (bio)organic molecule of interest, was used in a metal-insulator-metal junction configuration. While indeed IETS measurements on molecular junctions provide information on the vibrational modes of the molecules involved in transport, the ill-defined coverage and random orientations of molecules in the junctions of these pioneering studies made them act as dopants of the main (inorganic) tunneling barrier. Thus, it was difficult to draw conclusion about the ETp properties of the molecules. Recently, the IETS of an azurin-based molecular junction was measured where the most prominent inelastic transport peak was the C-H stretching mode around 2,900 $cm^{-1}$.[341] Because of the special conformation of the junction (azurin between two Au electrodes; cf. Figure 29a) this could be a double junction, in which case the applied voltage would be divided and cause the C-H stretching peak to appear at a higher applied voltage of 3000 $cm^{-1}$. Instead, IETS peaks around 1,600 $cm^{-1}$ were observed, corresponding to amide I and II vibrational modes with significant amplitudes, i.e., the intensity ratio is quite different from that seen in the IR (infrared) spectrum of a monolayer on Au (using PM-IRRAS). [341]The other peaks could be assigned to modes of the side groups of amino acids, such as the $NH_2$ in-plane bending and C=C or CN bonds in aromatic side groups present in tyrosine, phenylalanine, and tryptophan. These findings confirm that the side groups of the amino acids play a role in the inelastic part of the transport (which is at most a few percent of the total current in the case of the azurin junction) across proteins.

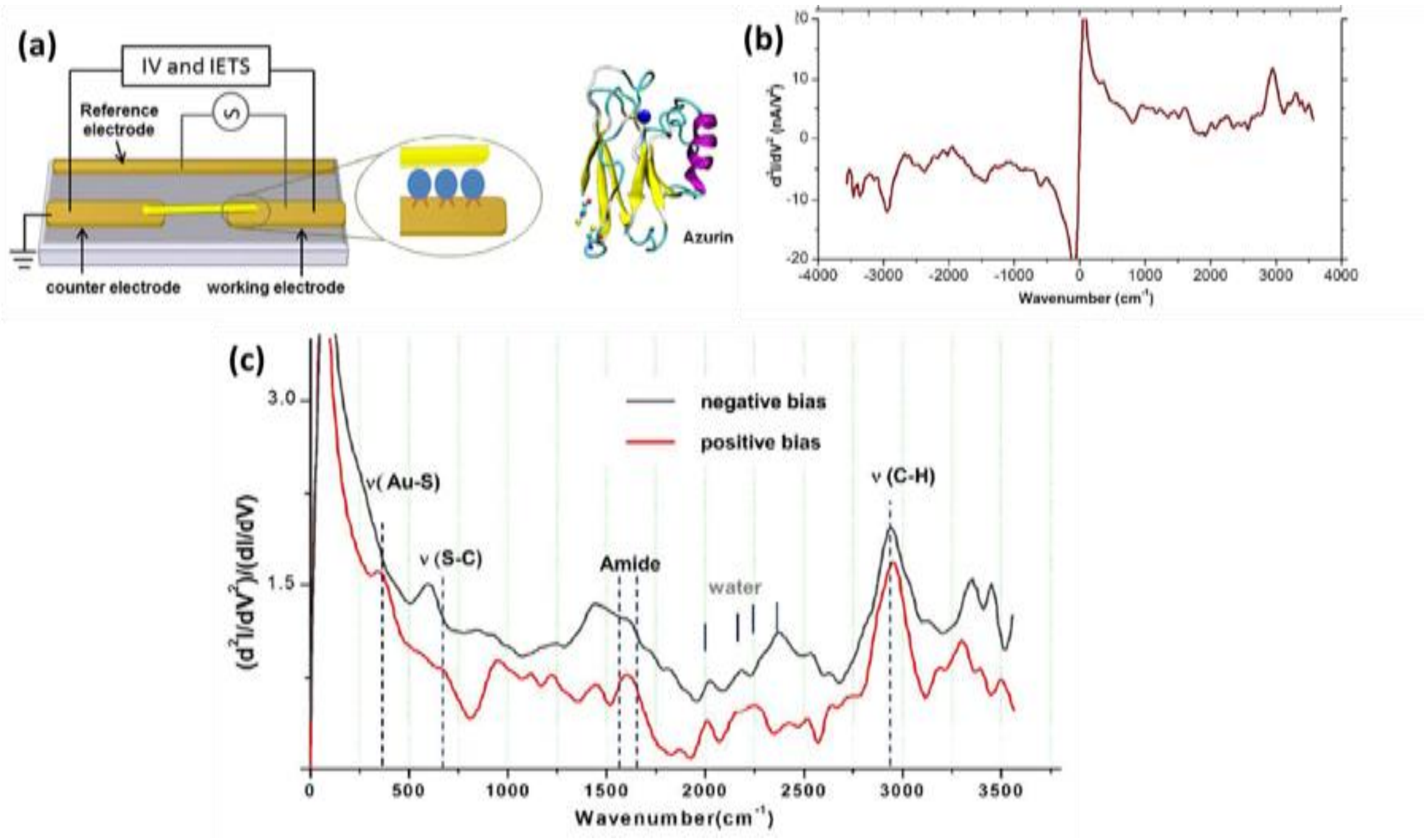

Figure 29: (a) Scheme of the azurin junction fabricated by trapping nanowires by ac electric field and I-V and IETS measurement. (b) IETS spectrum for the azurin junction (c) IETS

spectrum of azurin junction at around 5.5 K with 6 mV ac modulation amplitude. Tentative assignments of the peaks are based on previously reported IR spectra of the protein. Reprinted with permission from reference [341] Copyright © 2015, American Chemical Society.

#### 4.3.5 Opto-electronic properties of immobilized protein

The development of optoelectronic biomaterials has been driven by the convergence of biochemical techniques for peptide/protein engineering. Multilayers of photoactive rhodopsin families of (mutated) proteins have been explored over the last three decades as optical sensors and memory devices.[58] To date, much of protein-based optoelectronics research has focused on the photoresponse characteristics of the light-sensitive protein bacteriorhodopsin (bR) with the aim of developing light switches, or modulators using semiconductor-based photonic crystals.[461] No membrane protein has been studied as extensively as bR, where the reversible photocycle is triggered by light-induced trans → cis retinal double bond isomerization, as occurs in human visual pigments. The configuration alteration of the retinal chromophore, which is covalently bound to the protein and located at the center of the protein, results in a conformational change of the surrounding protein and the proton pumping action (Figure 30a).[462] Transient (photo-voltage/-current) and steady-state photoconduction with bR, acting as the light-sensitive material across an indium tin oxide (ITO)/bR/Au/GaAs heterostructures, has been reported (Figure 30).[463] Light-induced optical memory elements have been realized with a genetically engineered or chemically modified bR, with long excited state lifetimes (~ minutes).[464] Photoinduced surface potential alteration of bR patches has been employed as electrostatic memory using scanning surface potential microscopy (Figure 30d and 30e).[465,466] bR-FET-VCSEL monolithically integrated bio-photo-receivers and photo-transceiver arrays have been successfully designed, fabricated, and characterized for standard communication applications.[467] The large electro-optic effect in bR films can in principle be used in other novel all-optical light switches in bio-hybrid semiconductor photonic crystals.[468]

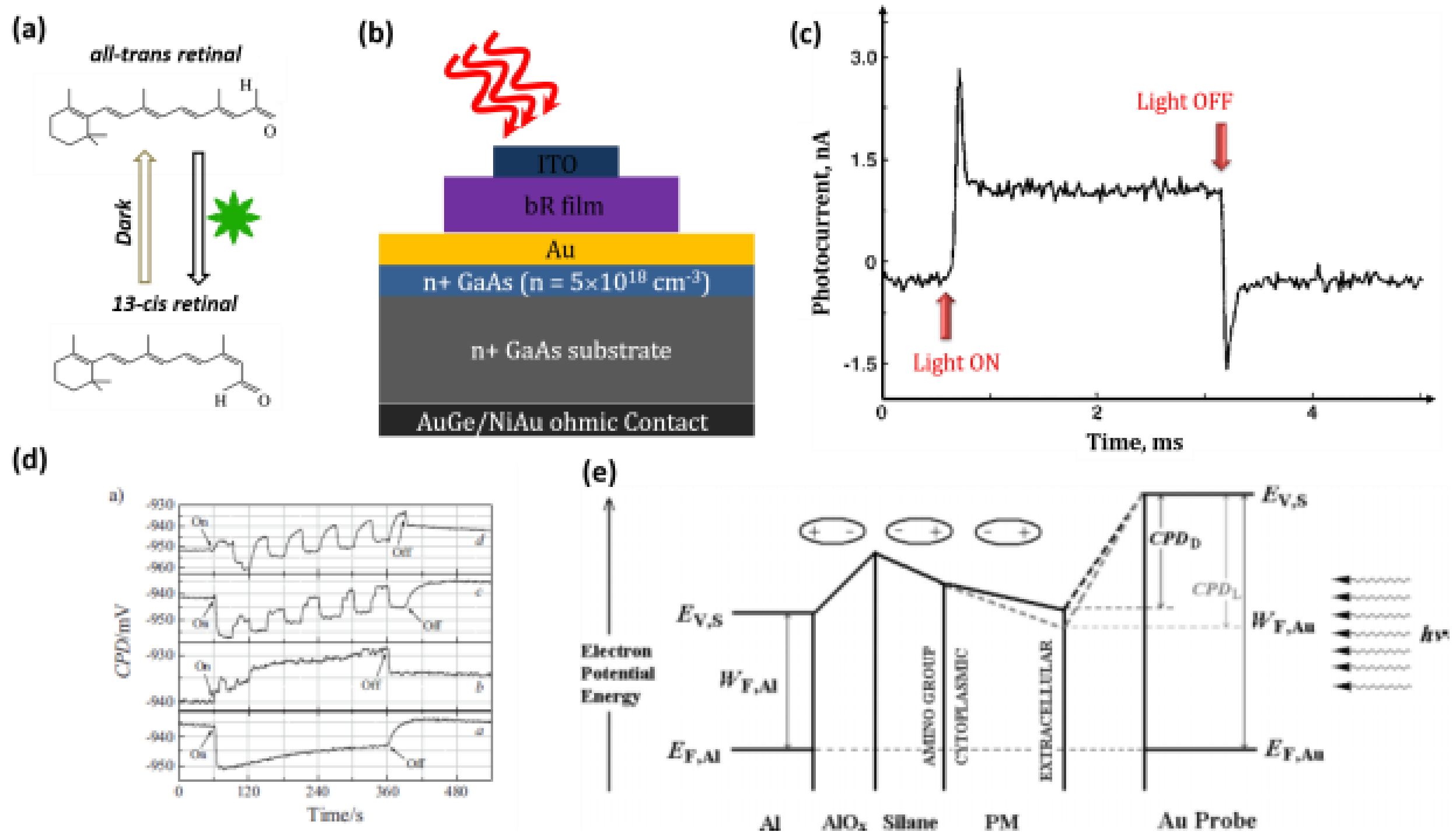

Figure 30: (a) Conformation alteration of retinal cofactor in bacteriorhodopsin upon light absorption. (b) Schematic device configuration for ITO/bR/Au/GaAs heterostructure showing terminal connection for measuring photoconduction; (c) Measured photoresponse with 630nm light modulation for devices above under ambient humidity conditions. Reprinted with permission from reference [463]. (d) Time trace of contact potential difference (CPD) from bR layer on solid support (Si/Al/AlO$_x$/APS/bR) with green light modulation. (e) Empirical energy level diagram explains the results from CPD measurement, where the polarity of the dipole for each layer shown above the diagram. Dashed lines indicate experientially observed shifts after irradiation with green light.

Monolayers of bR could be more suitable for optoelectronic transport studies than multilayers because protein structure and orientation in monolayers is better defined. The transient photocurrent of bR monolayers in an aqueous environment has been measured in a photo-electrochemical cell where a bR monolayer was created at the interface between a transparent conductive electrode and an aqueous electrolyte gel.[469,470] Systematic optoelectronic transport studies (at 293 K and 40% RH) were reported with monolayers of bR that were prepared as follows: bR was extracted from its natural purple membrane and reconstituted with a detergent to form membranes with the detergent, containing ~ 80 % bR in terms of area. Such membranes close on themselves to form spherical vesicles. These vesicles were self-assembled on a suitable solid support, such as surface-modified silicon or gold, and spontaneously opened when incubated in water (pH - 7). By using positively charged silicon surfaces as electrically conducting supports with semi-transparent (30 – 50 %) gold-pads as top electrodes, planar junction structures were fabricated, as shown schematically in the inset of Figure 31b.[471] Those junctions were then used to test if the so-called photocycle, i.e., the photochromic activity of bR that is central to its biological proton-pumping action, is preserved in a dry solid-state monolayer configuration.

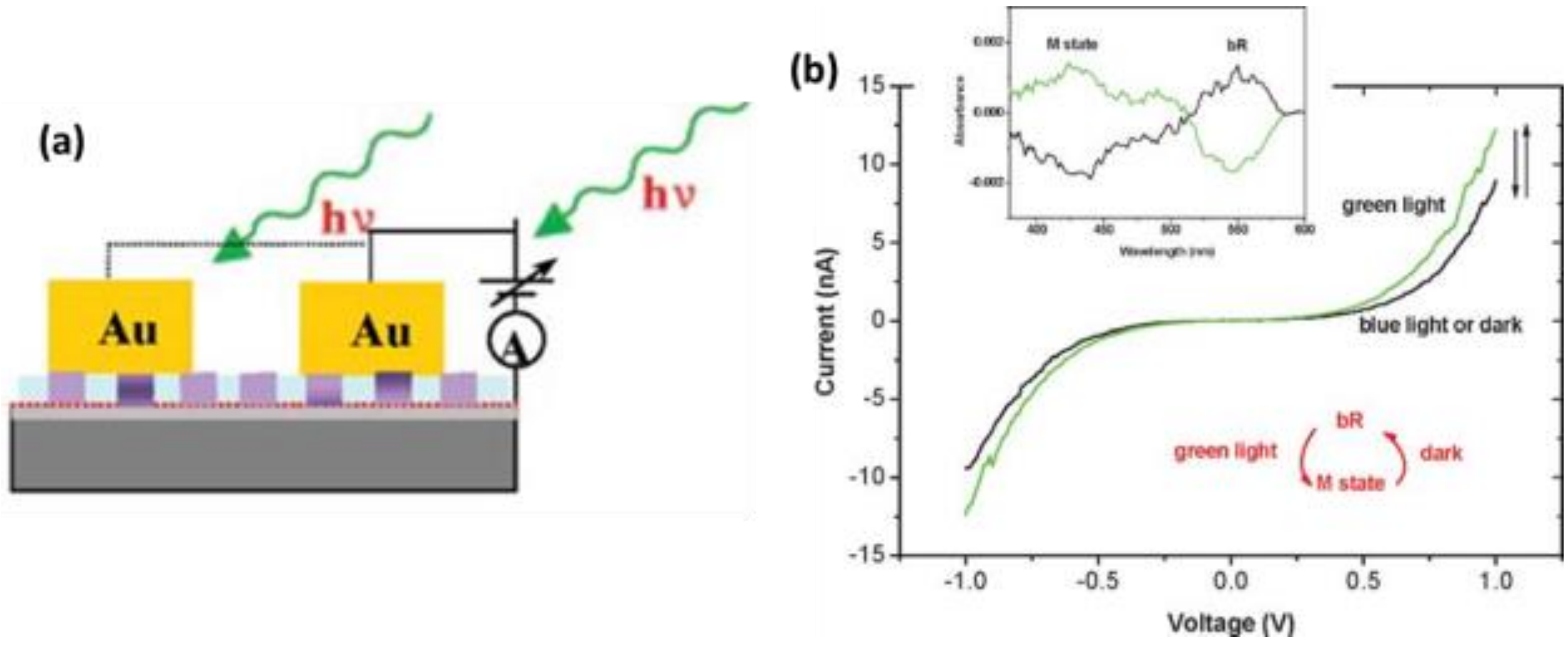

Figure 31: (a) Schematic of Au/WT-bR (oriented)/APTMS/$AlO_x$/Al junction, with geometric area ~ 2 × $10^{-3}$ $cm^2$. Measurements were carried out upon illumination with green light (λ > 550 nm; black), and after varying times in the dark, after this illumination, (red, blue, green) (b) Reversible I–V curves upon illumination at λ > 550 nm and upon illumination with (380 nm< λ < 440 nm) light. Inset represents differential-adsorption spectra of bacteriorhodopsin monolayers. The black curve shows the irradiated–dark measurements; the green curve shows the (dark after illumination)–irradiated measurements. Reprinted with permission from references [471,472].

As shown in Figure 31, the electrical current could indeed be optically modulated in a dry, solid-state bR monolayer junction with electrostatically favorable protein orientations, following the photocycle of the protein. The origin of photo-induced current enhancement was further confirmed via uv-visible absorption of bR monolayer in the presence of green light (abs. maximum around ~ 560 nm). Upon green light (> 495 nm) illumination the 560 nm absorption peak of bR disappeared and a new band appeared at ~ 420 nm, indicating the formation of the so-called M-state, which results from the retinal in the bR undergoing the trans-cis isomerization that triggers the photocycle. (Figure 31b, inset).

The photochemically induced intermediate M-state thermally decayed to the ground state in a few msec, as deduced from the re-appearance of the 560 nm band and the disappearance of the 420 nm band. The enhancement, induced by green light of the junction conduction can then be directly associated with the bR photocycle, implying more efficient conduction via the photochemically induced M intermediate state protein conformation which occurs upon irradiating the junction with green light. M-state involvement in photoconduction was further confirmed through control experiments with apo-retinal bR, where no photocurrent enhancement was obtained, which at least correlates with the known lack of M-state formation in the apo-form of bR.[472]

Davis, Watts and coworkers studied largely lipid-free (25% of endogenous lipid remaining) bR and reported reproducible, robust and controlled interfacial assembly.[473–475] They explored both partial delipidation and single-site mutation of bacteriorhodopsin ($bR_{cys}$) where one of the amino acids, methione (Met-163), located on the cytoplasmic side of bR (i.e., the side exposed to the bacteria's inside cytoplasm in the purple membrane), was replaced with a cysteine. This allows for direct binding of the protein to Au via a Au-S bond, resulting in a unidirectional uniform orientation of the protein on a gold substrate (Figure 31). Along with the photo-cycle, proton flux movement across protein structure-induced potential changes was mapped out by scanning probe microscopy.[466] The magnitude of the surface potential switch upon illumination is consistently larger for $bR_{cys}$ (delipidated or PM embedded) than with the WT-type bR. Strikingly, the photoinduced surface potential switch was solely unidirectional (positive direction), which confirms the designed and confined surface molecular orientation (Figure 32).

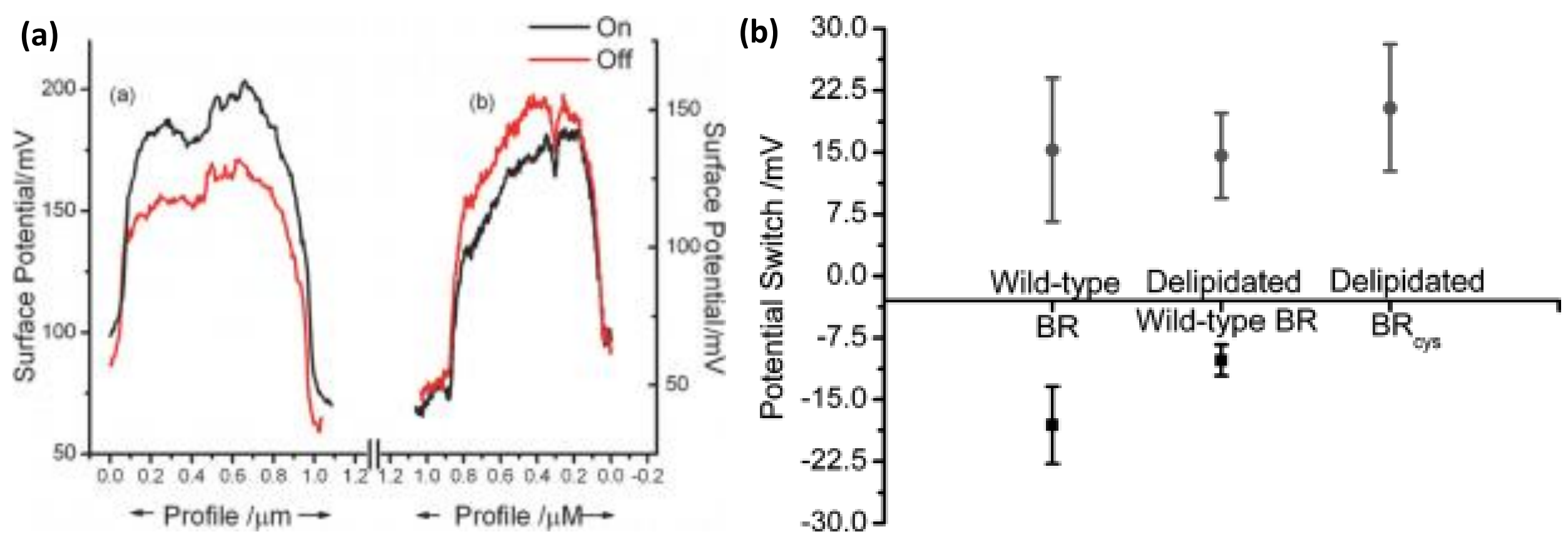

Figure 32: (a) Surface potential cross-section of wild-type purple membrane, PM, patches, immobilized on gold, as mapped by KPFM, in the dark (red line) and under illumination of λ > 495 nm (black line). (b) Comparison of light-induced surface potential modulation for wild-type bR in PM, delipidated wild-type bR and delipidated bR with cysteine mutation. Reprinted with permission from reference [474].

The study concludes that the process of freeing the protein from its associated lipid and introducing a reactive surface cysteine residue preserves the protein's photophysical characteristics, while enabling a more intimate interfacial sampling on gold electrodes compare to purple membrane. The modulated, wavelength-specific irradiation of a $bR_{cys}$ monolayer generates current transients similar to those naturally observed during the native membrane's proton pumping activities.[43] The chemical binding of functional, delipidated $bR_{cys}$ onto solid surfaces allowed analysis of potential photoswitching, which may be relevant to the development of derived nanoscale photoresponsive devices (the "molecular pixel").[473]

To enhance photoconduction across bR-containing monolayers, ETp was probed across higher protein concentrations in partially delipidated WT-bR, where ~75% of the PM lipids were removed.[365] bR was immobilized as a monolayer, taking advantage of interactions of the WT surfaces, which are normally in contact with lipids, with the surface of highly ordered pyrolytic graphite (HOPG) used as a conducting substrate (Figure 33a). The hydrophobic interaction induces bR in 'lying down' configuration rather than 'standing up' like on gold or silicon substrate. This junction configuration reduces tunneling barrier height from 6 nm, as for $bR_{cys}$, to 2.5 nm, thus enhancing junction current and light-induced conductance variations with electro-optical modulation.

The conductance of bR monomer junctions was significantly enhanced under green light illumination compared to that measured in the dark, as long as junctions were probed with an applied force within the elastic force range (2-10 nN) (Figures 33a and 33b). In the elastic force regime, the relative photo-conductance, $\Delta G/G_{dark}$, induced by green illumination does not depend on tip force and $G_{green}/G_{dark} = 2$ (i.e., $\Delta G/G_{dark} = 1$).

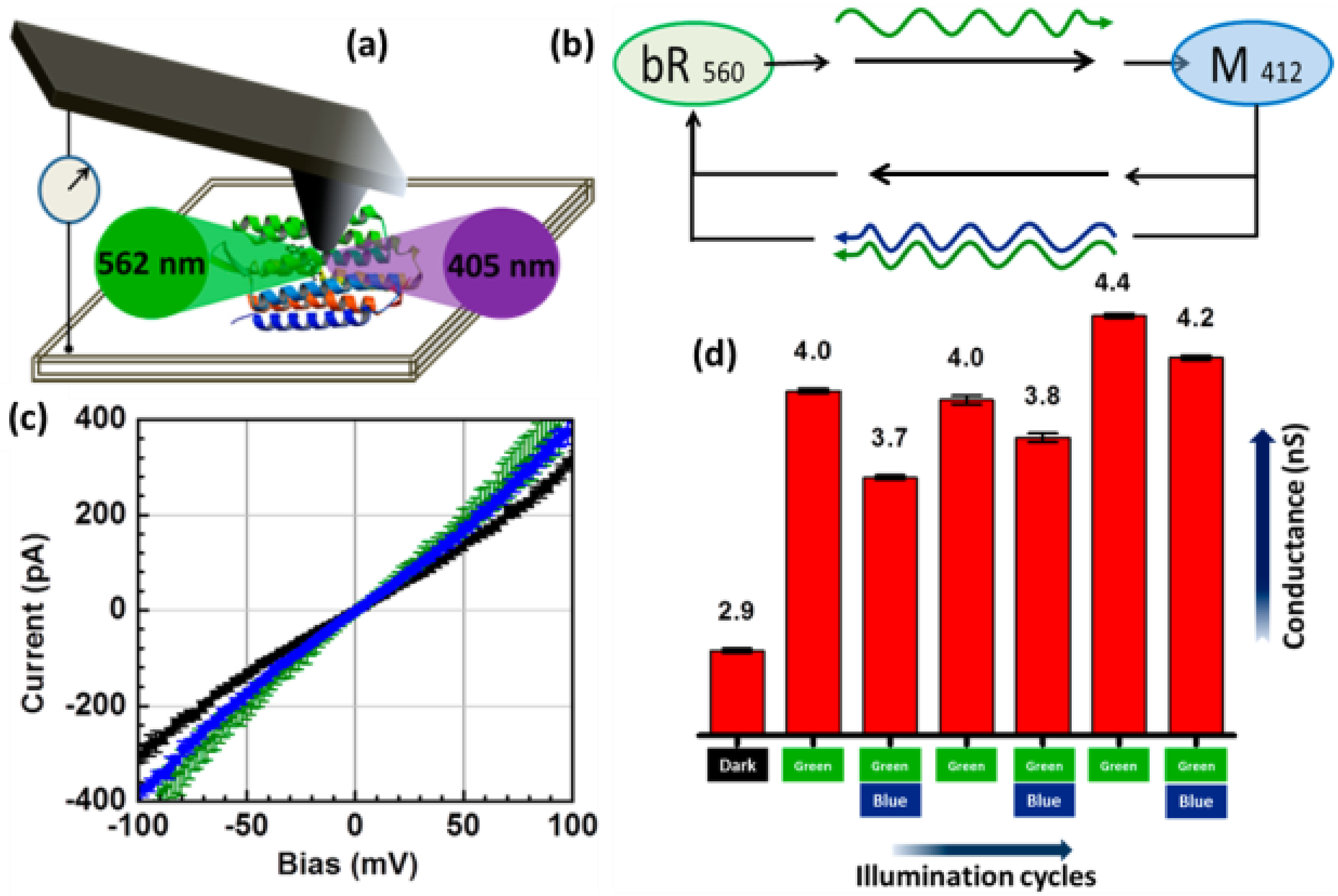

Figure 33: (**a**) Schematic of electron transport measurement across a bR momoners with conducting probe atomic force microscopy. (**b**) Schematic of bR photocycle in dry dehydrated condition, (**c**) Typical light-induced conduction modulation of a dLbR junction. I-V characteristics of a dLbR junction in the dark (black), under green illumination (green), and with blue+green illumination (blue). (**d**) Conductance modulation of a dLbR junction (numbers on top of bars, in nS) under different illumination conditions (shown at bottom). Reprinted with permission from reference [365] Copyright © 2014, American Chemical Society.

At higher applied forces (> 10 nN), in the plastic regime, the change in the relative photo-induced conductance decreases, and $\Delta G/G_{dark}$ reduces to as little as 0.2, decreasing monotonically with force (see section 3.6.2). This behavior can be interpreted as follows: at low forces, there is efficient conversion of the ground state to the M-like intermediate under green illumination and minor structural perturbation of the bR monomer (Figure 33b). At higher forces, in the plastic range, bR monomers are under higher applied force, which hinders the light-induced conformational changes and impedes M-like state formation (Figure 33b), leading to a decrease in photoconductance. During successive green and (blue + green) illumination cycles, a reproducible conductance variation was observed over typically three to four full cycles (Figures 33c and 33d). These results rule out junction heating upon illumination as the cause of the conductance changes and strongly support the interpretation of photochemically induced M-like intermediate accumulation during green light illumination, which increases the conductance (refer to Figure 31a). Exposure to blue light, on top of the green light, decreases the fraction of M-like intermediate in the mixture by accelerating its conversion back to the initial dark bR state, which further decreases the conductance (Figure 33c and 33d). Humidity-dependent light responses at two distinct applied forces, in the elastic (9 nN) and in the plastic (23 nN) regimes, demonstrate relative current enhancement with increasing humidity followed by saturation above 70% relative humidity.

Like bacteriorhodopsin, a dry photosystem I protein (PS I) monolayer on a linker modified gold surface retains its fundamental optoelectronic properties, as deduced from the optical absorption spectra and the surface photovoltage spectral response (SPV).[445] However, the SPV spectrum was broad and slightly blue-shifted (~ 10 - 20 nm) when compared to the absorption spectrum, which can be due to electronic coupling of PS I to the electronic energy states of the gold, i.e. substantial mixing between the molecular and substrate wave functions.[476] Illumination of the PS I monolayer with 632 nm (some 50 nm below the peak, but within the absorption band of PS I) caused a dramatic reversible increase of ~ 0.45 V in the contact potential difference (Figure 34). This photovoltage modification was explained by the light-induced charge separation that drives electron transfer across the reaction center, resulting in the appearance of a negative charge at the reducing end of the protein away from the gold surface. The explanation implies that PS I protein was highly orientated on the linker modified gold substrate, possibly due to the presence of intrinsic dipole moment of protein itself. In the dark, charge recombination takes place and the photovoltage nulls to original levels.

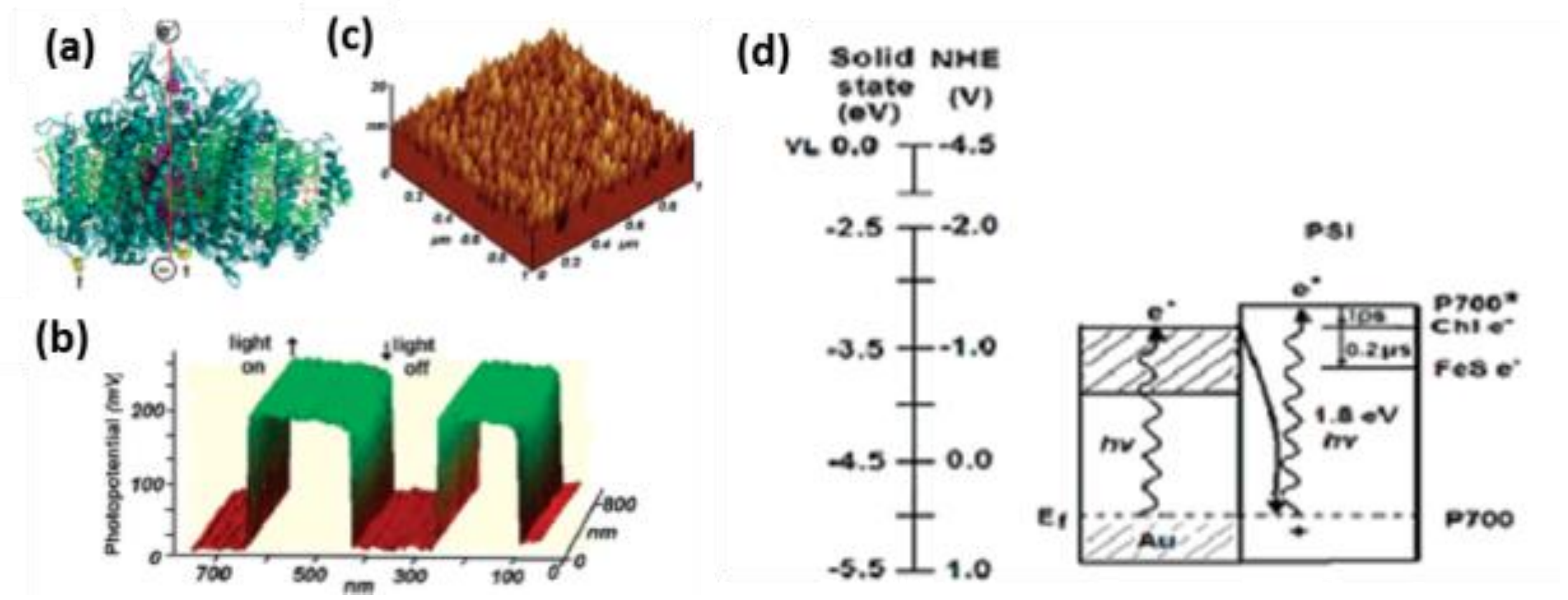

Figure 34: (a) Structure of photosynthetic reaction center I (RC I) based upon crystallographic data (PDB 2O01). (b, c) scanning Kelvin probe images of the PS I monolayer on a gold surface and corresponding topography images of oriented PS I monolayer. Reprinted with permission from reference [445] Copyright © 2007, American Chemical Society. (d) Schematic energy level diagram of the gold-PS I junction which were experimentally defined through contact potential difference. Reprinted with permission from references [477] Copyright © 2000, American Chemical Society.

Scanning Kelvin force probe microscopy was used to study the electrostatic potentials generated at single PS I reaction centers (RC I), immobilized on liker modified atomically flat gold (Figure 35).[477] The photovoltage at the central region of the PS I domains was more positive than at the peripheral envelop area by 240 – 360 meV. This energy difference suggests a possible mechanism following the Calvin-Benson cycle, whereby negatively charged ferredoxin, the soluble electron carrier from RC I, is anchored and positioned at the reducing end of RC I for electron transfer. Under illumination, the electric potentials of RC I acceptors develop a negative voltage following electron capture, whereas in the dark, the potential is positive.

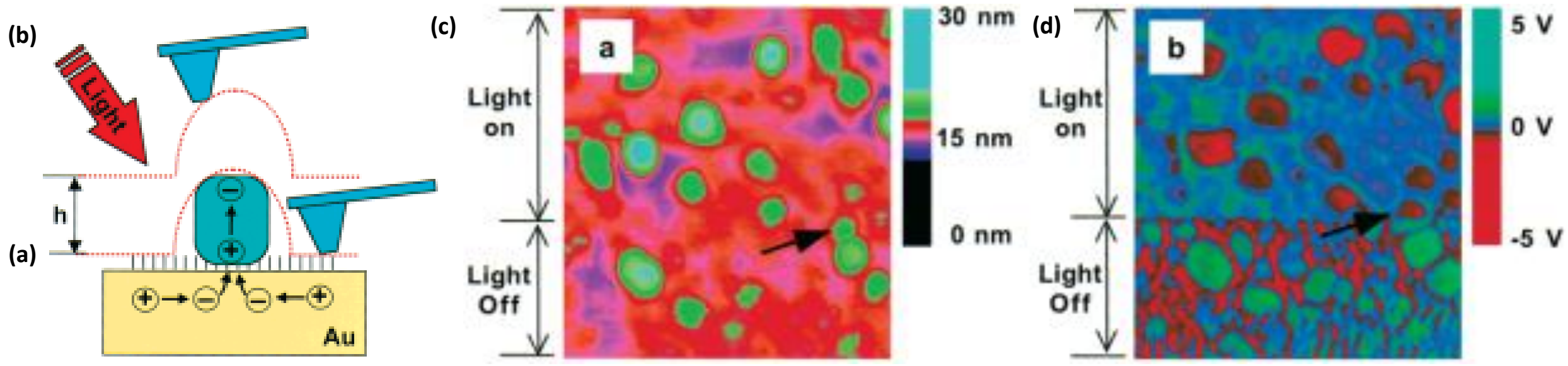

Figure 35: Schematic illustration of KPFM technique used for photovoltage measurement of single reaction centers RC I. Dual measurements employed (a) tapping mode AFM for topographic imaging and (b) lift-mode KFM for voltage imaging. (c) Topographic and (d) electric potential image maps (500 × 500 nm) of the same set of isolated and oriented PS I reaction centers on a 2-mercaptoethanolmodified gold surface. (d) Demonstrates a clear light-induced PS I electric potential reversal from positive voltage to negative upon illumination. The scanning directions for each raster of the constructed images were from left to right and top to bottom. Reprinted with permission from reference [477] Copyright © 2000, American Chemical Society.

Unidirectional photocurrent of RC I on π- system-modified graphene electrodes has been pursued in the quest for developing functional photobiohybrid systems(Figure 36).[478] Naphthalene derivatives provide a suitable surface for the adsorption of RC I, and their combination yield, already at open-circuit potential, a high cathodic photocurrent output of 4.5±0.1 μA/cm$^2$, which is significant for bio-hybrid device applications.

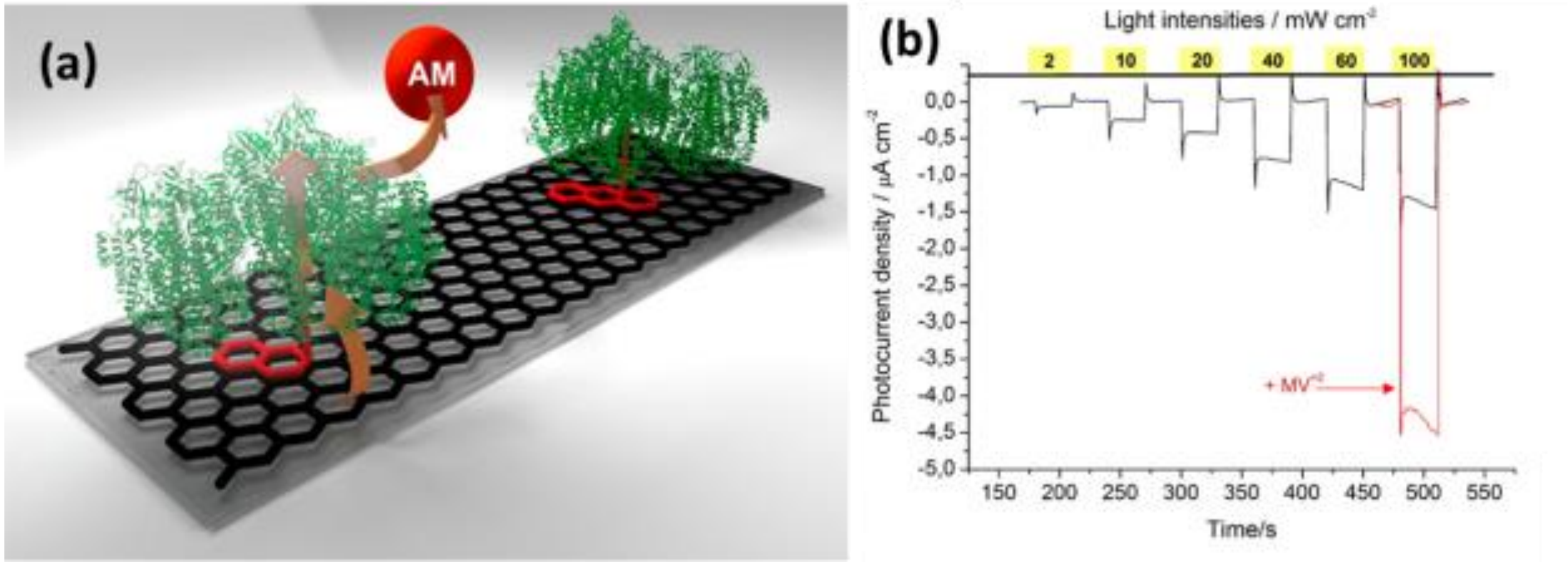

Figure 36: (a) Construction of graphene-based biohybrid light-harvesting architectures consisting of RC I adsorbed onto π-system-modified graphene interfaces. (b) Photochronoamperometric examination of a graphene·dicarboxylic-anthracene-PSI electrode, recorded with different light intensities (2, 10, 20, 40, 60, 100 mW cm−2). Cathodic photocurrents were shown in figure; all measurements have been performed under air saturation. Reprinted with permission from reference [478] Copyright © 2015, American Chemical Society.

What were likely close to monolayers of cytochrome c, with Sn instead of Fe, were found to act as reversible and efficient photo-switches in hexane immersed solid-state junctions. Excitations in either the longer wavelength (Q-)band, @ 535 nm, or in the Soret band @ 405 nm les to significant on/off ratios of up to 25. While no mechanistic studies were done, it was suggested that intramolecular excitations, rather than photo-induced conformational changes, opened new transport channels,.[258]

Conductance switching in the photoswitchable protein dronpa (similar to GFP, green fluorescent protein) self-assembled onto gold substrates (111), was measured using scanning tunneling microscopy (STM) and scanning tunneling spectroscopy (STS) methods before and after the protein was reversibly switched to a non-fluorescent dark state and to the fluorescent bright state in a cyclic fashion using 488 and 405 nm light (Figure 37).[479]

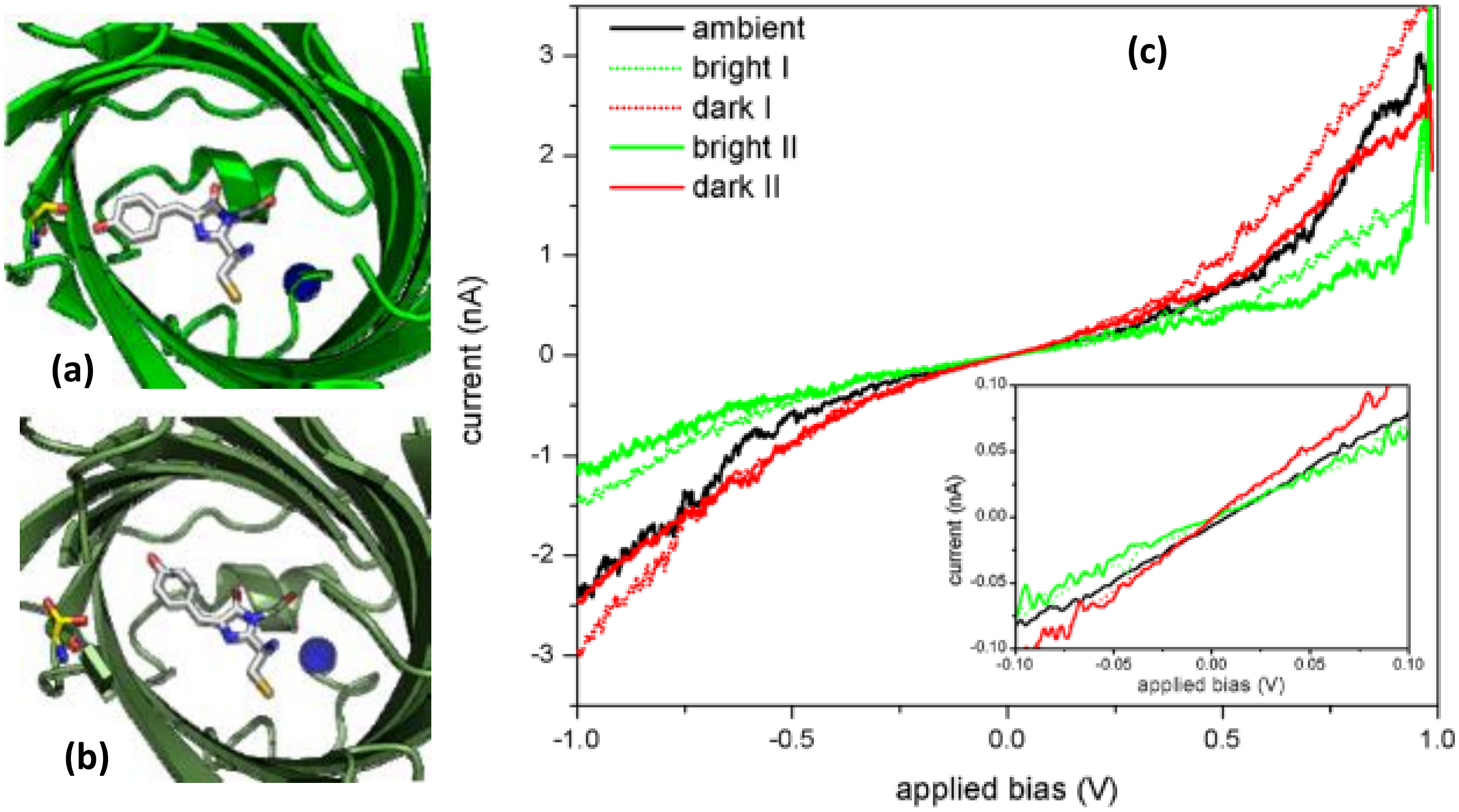

Figure 37: (a, b) Structure of bright state and dark state of Dronpa (PDB ID: 2Z1O and 2POX) with N-terminus highlighted in blue. The chromophore is representing in white, resides at different position of the β-barrel in different state, c) Averaged STS I−V spectra of His-tagged Dronpa as the protein is cycled through dark and bright states. Inset – showing Ohmic region (−0.1 V to +0.1 V) of I−V spectra. An increase in conductance is visible upon switching from the bright state to the dark state. Reprinted with permission from reference [479] Copyright © 2012, American Chemical Society.

ETp studies on photoreceptor proteins containing a flavin chromophore have shown that chromophore-protein interactions, and the electronic environment (which mostly originates from protein residues) around the chromophore, plays an important in light-induced transport enhancements.[366]

Theoretical and experimental studies of optoelectronic properties of bio-molecule based junctions have become an active field of research over the past decade and a half.

Proteins appear quite suitable for biomolecular optoelectronics, based on results from electron transport by solid-state electronic conduction (ETp) measurements of monolayers of "dry" proteins, i.e., with only tightly bound water retained, so that they keep their natural conformation. Although the specific mechanism of photo-induced conduction enhancements across protein monolayers is still being debated, the modulation of transport efficiency upon illumination indicates an alteration of protein structure, conformation, and charge distribution that directly impact their ETp properties. Alteration in conformation could be also be associated with alteration of the hydrogen-bond network, including water rearrangement in case of photoactive membrane proteins.

To use proteins as tunable soft-matter building blocks for optoelectronic applications, they must be modified to generate properties beyond their native biological functions, which is possible using biochemistry and via mutations. A further level in functionality can be reached by combining proteins, as indeed occurs in several cases in nature (e.g., in the photosynthetic and mitochondrial electron transport chains). An example, demonstrated in an electrochemical set-up by Lee *et al.*,[480] is an artificial photosynthetic cascade, where a monolayer of the redox-active electron transfer, ET protein CytC is attached, by chemical modification, to a monolayer of the redox-active ET protein azurin. Future work may allow to extend this approach to the solid state by sandwiching the protein bilayer between an electron- and a transparent hole-selective electrode. In their natural environment, protein-protein docking enables electron transfer between these two proteins.

Until now, the challenge in bio-*opto*-electronics has been to be able to choose and chemically modify proteins to get predictable ETp modulation characteristics with illumination. To that end, the origin of dominant ETp modulation mechanisms must be identified and parameters crucial for tailoring the optoelectronic properties of these biomolecules, such as the relations between ETp and protein structure and composition, will have to be determined.

### 4.3.6 Spintronic properties

The possible role of the electron spin electronics, or spintronics, in chemistry and biology has received much attention recently (cf. also section 4.3.3. and discussion of Figs.

26 and 27). For example, it was recently discovered that the earth's magnetic field has a significant effect on bird and fish navigation.[481,482] A general spin-selectivity effect, due to electrons passing through chiral molecules, the so-called chiral-induced spin selectivity, CISS, was observed first in electron transmission (i.e. photoemission) through organic molecules and DNA attached to gold electrodes.[483–485] Later, a similar effect was observed in electrochemical ET experiments,[486] and in a few cases also in solid state (CP-AFM) ETp studies.[487,488] Spin filtering by chiral molecules has been suggested to be a result of evolution, which would increase the conductance of one spin channel (e.g., spin-up) while decreasing the conductance of the other spin channel (e.g., spin-down).[484,489,490] Spin selectivity in biological electron transfer was first observed in an oriented monolayer of PSI protein monolayers.[491] Spin polarization in ET across PSI was evaluated following photo-induced charge transport in PSI, from the primary electron donor, P700, to a tightly bound phylloquinone molecule, through the primary acceptor, chlorophyll a. Spin-selective measurements with different orientations of the PSI protein complex show that spin polarization is highly dependent on the ET path in PSI. Different substrate magnetizations also demonstrate that light-induced electron transfer through oriented PSI is highly spin selective and that the favorable electron spin is aligned parallel to the ET direction in the protein.[491]

Recent studies also demonstrate spin-dependent electron transmission through helical structured bR proteins embedded in their native membrane environment (Figure 38).[492] The results point to the possibility that the spin degree of freedom may play a role in ETp in biological systems. Both spin-filtered photoelectron transmission through an adsorbed protein membrane film and spin-dependent cyclic voltammetry across patches of purple membranes, deposited on externally magnetized Ni substrates, showed efficient spin filtering by a bR layer.

The efficiency of electron spin filtering through purple membranes films and its mutant could be controlled by illumination with green light, in line with the bR photocycle.[493,494] Whereas significant spin-dependent electron transmission through the native membranes was observed, illumination with green light dramatically reduced the spin filtering of a bR mutant. These studies are consistent with the idea that the accumulation of the M-intermediate state in films upon illumination opens, or facilitates transport along pathways for spin-down electrons, thus resulting in lower electron spin filtering efficiency.

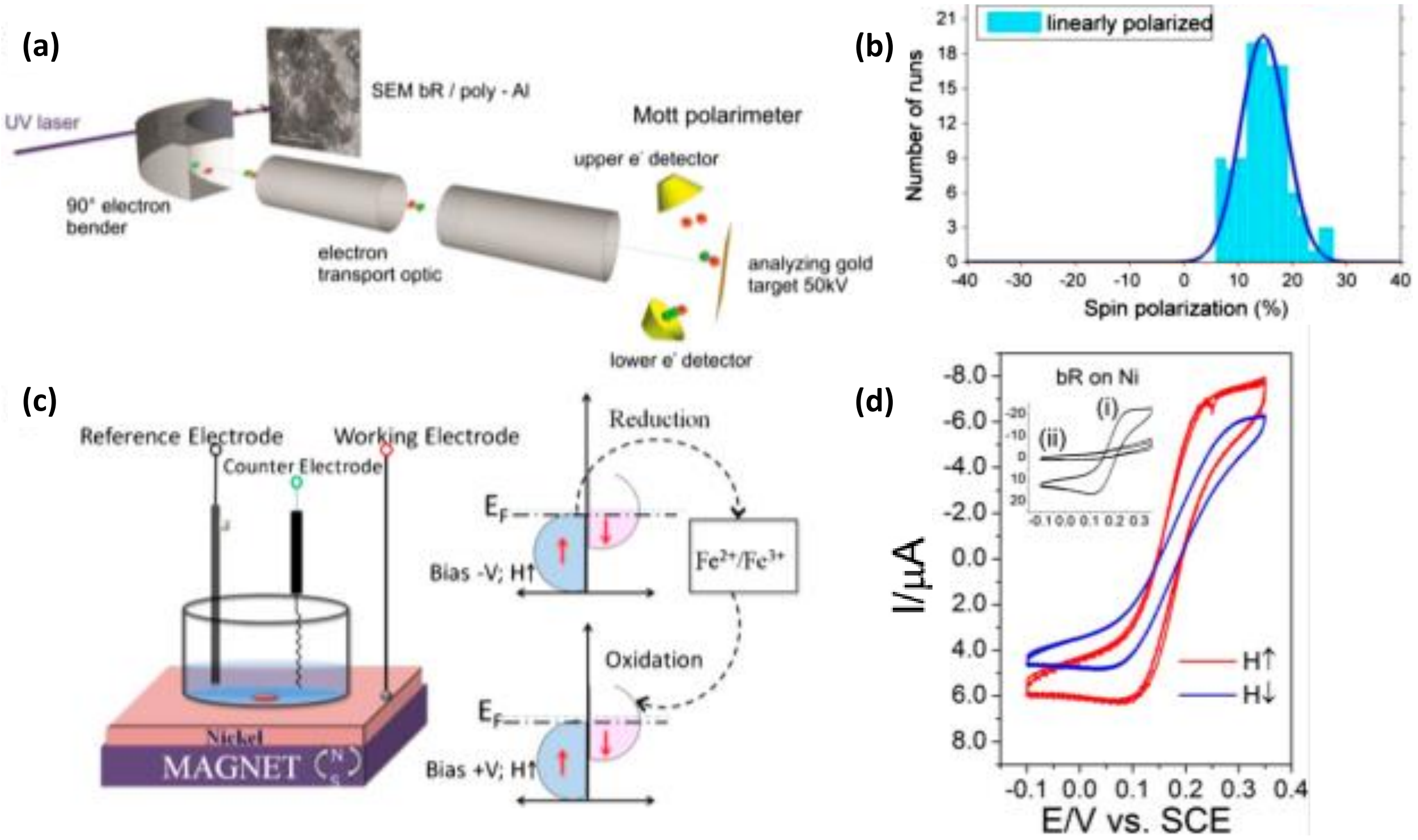

Figure 38: Schematic of Spin-dependent photoelectron' experimental setup and (b) percentage of spin polarization when sample was illuminated with linearly polarized UV excitation. (c) Spin-dependent electrochemistry setup where the working electrode was made from nickel; a permanent magnet was used to control the magnetic direction of substrate. (d) Spin dependent electrochemical studies as obtained, using the bR thin film, physisorbed on the Ni working electrode; arrows in the figures indicate the two possible directions (conventionally UP and DOWN) of the magnetic field (H = 0.35T), which is orthogonal to the surface of the working electrode. (Inset) CVs of a bR/Ni thin film without magnetic field (i) freshly deposited bR on Ni and (ii) after the electrochemical burning of the bR. Reprinted with permission from references[492–494].

# 5. Summary and Future Outlook

In this report, we have we have summarized what we view as some of the important new findings in the field of protein-based biomolecular electronics, in recent years. While formidable progress has been made in towards a fundamental understanding of ET, the understanding of the solid-state analogue, ETp, is far less advanced. Further measurements that incorporate more detailed interrogation at the single molecule level, and more advanced control of variables, related to protein conformation and orientation, may provide parameters needed to optimize the use of proteins in bioelectronics. The remarkable electronic transport efficiencies of proteins might be an unintended consequence of properties that evolved, over billions of years of evolution, for other purposes. The solid-state junction (ETp) approach to contacting proteins might be more similar to what happens in nature (i.e., ET processes) than would seem to be the case at first sight (cf. section 4.1., discussion of Fig. 21). If this is not accidental, then the reasons for it are not (yet) clear. Understanding of ETp and any fundamental relationships with ET, and finding conditions when these two can be combined, may well become possible in the near future.

Proteins already represent refined structures capable of complex and specific reactions. By exploiting the electronic properties of protein-based molecular junctions in various environments, we foresee the creation of advanced solid-state biosensing devices and bio-circuits that integrate such sensing devices. Implantable electronics for real-time monitoring has been a reality for some time (e.g., heart pace makers), and, using organic electronics, flexible electronic implants are rapidly developing towards widespread use. The ability to use proteins as electronic components can avoid biocompatibility concerns that are present with artificial electronic components, and may allow easier integration into biological systems. However, before proteins can be used in these applications, it is crucial to control how they interact with the solid-state platforms to which they will likely be coupled to communicate with the outside world. Such connections should preserve their functionality in downstream platforms, and this requires understanding the fundamental electronic transport properties.

There are many aspects in the selection of the solid-state substrate and binding methodologies that need to be considered for successful immobilization of active proteins. An important conclusion is that protein conformation plays a key role in maintaining protein stability and function after immobilization and can play a decisive role in the measurement of reproducible electron transport characteristics. In nature, proteins have evolved to function in very specific ways. To use these functionalities, protein function must be preserved on solid supports, and specifically must preserve the integrity of what are known or assumed to be important functional groups. Although maintaining protein structural integrity is challenging in the experimental work discussed here, the results reveal that immobilization methodology can be tailored to obtain useful results.

Our review of the literature shows that immobilization schemes that mimic the protein's biological attachment (e.g., Figures 6 and 10) and which allow for interrogation of desired aspects is an effective way to retain protein function and to obtain reproducible results. Different proteins assume a variety of conformations that are necessary for biological functionality, and using this information to immobilize the proteins can yield optimal protein function. For example, if the goal of an experiment is to measure transport involving a particular site of the protein known to participate in ET with a partner (small molecule, ion etc.), immobilization onto the electrode for *in situ* electrochemical ET measurements or solid-state ETp measurements after *ex situ* binding should be optimized to assure that the natural binding surface remains exposed for interaction. If the protein of interest is membrane-bound in nature, it is likely to retain its function if covalently tethered using a SAM, rather than directly bound to a solid surface. Taking these steps allows for retention of protein activity, which can then be used in bioelectronic devices. By determining protein orientation and conformation when immobilized, the ability to correlate ETp with biological activity will be enhanced. For example, when optically active cofactors of the membrane-bound proteins bacteriorhodopsin and halorhodopsin, were modified through chemical reactions, the alteration in their optical absorptions was directly correlated to their ETp efficiencies. Altering or removing metal-ion containing cofactors was directly manifested in both their ET and ETp behavior and in their optical absorption properties. Bacteriorhodopsin, immobilized on a ferromagnetic substrate, serves as a spin-filter for electrons that are transmitted across it, which has been ascribed to the 7 left-handed $\alpha$-helices that define its secondary conformation. Combining spin-filtering with optoelectronic properties of proteins, which rely strongly on their conformations, is a fascinating opportunity to understand the role of spin in biology and, possibly, to future "biospintronic" devices.

Quite a few studies of protein ETp both at the nanoscale and macroscale have been made, as discussed in this report. Large area monolayer structures can in principle be integrated in future devices, in contrast to what seems (at least presently) possible for single proteins. Therefore, the fabrication of macroscopic electrical junctions, described in this report, which use protein monolayers of geometric areas ranging from $10^2$ to $2\times10^5$ $\mu m^2$ (as defined by the smallest electrode), constitute an important step towards using proteins as electronic components. In the future, further miniaturization to a few $\mu m^2$ areas may well become possible. At the same time working at the nanoscale and probing ETp of single (or few) proteins can elucidate details about protein function that may be averaged out in macroscopic measurements of the ensemble. In a biological system, proteins are usually not in a monolayer structure, and are often isolated (non-aggregated), implying that single molecule studies could offer more accurate information on actual protein function if it is present in an environment that mimics their natural state (e.g., a membrane protein in low concentrations in a monolayer of detergent). Thus, while unlikely to become part of bioelectronic devices in the near future, studies at the single protein level are important for obtaining fundamental understanding to make bioelectronic devices.

Presently most single molecule studies rely on nanoscale SPM techniques, apart of some studies on break junctions described above. However, the conductance data for proteins (and those for conjugated organic molecules) have a much larger degree of variability for nanoscopic junctions than for macroscopic ones,[47] even though both show the same trends. ETp efficiency of proteins is remarkably high when compared to ET rates in proteins and to the transport efficiency through conjugated molecules.

While the ETp studies reviewed here provide information that may well be necessary for incorporating proteins into solid-state platforms, the question remains what have we learned, and what can we learn about the transport mechanism across these proteins from these measurements. Compared to organic molecules, what allows electron transport across proteins to be more efficient than expected? How does electrode-protein (and protein-protein) coupling affect solid-state electronic transport? How can electron transport across relatively long distances (≥ 6 nm) observed in proteins be temperature-independent? Although it is too ambitious to hope that solid state bioelectronics research can answer all these questions in the short term, progress towards understanding the connections between transport efficiency, protein-electrode coupling, and temperature-independent transport are good places to start.

While there is a clear difference between the ionic transport-coupled electron transfer in biology and all-electronic transport in the artificial solid-state - like systems for future bioelectronics, an understanding of ETp may be relevant to ET because the presence of efficient transport channels in a protein can be expected to affect both. This possibility is intriguing within the context of the earlier mentioned (section 4.1), surprising long-distance electron transport in pili of *G. sulfurreducens* and in nanowires *of S. oneidensis*.

Findings in various report on ETp in/through proteins can be linked back to the proteins' innate function, strengthening the notion that at least some of the biological functionality can be studied with ETp. The importance of cofactors has been demonstrated in redox-active and redox-inactive proteins, illustrating that cofactors are not simple hopping sites. Possibly, the cofactors' effects are related to the conformation and/or charge envelope, as well as to the electronic energy levels of the protein, relevant to transport. The interactions of cofactors with proteins that affect ETp may be relevant to protein function, and understanding how these changes occur may shed light on transport mechanisms and pathways. An important step to understand the effects of cofactors will be to develop the ability to apply gate voltages to protein electronic junctions. While some reports of this exist (and are discussed in this report), it has not yet been possible to achieve gating in a manner that is clear enough so that it can be readily reproduced.

We have experimentally demonstrated that the interaction of proteins with biological binding partners influences ETp characteristics. A possible next step is to determine if ETp changes are related to the degree of conformational change caused by binding, and to determine how ETp pathways are altered.

Although high-resolution SPM imaging allows observing single proteins, it does not always guarantee isolation. This is important not only to prevent aggregation, but also to enable future studies where the same protein can be studied under different conditions, as is usually the case in biology. Nanolithography has helped create nanoarrays that allow both isolation and indexing for enhanced studies, but control of protein conformation remains a challenge. In-depth understanding of protein structure and conformation after immobilization is indeed a great challenge. Protein conformation is highly dynamic around room temperature, and movements of different structural elements have been demonstrated to affect ETp. Currently, spectroscopic techniques can offer snapshots of some general structural elements, but this is far from the ability to trace pathways of electronic transfer or transport. To understand real-time conformational changes, and how they relate to ETp, more quantitative molecular modeling of the protein conformation in the monolayer environment will be required.

In closing, we note that, although the worlds of molecular life sciences and electronics are often separated by a large divide, the growing mutual interest of these areas in each other to address common challenges, is contributing to the creation of new scientific and technological disciplines. Harnessing the unique chemical and physical properties of proteins could allow devices that can be fully integrated with biological systems. We hope that this report will boost interdisciplinary discussions and studies of topics that include bio- and bio-inspired -electronics, with an eye also towards future implantable bio-compatible integrated circuits and innovative healthcare applications.

## Acknowledgements

DC, MS and IP thank the Minerva foundation (Munich), the Israel Science Foundation, the Nancy and Stephen Grand Centre for Sensors and Security, the Kimmelman Center for Biomolecular Structure and Assembly, the GMJ Schmidt Minerva Centre for Supramolecular Architecture and the Benoziyo Endowment Fund for the Advancement of Science, for partial support, the students and postdoctoral fellows that they had the pleasure to work with over the past decade and a half and many colleagues, esp. H. B. Gray (Caltech), for stimulating discussions. DC's research is made possible in part by the historic generosity of the Harold Perlman family. DL thanks the US National Institutes of Health, the West Virginia Shared Research Facilities, and the US National Science Foundation for past support, which allowed him to embark on research in bioelectronics and supported CB's research. The support of the Rosi and Max Varon Visiting Professorship at the Weizmann Institute of Science, which allowed DL to spend time at the WIS exploring some of the ideas discussed in this article, is also gratefully acknowledged. SM thanks the Council for Higher Education (Israel) for a postdoctoral research. CB thanks the US National Institutes of Health and National Science Foundation IGERT program, which supported his work on protein electronics. M.S. holds the Katzir-Makineni Chair in Chemistry; D.C. holds the Rowland and Sylvia Schaefer Chair in Energy Research.